\documentclass[twocolumn]{aastex631}

\usepackage{graphicx}
\usepackage{amsmath}
\usepackage{natbib}
\usepackage{amssymb}

\newcommand{\pfrac}[2]{\left( \frac{#1}{#2} \right)}
\newcommand{\be}{\begin{equation}}
\newcommand{\ee}{\end{equation}}

\defcitealias{Linial&Metzger23}{LM23}

\newcommand{\uph}{u_{\rm ph}}
\newcommand{\nph}{n_{\rm ph}}
\newcommand{\ovN}{\tilde{N}_{\rm ph}}

\newcommand{\nel}{n_{\rm e}}
\newcommand{\nI}{n_{\rm I}}
\newcommand{\nb}{n_{\rm b}}

\newcommand{\me}{m_{\rm e}}
\newcommand{\mpr}{m_{\rm p}}
\newcommand{\vej}{v_{\rm ej}}
\newcommand{\Mej}{M_{\rm ej}}

\newcommand{\tauT}{\tau_{\rm T}}
\newcommand{\sigmaT}{\sigma_{\rm T}}
\newcommand{\kappaT}{\kappa_{\rm T}}
\newcommand{\alphaT}{\alpha_{\rm T}}
\newcommand{\alphaff}{\alpha_{\rm ff}}
\newcommand{\epsff}{\epsilon_{\rm ff}}

\newcommand{\Rstar}{R_{\star}}
\newcommand{\vstar}{v_{\star}}

\newcommand{\tdiff}{t_{\rm diff}}
\newcommand{\tdyn}{t_{\rm dyn}}

\newcommand{\Erad}{E_{\rm rad}}
\newcommand{\Ekin}{E_{\rm kin}}

\newcommand{\thetaBB}{\theta_{\rm BB}}
\newcommand{\TBB}{T_{\rm BB}}
\newcommand{\Te}{T_{\rm e}}
\newcommand{\TC}{T_{\rm C}}
\newcommand{\afs}{\alpha_{\rm fs}}
\newcommand{\gff}{\overline{g}_{\rm ff}}

\newcommand{\tauff}{\tau_{\rm ff}}

\newcommand{\tCompt}{t_{\rm Compt}}

\newcommand{\Msq}{{\cal M}_{\rm u}^2}

\shorttitle{X-ray Eruptions from Star-Disk Collisions}
\shortauthors{Vurm, Linial \& Metzger}

\begin{document}

\title{Radiation Transport Simulations of Quasi-Periodic Eruptions from Star-Disk Collisions}

\correspondingauthor{Indrek Vurm}
\email{indrek.vurm@gmail.com}

\author[0000-0003-1336-4746]{Indrek Vurm}
\affil{Tartu Observatory, University of Tartu, T\~oravere, 61602 Tartumaa, Estonia}

\author[0000-0002-8304-1988]{Itai Linial}
\affil{Institute for Advanced Study, 1 Einstein Drive, Princeton, NJ 08540, USA}
\affil{Department of Physics and Columbia Astrophysics Laboratory, Columbia University, New York, NY 10027, USA}

\author[0000-0002-4670-7509]{Brian D.~Metzger}
\affil{Department of Physics and Columbia Astrophysics Laboratory, Columbia University, New York, NY 10027, USA}
\affil{Center for Computational Astrophysics, Flatiron Institute, 162 5th Ave, New York, NY 10010, USA} 

\begin{abstract}
Periodic collisions between a star on an inclined orbit around a supermassive black hole and its accretion disk offers a promising explanation for X-ray ``quasi-periodic eruptions'' (QPEs). Each passage through the disk midplane shocks and compresses gas ahead of the star, which subsequently re-expands above the disk as a quasi-spherical cloud. We present spherically symmetric Monte Carlo radiation transport simulations which follow the production of photons behind the radiation-mediated shock, Comptonization by hot electrons, and the eventual escape of the radiation through the expanding debris. Such one-dimension calculations are approximately justified for thin disks (scale-height $h \lesssim few \times R_{\star}$), through which the star of radius $R_{\star}$ passes faster than the shocked gas can flow around the star.  For collision speeds $v_{\rm coll} \gtrsim 0.15 c$ and disk surface densities $\Sigma \sim 10^{3}$ g cm$^{-2}$ characteristic of those encountered by stellar orbits consistent with QPE recurrence times, the predicted transient light curves exhibit peak luminosities $\gtrsim 10^{42}$ erg s$^{-1}$ and Comptonized quasi-thermal (Wien-like) spectra which peak at energies $h\nu \sim 100$ eV, broadly consistent with QPE properties. For these conditions, gas and radiation are out of equilibrium, rendering the emission temperature harder than the blackbody value due to inefficient photon production behind the radiation-mediated shock. The predicted eruptions execute counterclockwise loops in hardness-luminosity space, a hallmark of QPE observations.  Alternatively, for higher disk densities and/or lower shock velocities, QPE emission could instead represent the comparatively brief phase shortly after shock break-out, though in this case the bulk of the radiation is thermalized and occurs in the ultraviolet instead of the X-ray band.  In either scenario, reproducing the observed eruption properties (duration, luminosity, temperature) requires a large radius $R_{\star} \gtrsim 10R_{\odot}$, which may point to inflation of the star's atmosphere from repeated collisions.
\end{abstract}


\section{Introduction}

Optical and X-ray time-domain surveys have over recent years uncovered a growing sample of regular or periodic flaring sources spatially coincident with the nuclei of distant galaxies. Among these new event classes are the X-ray ``quasi-periodic eruptions'' (QPEs), which recur over timescales from a couple hours to a couple days (e.g., \citealt{Miniutti+19,Giustini+20,Arcodia+21,Chakraborty+21,Arcodia+24,Nicholl+24}), as well as flaring phenomena with longer recurrence periods ranging from weeks (e.g., \citealt{Guolo+23}) to years (e.g., \citealt{Payne+21,Wevers+22,Liu+22,Malyali+23}). While the physical origins of these different types of periodic nuclear transients remains under debate, unlocking their mysteries may offer fresh probes of the dynamics of stars and compact objects in close proximity to the supermassive black hole (SMBH) and its accretion flow.

The recurrent eruptions from QPE systems are characterized by durations of hours and peak luminosities $\sim 10^{41}-10^{43}$ erg s$^{-1}$, making them visible over the frequently more luminous but softer ``quiescent'' X-ray emission of the SMBH accretion disk (the more slowly varying emission from which is also present in many QPE systems). The spectra of QPE eruptions are quasi-thermal, with temperatures $\approx 100-200$ eV \citep{Miniutti+19,Giustini+20,Arcodia+21,Chakraborty+21,Arcodia+22,Miniutti+23,Webbe&Young23}.  A unifying feature of all QPEs to date are the counterclockwise ``hysteresis loops'' the eruptions traverse in the space of temperature (hardness) and luminosity (e.g., \citealt{Arcodia+22}). The modest stellar masses of QPE host galaxies point to SMBH with relatively low masses, $M_{\bullet} \lesssim 10^{6.5}M_{\odot}$ (e.g., \citealt{Wevers+22}).

Most theoretical models for QPEs invoke invoke a star or compact object on a tight orbit around the SMBH, which once or twice per orbit interacts in some way with the SMBH, its accretion disk, or other stars in the galactic nucleus
\citep{Zalamea+10,Metzger&Stone17,King20,Sukova+21,Metzger+22,Zhao+22,King22,Krolik&Linial22,LuQuataert23,Linial&Sari23,Linial&Metzger23,Franchini+23,Tagawa&Haiman23}. However, an important clue unaddressed by many of these models is the regular alternating behavior, observed in at least two QPE sources \citep{Miniutti+19,Arcodia+21,Miniutti+23}, in which the temporal spacing between consecutive bursts varies back and forth by around 10\%. Flares that precede longer recurrence intervals are also distinguished by appearing brighter than those flares preceding shorter intervals (e.g., \citealt{Miniutti+23}).

\citet[hereafter \citetalias{Linial&Metzger23}]{Linial&Metzger23} show using analytic arguments that many of the observed properties of X-ray QPEs (period, luminosity, duration, emission temperature, occurrence rates in galactic nuclei) can plausibly be reproduced in a scenario in which a main-sequence star on a mildly eccentric inclined orbit collides twice per orbit with a gaseous accretion disk (see also \citealt{Xian+21,Tagawa&Haiman23,Franchini+23,Zhou+24}). Such stellar ``extreme mass-ratio inspiral'' (EMRIs), which migrate towards the SMBH via gravitational wave emission, are predicted to be relatively common in galactic nuclei on orbital periods of several hours (e.g., \citealt{Linial&Sari22}). The gaseous accretion disk with which the star interacts is either produced by mass stripped from the star itself during its interaction with the disk (e.g., \citealt{LuQuataert23,Linial&Metzger24,Yao+24}) or was created by a recent but otherwise unrelated tidal disruption event (TDE) involving a different star (e.g., \citealt{Miniutti+19,Chakraborty+21,Quintin+23,Nicholl+24}). The oscillating long-short recurrence behavior seen in some QPE sources results from the longer time the star spends between disk collisions on the apocenter side of the disk. 

\citetalias{Linial&Metzger23} postulate that the eruptions themselves are powered by the hot disk material shocked by the star, which due to its high pressure expands outwards from either or both sides of midplane, akin to dual miniature supernova explosions (e.g., \citealt{Ivanov+98}).  The eruption duration is set by the photon diffusion time through the expanding debris cloud, while the radiated energy is set by the thermal energy deposited by the star-driven shock, accounting for adiabatic losses from the collision site near the midplane to larger radii in the outflow where radiation is no longer trapped. 

As in the case of supernova shock breakout (e.g., \citealt{Weaver76}), gas and radiation in the expanding shocked disk debris may not be in equilibrium. In particular, because of the low gas density and rapid expansion rate of the debris, inefficient photon production can result in harder emission temperatures than would be obtained for blackbody emission (e.g., \citealt{Nakar&Sari10}). Such high temperatures could help explain how X-ray QPEs become detectable over the luminous but softer quiescent disk emission (e.g., \citealt{Miniutti+23}). 

Photon production occurs via free-free emission in the hot ejecta and is dominated by the earliest times after the star/shock passage, when the gas density is highest (\citetalias{Linial&Metzger23}). This renders the emission properties sensitive to the detailed post-shock structure and the hydrodynamics of the earliest phases of the debris compression/decompression, which must be modelled more carefully than allowed by simple analytic estimates to obtain reliable predictions. Compton scattering of the radiation by hot electrons in the expanding fluid can also alter the photon spectrum in a way that is challenging to fully account for analytically. Motivated by the need for quantitative light curve and spectral models of star-disk collisions to compare to QPE observations, here we present Monte Carlo radiation transfer (MCRT) hydrodynamic simulations, which follow self-consistently photon creation and Comptonization processes in the shocked stellar debris and the escape of radiation as this material decompresses out of the disk midplane.

\begin{figure*}
\begin{center}
\includegraphics[width=0.99\textwidth]{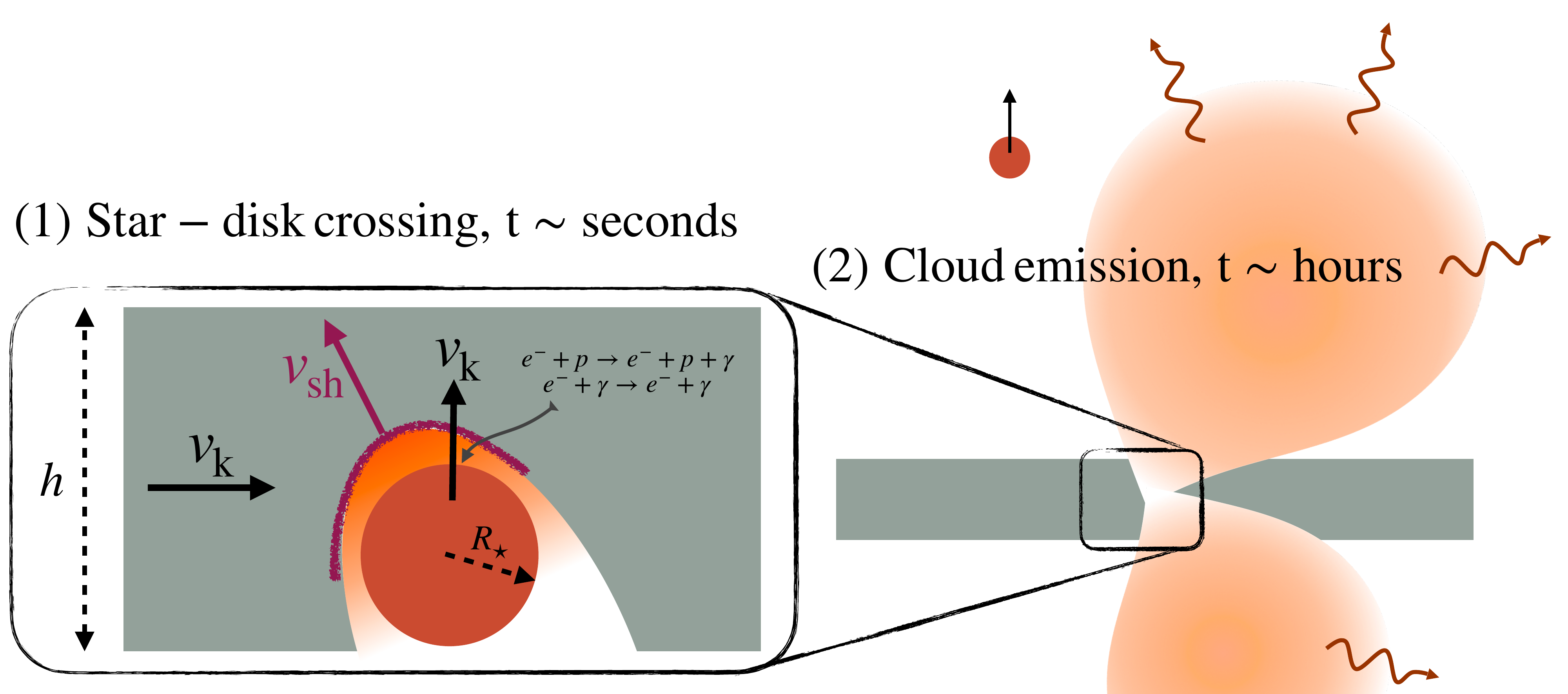}
\end{center}
\caption{Schematic illustration of the phases of disk-star collisions explored in this paper: (1) star of effective radius$^{\dagger}$ $R_{\star}$ passes supersonically through the disk midplane of vertical thickness $\simeq h$, generating a radiation-mediated shock ahead of the star which heats the disk material and collects it into a narrow cap of thickness $\ll h$. Photon production in the shocked gas occurs as a result of free-free emission at $h\nu \sim kT$ and Compton upscattering of free-free photons with $h \nu \ll kT$. (2) After the star emerges from the midplane, the now over-pressurized shocked disk material expands freely into space above (and, potentially, also below) the midplane in a quasi-spherical outflow. Photon production during most of this expansion phase has frozen out but adiabatic losses are important. The optical depth through the expanding debris eventually decreases sufficiently to allow radiation to escape to a distant observer, typically on a timescale of hours. $^{\dagger}$The effective cross section for purposes of interacting with the disk may exceed the original size of the unperturbed star as a result of disk-collision heating puffing up its surface layers and the streams of unbound mass from these layers that follow the star's original orbit \citep{Yao+24}.}
\label{fig:schematic}
\end{figure*}

This paper is organized as follows. In Sec.~\ref{sec:LM23} we overview the physical picture of disk-star collisions and provide analytic estimates of the star and disk properties which motivate parameters entering the calculations to follow. In Sec.~\ref{sec:expansion} we present MCRT calculations of the light curve and spectral evolution, first under the idealization of a uniform freely-expanding (homologous) ejecta cloud, as approximately satisfied by the late-time evolution of shocked disk debris. While this calculation provides a baseline description of the key physical processes at work, the predicted emission turns out to be sensitive to the initial photon distribution as determined by the earlier shock-crossing phase. To explore the latter, in Sec.~\ref{sec:shock} we present 1D plane-parallel MCRT calculations of radiation-mediated shocks which quantify the photon creation process; in addition to motivating the initial conditions for the subsequent ejecta expansion phase, these calculations are directly amenable to analytic estimates which verify their accuracy (Appendix \ref{sec:app:sh}, \ref{sec:app:pprod}). Finally, in Sec.~\ref{sec:model} we present MCRT hydrodynamic simulations which follow both the initial compression and subsequent re-expansion of, and the escape of radiation from, the shocked disk material, in idealized 1D spherical geometry (as is approximately justified if the star passes through the disk faster than the shocked gas can flow laterally around the star). The reader mainly interested in the results should skip ahead to this section. In Sec.~\ref{sec:discussion} we discuss the implications of our results for observed QPE and compare them to previous simpler analytic estimates \citetalias{Linial&Metzger23}. In Sec.~\ref{sec:conclusions} we summarize our conclusions.
Figure~\ref{fig:schematic} illustrates the phases of star-disk collisions explored in this work (see also \citetalias{Linial&Metzger23}).

\section{Disk-Star Collisions}
\label{sec:LM23}

We consider a central SMBH of mass $M_{\bullet} = 10^{6}M_{\bullet,6}M_{\odot}$ accreting steadily at a rate $\dot{M} = \dot{m}\dot{M}_{\rm Edd}$, where
$\dot{M}_{\rm Edd} \equiv L_{\rm edd}/(\epsilon c^{2})$ is the Eddington accretion rate for a characteristic radiative efficiency of $\epsilon = 0.1$ and $L_{\rm Edd} \simeq 1.5\times 10^{44}M_{\bullet,6}$ erg s$^{-1}$.  In the inner regions of the disk of greatest interest, radiation pressure dominates over gas pressure. The aspect ratio of the disk at radii $r \gg R_{\rm g} \equiv GM_{\bullet}/c^{2}$ can then be written (e.g., \citealt{Frank+02}) 
\be
\frac{h}{r} \simeq \frac{3}{\epsilon}\frac{R_{\rm g}}{r}\frac{\dot{M}}{\dot{M}_{\rm Edd}} \approx 0.3\; \dot{m}\left(\frac{r}{100R_{\rm g}}\right)^{-1},
\label{eq:hoverr}
\ee
where $h$ is the disk thickness.  For steady accretion, the surface density of the disk is given by
\be
\Sigma \simeq \frac{\dot{M}}{3\pi \nu} \approx 5\times 10^{2}\,{\rm g\,cm^{-2}}\alpha_{-1}^{-1}\dot{m}^{-1}\left(\frac{r}{100R_{\rm g}}\right)^{3/2},
\label{eq:tauc}
\ee
where $\nu = \alpha (GM_{\bullet} r)^{1/2}(h/r)^{2}$ is the kinematic viscosity \citep{Shakura&Sunyaev73}, and we scale $\alpha = 0.1\alpha_{-1}$ to a characteristic value.

There are reasons to suspect that Eq.~\eqref{eq:hoverr} underestimates the disk thickness and Eq.~\eqref{eq:tauc} overestimates the disk surface density at a given accretion rate (i.e., at a given quiescent disk luminosity). GRMHD simulations of radiation-dominated SMBH accretion disks typically find lower midplane densities than predicted by the \citet{Shakura&Sunyaev73} model (e.g., \citealt{Jiang+16}), as would highly magnetized disks supported by magnetic pressure (for which the effective value of $\alpha \sim 1$; e.g., \citealt{Squire+24}). Late-time observations of TDE accretion disks (which sometimes host QPEs; e.g., \citealt{Miniutti+19,Nicholl+24}) also motivate higher rates of viscous evolution and accretion than predicted for a disk as thin as implied by Eq.~\eqref{eq:hoverr} (e.g., \citealt{vanVelzen+19,Mummery+24}). Furthermore, if the disk is not in steady-state as a result of thermal instabilities or energy injected from the stellar collisions \citep{Linial&Metzger24}, the gas surface density near the collision radius can also differ significantly from the steady-state prediction of Eq.~\eqref{eq:tauc}.

The midplane temperature is given by
\begin{eqnarray}
k T_{\rm c} &=& k \left(\frac{3\Sigma}{4a}\frac{GM_{\bullet}}{r^{2}}\frac{h}{r}\right)^{1/4} \nonumber \\
&\approx& 36 \,{\rm eV}\, \,\alpha_{-1}^{-1/4}M_{\bullet,6}^{-1/4}\left(\frac{r}{100R_{\rm g}}\right)^{-3/8},
\label{eq:Tc}
\end{eqnarray}
where $k$ is the Boltzmann constant and $a$ is the radiation constant. The ratio of photons to baryons in the midplane is approximately given by
\be
\frac{n_{\rm ph,0}}{n_{\rm 0}} \approx 64 \, \alpha_{-1}^{1/4}M_{\bullet,6}^{1/4} \left(\frac{\dot{m}}{0.1}\right)^{2}\left(\frac{r}{100R_{\rm g}}\right)^{-21/8},
\label{eq:photonbaryonratio0}
\ee
where $n_{\rm ph,0} \approx aT_{\rm c}^{4}/(2.7 kT_{\rm c})$ and $n_0 \simeq \rho_{\rm c}/m_p = \Sigma/(h m_p)$.

Absent interaction with orbiting bodies (i.e., when in ``quiescence''), the disk emission is dominated by radii near the innermost circular orbit $R_{\rm isco}$, with total luminosity
\be
L_{\rm Q} = \dot{m}L_{\rm Edd} \simeq 1.5\times 10^{44}\,{\rm erg\,s^{-1}}\dot{m} M_{\bullet,6}
\label{eq:LQ},
\ee
and characteristic emission temperature
\be
kT_{\rm Q} \approx k\left(\frac{3GM_{\bullet}\dot{M}}{8\pi \sigma R_{\rm isco}^{3}}\right)^{1/4} \simeq 105\,{\rm eV}\,\frac{\dot{m}^{1/4}}{M_{\bullet,6}^{1/4}}\left(\frac{R_{\rm isco}}{4R_{\rm g}}\right)^{-3/4}.
\label{eq:TQ}
\ee

The colliding body is fiducially taken to be a main-sequence star on a circular orbit. Collisions between the star and gaseous disk happen twice per orbit, such that the average interval between observed eruptions (the QPE period) is given by $P_{\rm QPE} = P_{\rm orb}/2$, neglecting orbital eccentricity. The collision radius within the disk corresponds to the star's semi-major axis,
\be
r_0 \approx 1.4 \times 10^{13}\,\,{\rm cm} \; \mathcal{P}_{\rm QPE,4}^{2/3}M_{\bullet,6}^{1/3} \approx 95 \, R_{\rm g} \, \pfrac{\mathcal{P_{\rm QPE,4}}}{M_{\bullet,6}}^{2/3} \,,
\ee
where $\mathcal{P}_{\rm QPE,4} \equiv P_{\rm QPE}/(4\,{\rm hr})$. Across the observed range of QPE periods $\mathcal{P}_{\rm QPE} \sim 2-40$ hr, we thus expect $r_0 \approx 50-500 R_{\rm g}$ for $M_{\bullet} \approx 10^{6}M_{\odot}.$

For low-mass stars that likely dominate the EMRI population, the stellar radius is generally comparable to a solar radius. However, for purposes of interacting with the disk, the effective radius can be substantially larger than that of its physical surface as the result of several effects. These include inflation of the star as a result of heating due to previous disk-star collisions (e.g., \citealt{Yao+24}) or tidal squeezing by the SMBH (e.g., \citealt{Linial&Quataert24}). A natural size scale for a gravitationally-bound cloud around a star of mass $M_{\star} = M_{\star,1}M_{\odot}$ is the Hills radius:
\be
r_{\rm H} = r_0 \left(\frac{M_{\star}}{M_{\bullet}}\right)^{1/3} \approx 2.0R_{\odot}\,\, \mathcal{P}_{\rm QPE,4}^{2/3}M_{\star,1}^{1/3}.
\label{eq:rH}
\ee
Even solar-mass stars can thus have effective radii $R_{\star} \approx r_{\rm H} \sim 1-10R_{\odot}$ for $P_{\rm QPE} \approx 2-40$ hr. Alternatively, the star's effective cross section could be augmented by a comet tail-like streams of stripped stellar debris from previous collisions that continue to follow the trajectory of the star (e.g., \citealt{Yao+24}). 

Both the star and the disk orbit the SMBH at roughly the Keplerian velocity $v_{\rm K} = (GM_{\bullet}/r_0)^{1/2}$ (phase 1 in Fig.~\ref{fig:schematic}). For a roughly head-on collision (inclination angle $\pi/2$), their collision speed is given by
\be
v_{\rm coll} \approx \sqrt{2}v_{\rm K} \approx 0.14c\,\left(\frac{r_0}{100R_{\rm g}}\right)^{-1/2}.
\label{eq:vc}
\ee
The mass of the disk material heated by the shock is approximately that intercepted by the star, viz.~
\be M_{\rm ej} \simeq 2\pi R_{\star}^{2}\Sigma \approx 3\times 10^{-5} M_{\odot}R_{\star,12}^{2}\left(\frac{\Sigma}{10^{4}\,{\rm g\,cm^{-2}}}\right),
\label{eq:Mej}
\ee
where the prefactor of 2 accounts for the additional mass swept up by the disk rotating into the path of the star. One can define an effective initial volume $V_{\rm sh} \approx \pi R_{\star}^{2}(h/7)$ of the layer of shocked gas ahead of the star, where the factor of $7$ is the compression ratio for a strong radiation-dominated shock. Equating $V_{\rm sh}$ to a spherical volume $(4\pi/3)R_0^{3}$ gives an estimate of the initial size of the shocked ejecta after the star emerges from the disk:
\be
R_0 \simeq \left(\frac{3}{28}hR_{\star}^{2}\right)^{1/3} \approx
11R_{\odot}M_{\bullet,6}^{1/3}\dot{m}^{1/3} R_{\star, 12}^{2/3}
\label{eq:R0}
\ee
For typical star and disk parameters, the values of $R_{\star}, h,$ and $R_0$ are thus of the same order of magnitude. 

The shocked layer of gas ahead of the star begins highly optically-thick ($\tau_0 \sim \rho\kappa_{\rm T}R_0 \gg 1$) and over-pressurized relative to its surroundings. However, as the shocked debris decompresses above the disk midplane, it will spread out roughly spherically (e.g., \citealt{Ivanov+98}; phase 2~in Fig.~\ref{fig:schematic}), quickly achieving an asymptotic speed comparable to the initial shock speed, $v_{\rm ej} \approx v_{\rm coll}$, as a result of the internal energy being reconverted into bulk kinetic energy via adiabatic expansion. This occurs on the expansion timescale,
\be
t_0 \approx \frac{R_0}{v_{\rm ej}} \approx
185
\,{\rm s}\,M_{\bullet,6}^{1/3}\dot{m}^{1/3}R_{\star,12}^{2/3}\left(\frac{r_0}{100R_{\rm g}}\right)^{1/2},
\label{eq:t0}
\ee
where we have used Eq.~\eqref{eq:vc} for $v_{\rm ej} = v_{\rm coll}$.

As the ballistic debris cloud spherically expands, its optical depth $\tau_{\rm T} \equiv \Sigma \kappa_{\rm T} \propto t^{-2}$ decreases with time, where $\kappa_{\rm T} \simeq 0.34$ cm$^{2}$ g$^{-1}$ is the Thomson scattering opacity. The bulk of the trapped radiation escapes once the photon diffusion and expansion timescales become comparable, once $\tau_{\rm T} \sim c/v_{\rm ej}$. This occurs on the characteristic diffusion timescale (e.g., \citealt{Arnett82})
\begin{eqnarray}
t_{\rm diff} &\approx& \left(\frac{\kappa_{\rm T} M_{\rm ej}}{4\pi c v_{\rm ej}}\right)^{1/2} \nonumber \\
&\approx& 3.7\times 10^{3}\,{\rm s}\,R_{\star,12}\left(\frac{\Sigma}{10^{4}\,{\rm g\,cm^{-2}}}\right)^{1/2}\left(\frac{r_0}{100R_{\rm g}}\right)^{1/4} \nonumber \\
&\approx& 
1.5\times 10^{3}\,{\rm s}\,\frac{\mathcal{P}_{\rm QPE,4}^{2/3}R_{\star,12}}{\alpha_{-1}^{1/2}\dot{m}^{1/2}M_{\bullet,6}^{2/3}}. \,
\label{eq:tpk}
\end{eqnarray}
The size of the ejecta cloud at the diffusion time is still much smaller than the collision radius within the disk, i.e., $R_{\rm diff} \approx v_{\rm ej}t_{\rm diff} \ll r_0$, thus motivating why the gravity of the SMBH has a negligible impact on the debris cloud evolution at times relevant to the observed emission.  As we shall discuss, both $t_0$ and $t_{\rm diff}$ play an important role in the light curve of star-disk collisions.

In summary, the star-disk collision scenario for QPEs predicts characteristic parameter values $R_{\star} \sim h \sim R_0 \sim 1-10R_{\odot}, M_{\rm ej} \sim 10^{-7}-10^{-5}M_{\odot}$, and $v_{\rm coll} \sim 0.03-0.3 c$, motivating those adopted in our numerical simulations to follow.

\section{Light Curve and Spectral Formation in Expanding Spherical Ejecta Cloud}
\label{sec:expansion}

As a means to introduce the basic physical processes at work, we first consider the emission from a homologous expanding spherical ejecta cloud of total mass $\Mej$ and expansion velocity $\vej$ at its outer boundary. This idealized set-up roughly approximates the expected state of the shocked disk material after it has begun to decompress above or below the disk midplane (phase 2 depicted in Fig.~\ref{fig:schematic}). After the initial shock passage, the subsequent evolution of the ejecta is passive with no further energy generation. The dissipated heat is carried mainly by radiation and will be degraded by adiabatic cooling in the initially highly opaque material, before gradually leaking out as the optical depth declines.

We further assume uniform radiation energy density $u_{\rm ph, 0}$, matter density $\rho_0$, and dimensionless temperature $\theta_0 = kT_0/\me c^2$ in the initial state (i.e., soon after the shock crossing). Even in the upstream medium of the unshocked disk, the number density of photons is much larger than the ion density, $n_{\rm ph, 0} \gg n_0 = \rho_0/(\mu \mpr)$ (Eq.~\eqref{eq:photonbaryonratio0}). The pressure and heat content are thus dominated by radiation, for which one would naively adopt an initial temperature $\theta_0 \sim \rho_0\vej^2/(n_{\rm ph, 0} \me c^2)$. However, this expression neglects photon production during the shock crossing phase itself (which we shall find are important; Sec.~\ref{sec:shock}).

\subsection{Bolometric luminosity}

Define an initial radius of the sphere to be $R_0$ (Eq.~\eqref{eq:R0}) and $t_0 = R_0/\vej$ as the initial time after which homologous expansion has roughly been achieved (Eq.~\eqref{eq:t0}). The initial scattering optical depth through the sphere obeys $\tau_{0} \gg c/\vej \gg 1$.

At some later time $t > t_0$ photons will have begun to diffuse out of the sphere from an outer shell of thickness $\Delta R \ll R$, which can be obtained as follows.
The characteristic escape time of photons from the shell is
\begin{align}
t_{\rm esc}(\Delta R) \approx \frac{\Delta R^2}{D} \,, 
\end{align}
where $D = c\lambda_{\rm T}/3 = c/(3\kappaT\rho)$ is the diffusion coefficient.
Equating $t_{\rm esc} = \tdyn = t$ and assuming $\rho = \rho_0 (t/t_0)^{-3}$ and $R = R_0 t/t_0$, one obtains
\begin{align}
\Delta R = \left(\frac{ct}{3\kappaT\rho}\right)^{1/2} = R_0 \left(\frac{c}{3\tau_0\vej}\right)^{1/2} \left(\frac{t}{t_0}\right)^{2}.
\label{eq:DeltaR}
\end{align}

The normalized radius $\xi \equiv r/R$ within the ejecta above which photons are able to diffuse out is
\begin{align}
\xi_{\rm diff} \equiv 1 - \frac{\Delta R}{R} \approx 1 -  \left(\frac{c}{3\tau_0\vej}\right)^{1/2} \frac{t}{t_0}
= 1 - \frac{t}{3\tdiff} \,,
\end{align}
where the diffusion time is (Eq.~\eqref{eq:tpk})
\begin{align}
    \tdiff =
    \left(\frac{\tau_0\vej}{3c}\right)^{1/2} t_0 = 
    \left(\frac{\kappa M_{\rm ej}}{4\pi c\vej}\right)^{1/2} \,.
    \label{eq:tdiff}
\end{align}

The escaping luminosity can be written as
\begin{align}
L_{\rm esc} = \frac{dE_{\rm esc}}{dt} \approx -\frac{d\Erad(<\xi_{\rm diff})}{d\xi_{\rm diff}} \frac{d\xi_{\rm diff}}{dt},
\end{align}
where
\begin{align}
\frac{d\Erad}{d\xi_{\rm diff}} = 4\pi R^3 \uph \, \xi_{\rm diff}^2.
\end{align}
This yields
\begin{align}
L_{\rm esc} \approx \frac{ 4\pi R^3 \uph \, \xi_{\rm diff}^2}{3\tdiff} = \frac{E_{\rm rad, 0}}{\tdiff} \, \xi_{\rm diff}^2 \left(\frac{t}{t_0} \right)^{-1},
\label{eq:Lbol}
\end{align}
where we have used $\uph \propto t^{-4}$ (adiabatic cooling) and $E_{\rm rad} = 4\pi R^3 \uph/3$.

\begin{figure*}[h]
\begin{center}
\includegraphics[width=0.99\textwidth]{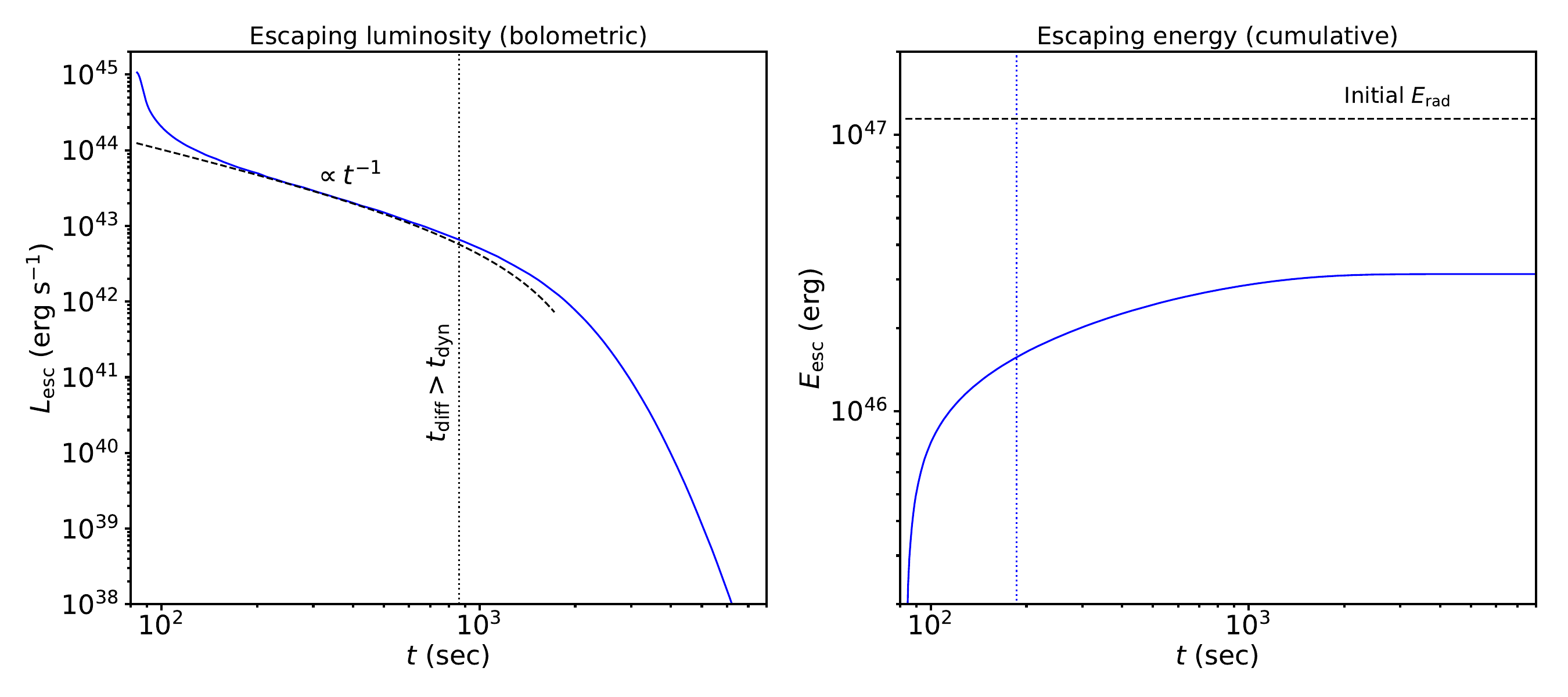}
\end{center}
\vspace{-5mm}
\caption{Bolometric luminosity (left panel) and cumulative escaping energy (right panel) versus time since ejection from a homogeneous expanding ejecta cloud (see text). The approximation for the light curve given by Eq.~(\ref{eq:Lbol}) is shown as a dashed line. The vertical line on the left panel shows the characteristic diffusion time after which photons can escape through the bulk of the ejecta (Eq.~\eqref{eq:tdiff}), while on the right panel it indicates when half of the total emitted radiation energy has escaped.}
\label{fig:bol}
\end{figure*}

\begin{figure*}[h]
\begin{center}
\includegraphics[width=0.49\textwidth]{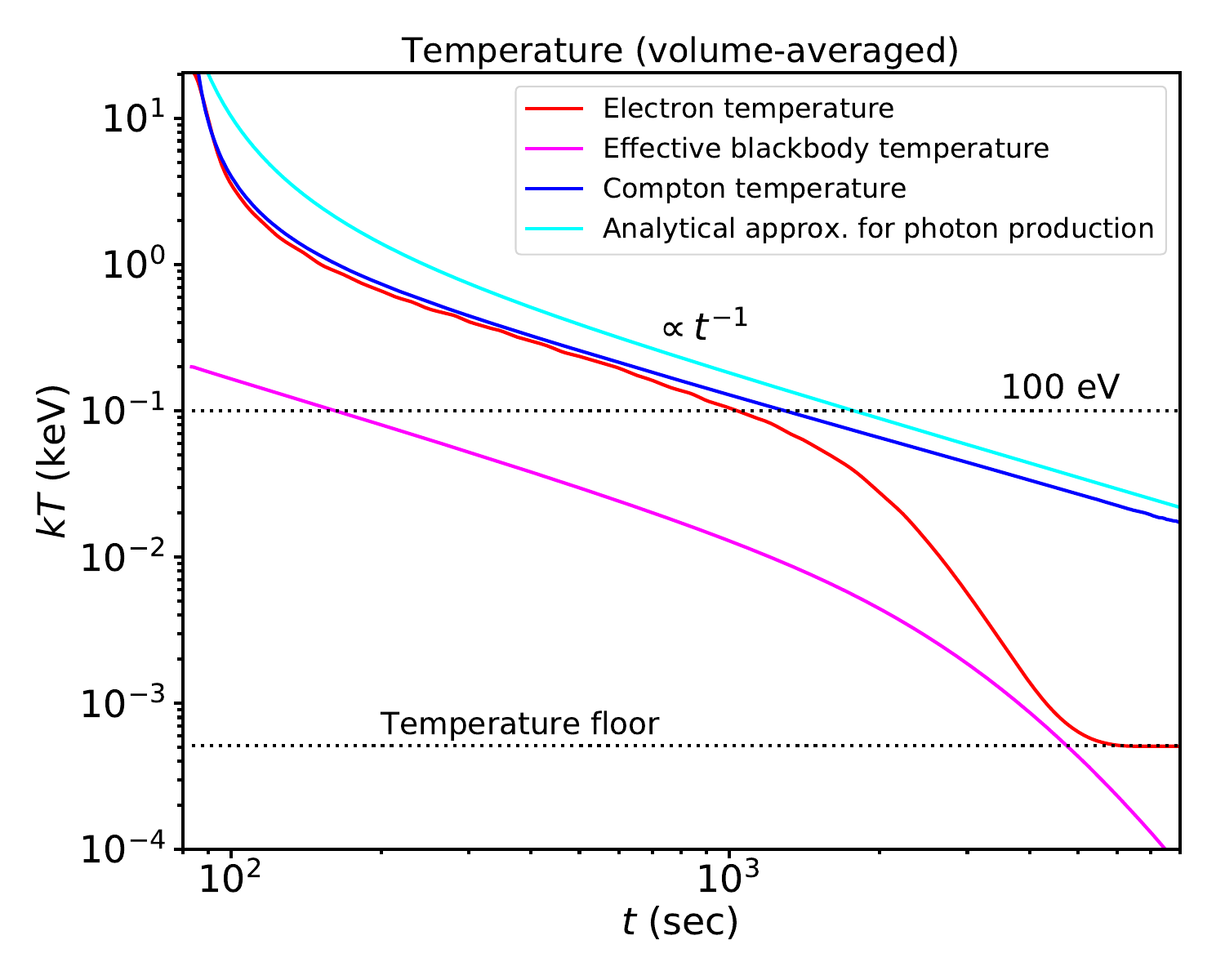}
\includegraphics[width=0.49\textwidth]{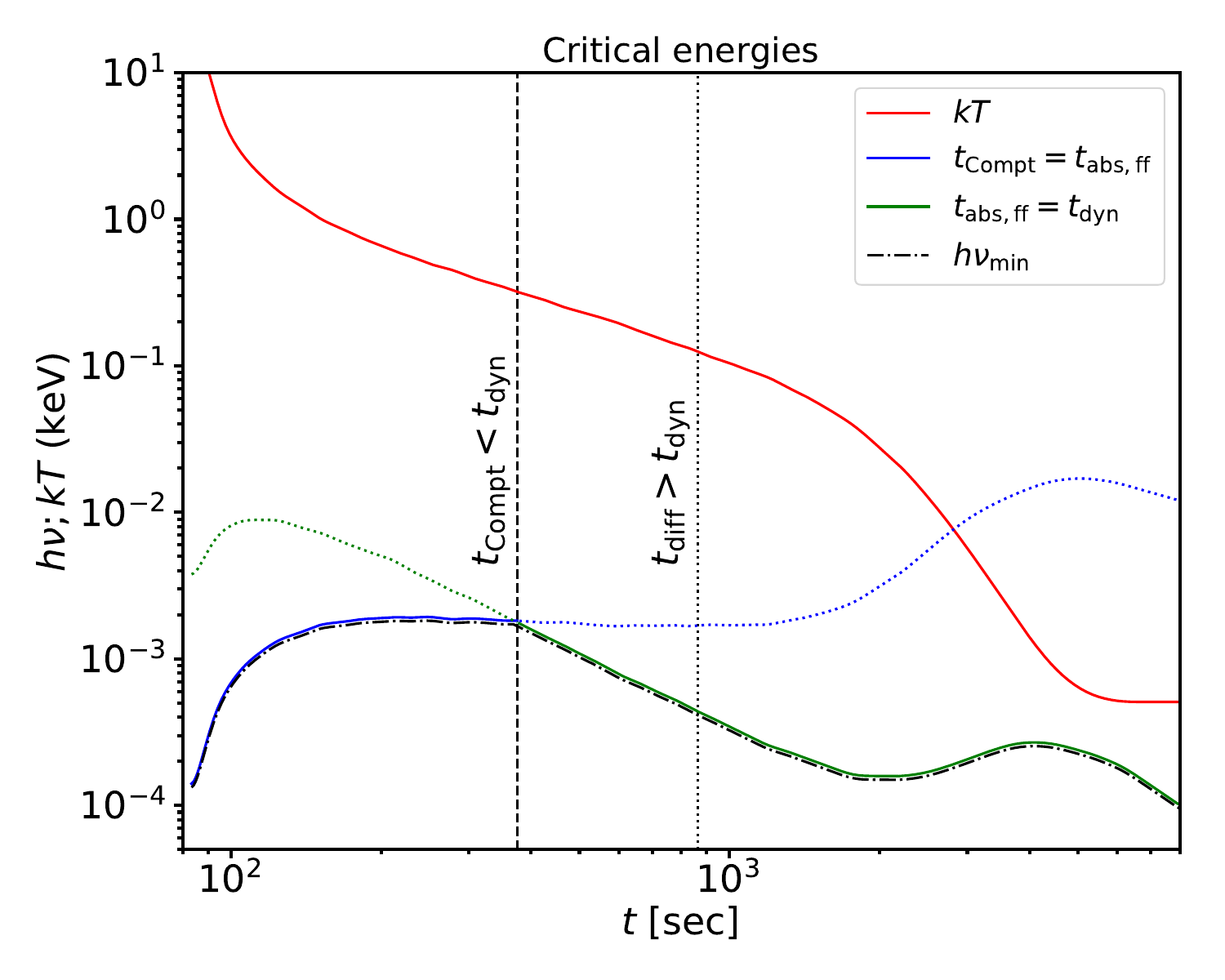}
\end{center}
\vspace{-5mm}
\caption{Evolution of temperature (left panel) and critical energies (right panel) within the expanding ejecta for the same model shown in Fig.~\ref{fig:bol}. The electron temperature $\Te$ (red) and Compton temperature $\TC$ (blue) are initially locked together and evolve rapidly over the first dynamical time due to efficient photon production, settling to a roughly constant ratio $\TC/\TBB$, where $\TBB= (\uph/a)^{1/4}$ is the blackbody temperature (magenta). The electron temperature falls out of equilibrium with radiation at $\sim 1000$~s. For comparison, the solution of Eq.~(\ref{eq:Nph}) with $\overline{E_1(x_{\rm min}/\theta)} = 5$ is shown by the cyan line (where $\theta = kT/m_{\rm e} c^2$). The right panel shows the energy range where photon generation can operate: photons above the blue line are Comptonized before being reabsorbed, above the green line the reabsorption timescale is longer than the dynamical time. Comptonization is efficient before the time marked by the vertical dashed line, diffusion out of the whole ejecta becomes efficient at the time shown by the vertical dotted line.}
\label{fig:theta}
\end{figure*}

The preceding argument is valid at $t \gg t_0$ but yet still sufficiently early that $\Delta R/R \lesssim 1$, i.e. before the nominal diffusion time $t_{\rm diff}$ when $\tau \approx c/\vej$ (Eq.~\eqref{eq:tpk}), after which radiation can freely escape the system and the light curve steepens/cuts off. For the assumed homogeneous ejecta, the bolometric light curve peaks at very early times $\sim t_0$ and declines monotonically thereafter, whereas since $t L_{\rm esc} \approx \mbox{const.}$, the bulk of the radiation energy is emitted over time span the order of the standard diffusion time (see \citealt{Nakar&Sari10} for a more detailed version of the above argument).

Figure~\ref{fig:bol} shows the bolometric light curve (left panel) and escaping total energy (right panel) for an example MCRT simulation of the above-described expanding homogeneous sphere of initial radius $R_0 = 5\times 10^{11}$~cm ($\simeq 7R_{\odot}$), mass $\Mej = 2.5\times 10^{-6} M_{\odot}$ (Eq.~\eqref{eq:Mej}), and maximum velocity $\vej = 0.2c$ (Eq.~\eqref{eq:vc}). The initial radiation energy $E_{\rm rad} \approx \Mej \vej^2/2 \approx 10^{47}$~erg
is comparable to the ejecta kinetic energy, as expected physically because the ejecta is accelerated mainly by internal PdV work. The initial Thomson optical depth through the sphere $\tau_0 \approx 1.5\times 10^3$ greatly exceeds $c/v_{\rm ej} \approx 5$, consistent with the radiation being initially trapped.

After an early transient phase $t \lesssim t_0$, the luminosity evolves as described above, whereby adiabatically-cooled radiation from successively deeper layers diffuses out of the ejecta with increasing time. The numerical luminosity profile decays slightly steeper than $t^{-1}$ predicted by the estimate \eqref{eq:Lbol}, with a turnover after the diffusion time when $\tdiff = \tdyn$. The total fraction of the initial radiation energy that is able to escape is $\sim 1/4$ (see Figure~\ref{fig:bol}, right panel), a factor of $\sim 3$ higher than the naive estimate obtained by accounting for adiabatic losses between the initial and diffusion times, i.e. $\Erad \propto (\tdiff/t_0)^{-1}$.
 
 A slightly more accurate estimate is obtained by intergrating Eq.~\eqref{eq:Lbol} over time, which yields (for $\tdiff \gg t_0$)
\begin{align}
E_{\rm esc} \approx \left(\frac{\pi c}{\kappa}\right)^{1/2}
 \Mej^{1/2} \vej^{3/2} R_0 \left[\ln\frac{3\tdiff}{t_0} - \frac{3}{2} \right],
\label{eq:Ebol}
\end{align}
resulting in values a factor $\sim 1.5 - 2$ below the numerical result.

\subsection{Photon production and thermalization}

Without photon production within the shock, the immediate post-shock temperature is determined by the shock velocity and the number of photons per material particle ahead of the shock. For fiducial parameters for QPEs (Sec.~\ref{sec:LM23}), one obtains $\nph/n \sim 10 - 100$
within the disk at the collision point prior to the star impact (Eq.~\eqref{eq:photonbaryonratio0}), which for collision speeds $v_{\rm coll} \approx v_{\rm ej} \sim 0.1 c$ implies post-shock temperatures of the order of $\sim 1$~MeV. We will see below that at high temperatures the photon production is much faster than the shock crossing time, in which case such high temperatures are likely never attained in reality behind the shock (Sec.~\ref{sec:shock}). In our setup, this fast photon production implies that the assumed initial photon number has little impact on the radiation field at later times of interest, such that we obtain the same results for any chosen (arbitrary) high value for the temperature at the start of our simulations. 

The photon number density in the three-dimensional expanding flow evolves according to:
\begin{align}
    \frac{1}{t^3} \frac{d}{dt} \left( t^3 \nph\right) = \dot{n}_{\rm ff},
    \label{eq:nph}
\end{align}
where the free-free photon-production rate is
\begin{align}
\dot{n}_{\rm ff}
= \left(\frac{8}{3\pi}\right)^{1/2} c\sigmaT\afs  Z^2 \nI \nel \, \theta^{-1/2} \, E_1\left(\frac{x_{\rm min}}{\theta}\right) \, \gff.
\label{eq:nff}
\end{align}
Here $\alpha_{\rm fs} \simeq 1/137$ is the fine-structure constant, $Z$ is the atomic charge ($Z \simeq 1$ for solar-composition gas), $\gff$ is the order-unity Gaunt factor, and $E_1$ the exponential integral. The minimum energy which contributes to the photon production, $x_{\rm min}\equiv h\nu_{\rm min}/\me c^2$, is determined by the condition $4c\alphaT\theta \gtrsim c\alphaff$, i.e. that emitted photons are Comptonized to higher frequencies before being reabsorbed (see \citealt{Nakar&Sari10}). Expressing temperature as $\theta = \uph/(3\me c^2\nph)$ and assuming $t\ll \tdiff$, i.e. that $\uph \propto V^{-4/3} \propto t^{-4}$ follows the adiabatic cooling law, one obtains $\theta = \theta_0 (t/t_0)^{-4} (\nph/n_{\rm ph,0})^{-1}$. Using this relation and defining the normalized photon number in a comoving volume element as $\ovN = (\nph/n_{\rm ph, 0})(t/t_0)^3$, one can recast Eq.~(\ref{eq:nph}) in the form
\begin{align}
    \ovN^{-1/2}\frac{d\ovN}{dt} = C n_{\rm ph, 0}^{-1} \, n_{\rm I, 0} \, n_{\rm e, 0} \, \theta_0^{-1/2}
    \left(\frac{t}{t_0}\right)^{-5/2} E_1\left(\frac{x_{\rm min}}{\theta}\right),
    \label{eq:Nph}
\end{align}
where $C \equiv (8/3\pi)^{1/2} c\sigmaT\afs  Z^2 \gff$. If $x_{\rm min}/\theta \ll 1$, the exponential integral on the right hand side is well approximated as $E_1(x_{\rm min}/\theta) \approx -\ln(x_{\rm min}/\theta) - \gamma$, where $\gamma$ is the Euler constant.

The $t^{-5/2}$ behaviour of the right hand side of Eq.~(\ref{eq:Nph}) implies that most of the photon-generation takes place at early times, roughly over the first dynamical time, after which the total photon number freezes out. In fact, provided that the condition $t_0 \dot{n}_{\rm ff,0} \gg n_{\rm ph, 0}$ holds, the final photon number (and the resulting temperature) is independent of the initial $n_{\rm ph, 0}$ and $\theta_0$. To see this, observe that at $t=t_0$, the right hand side is essentially $\dot{n}_{\rm ff,0}/n_{\rm ph, 0}$, so that if $t_0\dot{n}_{\rm ff,0}/n_{\rm ph, 0} \gg 1$ then the number of generated photons greatly exceeds the initial value, rendering the latter irrelevant. Also, using the definition of $\ovN$ and $\theta_0 = u_{\rm ph, 0}/(3\me c^2 n_{\rm ph, 0})$, one can show that the dependence on the initial photon density $n_{\rm ph, 0}$ and temperature $\theta_0$ in Eq~(\ref{eq:Nph}) can be replaced by the initial radiation energy density $u_{\rm ph, 0}$, independent of the initial number of available photons.

In the limit of small initial photon density, the asymptotic temperature at $t \gg t_0$ can be written by using the solution of Eq.~(\ref{eq:Nph}) as
\begin{align}
    \theta_{\star} \approx \frac{3\pi}{8(\afs\gff\overline{E}_1)^2}
    \left(\frac{\nel}{Z^2 \nI}\right)^2 \left(\frac{\vej}{c}\right)^2 \, \frac{l_{\rm rad, 0}^2}{\tau_{\rm T,0}^4}
    \left(\frac{t}{t_0}\right)^{-1},
    \label{eq:Tasymp}
\end{align}
where $l_{\rm rad, 0} = \sigmaT u_{\rm ph,0} R/(\me c^2)$ defines the initial radiation compactness and $\overline{E}_1$ is the average value of the logarithmic factor in Eq.~(\ref{eq:Nph}) within (approximately) the first dynamical time.

Equations $\eqref{eq:nph}-\eqref{eq:Tasymp}$ assume full thermalization is not achieved, as no compensating absorption term is included in Eq.~\eqref{eq:nph}. Whether this is justified can be checked {\it a posteriori} by comparing Eq.~(\ref{eq:Tasymp}) with the blackbody temperature 
\be
\thetaBB = \left(\frac{u_{\rm ph, 0}}{a}\right)^{1/4} \left(\frac{k}{\me c^2}\right)\left(\frac{t}{t_0}\right)^{-1},
\label{eq:thetaBB}
\ee
where $\theta > \thetaBB$ would imply incomplete thermalization and vice versa. Normalized to the fiducial parameters used in the simulation, one obtains
\begin{align}
    \frac{\theta_{\star}}{\thetaBB} \approx
    11 &\left(\frac{\gff}{2}\right)^{-2}\left(\frac{\overline{E}_1}{5}\right)^{-2} \zeta^{7/4}
    \left(\frac{\vej}{0.2c}\right)^{11/2} \nonumber \\
    &\left(\frac{\Mej}{2.5\times 10^{-6}M_{\odot}}\right)^{-9/4} \left(\frac{R_0}{5\times 10^{11}\,\mbox{cm}}\right)^{19/4},
    \label{eq:thetaratio}
\end{align}
where $\zeta \equiv E_{\rm rad, 0}/\Ekin$ is the initial ratio of thermal to kinetic energies (taken to be $\zeta = 1$ in the simulations\footnote{This choice is not entirely self-consistent, as the trapped radiation field would accelerate the ejecta by a factor $\sim 2^{1/2}$, a phase we have neglected by assuming ballistic expansion.}) and we have assumed solar composition material. Thus, we see that for high shock/ejecta speeds $v_{\rm ej} \gtrsim 0.1 c$, large effective collision radii $R_0$, and/or low ejecta masses $M_{\rm ej}$, the emission temperature can exceed the blackbody temperature as a result of inefficient photon production (see also \citetalias{Linial&Metzger23}).

\begin{figure*}
\begin{center}
\includegraphics[width=0.99\textwidth]{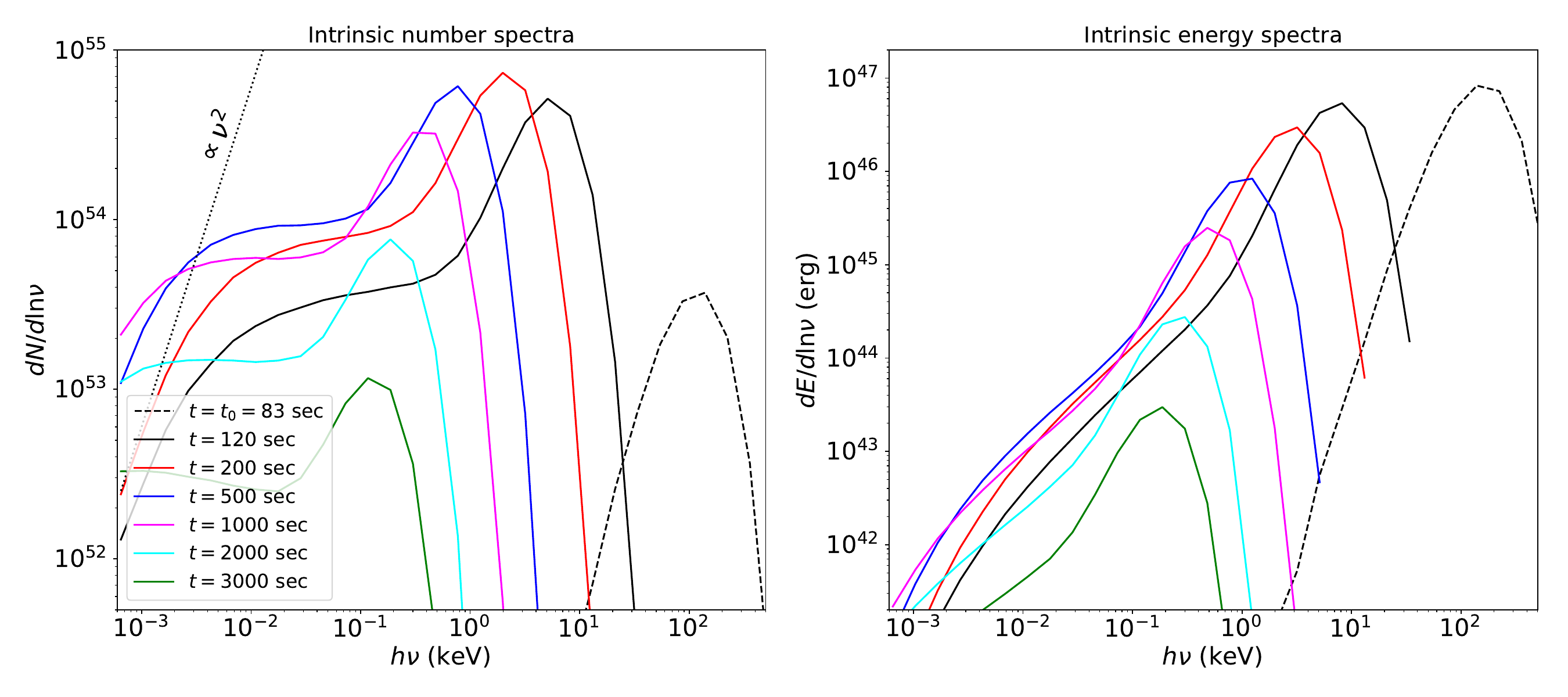}
\end{center}
\vspace{-5mm}
\caption{Evolution of volume-averaged (intrinsic) photon number (left panel) and energy spectra (right panel) for the same model shown in Fig.~\ref{fig:bol}. The initial spectrum (dashed line) follows a Wien shape with $kT=40$~keV. A dotted line indicates the Rayleigh-Jeans spectral slope $\propto \nu^2$ attained at the lowest photon energies.}
\label{fig:spec:intr}
\end{figure*}

\begin{figure}[h]
\begin{center}
\includegraphics[width=0.5\textwidth]{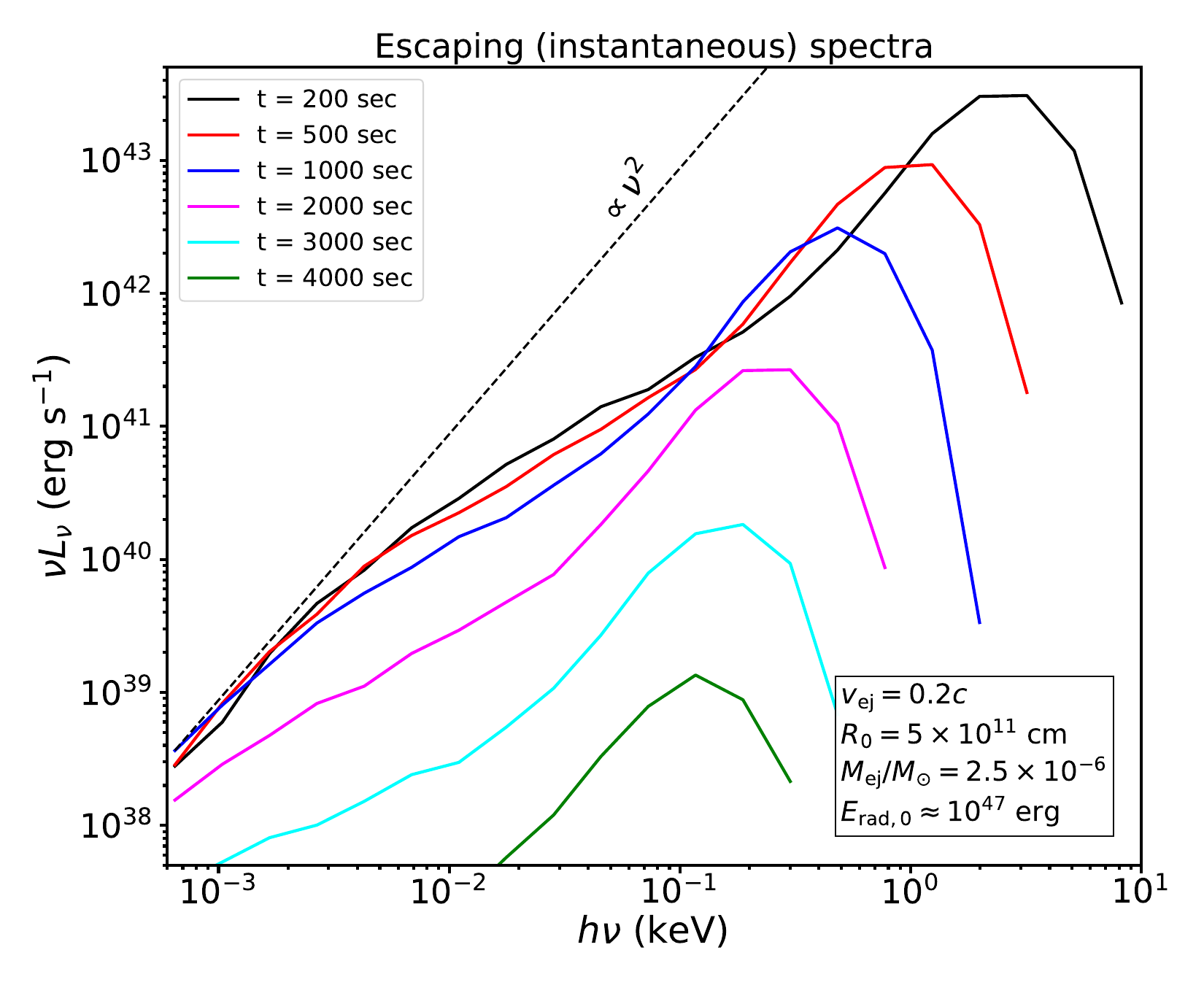}
\end{center}
\vspace{-5mm}
\caption{Evolution of the escaping spectrum for the same model shown in Fig.~\ref{fig:bol}. The dashed line indicates the low-energy spectral slope $\nu L_{\nu} \propto \nu^2$ at $\nu < \nu_{\rm min}$, in accordance with Eq.~\eqref{eq:Lnu}.}
\label{fig:spec:esc}
\end{figure}

Figure~\ref{fig:theta} (left panel) shows the evolution of the volume-averaged electron temperature $\Te$, Compton temperature $\TC$ of the radiation, as well as the blackbody temperature $\TBB = (\uph/a)^{1/4}$.
The initial rapid decline of the electron and Compton temperatures indicates efficient photon production during the first dynamical time. After a couple of dynamical times, the evolution of $\TC$ follows the adiabatic cooling law $\TC\propto t^{-1}$ with negligible further photon production. Note that this behavior persists even at $t>\tdiff$, indicating that the photons that have not yet escaped continue to experience energy losses in the diverging flow. With the chosen parameters in the simulation, full thermalization is not reached, i.e. $\Te > \TBB$ while the electrons and photons are still coupled (see below).

The Compton temperature, defined as the electron temperature at which the net energy transfer between matter and the radiation field vanishes, follows the actual $\Te$ closely for several dynamical times. The electrons eventually fall out of equilibrium with the radiation field when free-free energy losses overcome the net heating of the electrons by (direct) Compton scattering with the radiation field, after which $\Te$ drops rapidly down to the temperature floor set within the simulation\footnote{This regime is not treated self-consistently in our calculations, since only free-free cooling is considered (we neglect line-cooling, recombination, and other processes that become relevant at low temperatures).}.

The right panel of Figure~\ref{fig:theta} shows the critical energies and timescales relevant to photon production as well as spectral formation (below). Emitted free-free photons at $x\gtrsim x_{\rm min}$ (above the black dash-dotted line) avoid reabsorption either by being upscattered to higher energies where absorption probability is lower (above the blue line), or because the free-free absorption timescale is longer than the dynamical time (above the green line)\footnote{The relevant condition for the latter before the diffusion time $\tdiff = \tdyn$  is $\tauff c/\vej < 1$ rather than $\tauff < 1$, since scattering traps the photons within the ejecta so that they can only avoid reabsorption if $\alphaff c \tdyn < 1$, which is equivalent to the former condition.}. The condition $x_{\rm min} \lesssim x \lesssim \theta$ defines the effective photon production range, the corresponding factor $E_1({x_{\rm min}/\theta}) \approx 5$ at $t \approx t_0 + \tdyn$ in Eq.~(\ref{eq:Nph}). Comptonization of the emitted photons to the spectral peak is efficient as long as $\tCompt \ll \tdyn$ (vertical dashed line in Figure~\ref{fig:theta}, right panel), i.e. only for a few dynamical times.

\begin{figure*}
\begin{center}
\includegraphics[width=0.99\textwidth]{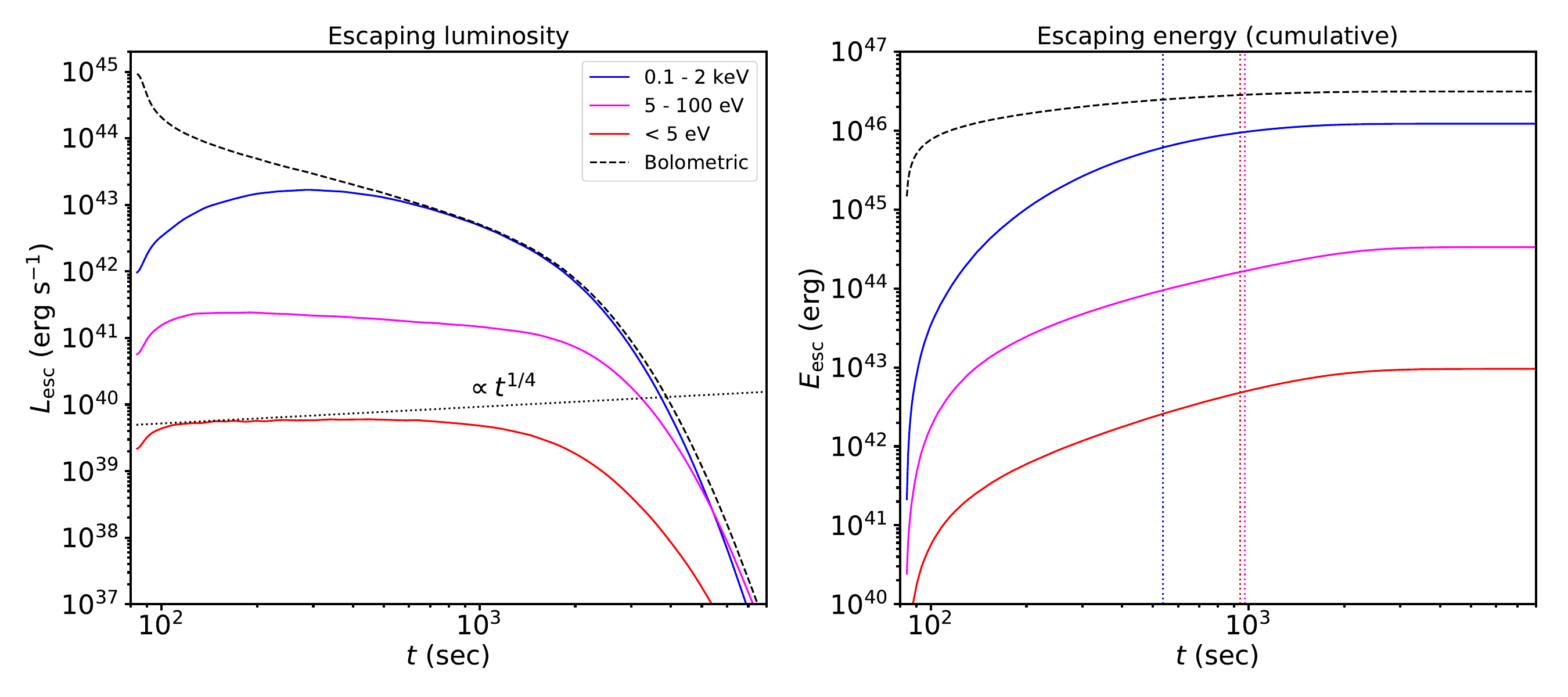}
\end{center}
\vspace{-5mm}
\caption{Light curves in different energy ranges, corresponding roughly to X-ray, UV and optical bands. The vertical dotted lines correspond to times by which half of the energy in a given band has been emitted. The $t^{1/4}$ slope indicated by Eq.~(\ref{eq:Lnu}) is shown by the dotted line on the left panel.}
\label{fig:LC:range}
\end{figure*}

\subsection{Spectra and light curves}

\label{sec:expansion:spectra}

The volume-averaged radiation spectra within the ejecta at different times are shown in Figure~\ref{fig:spec:intr}. The initial spectrum at $t = t_0 = 83$~s (dashed line) was chosen to have a Wien shape with a temperature $\sim 40$~keV, though our results are not sensitive to this assumption because rapid photon production and efficient Comptonization shift the spectral peak towards lower energies by almost two orders of magnitude during the first dynamical time. The classical spectral shape of saturated Comptonization of a low-energy photon source is attained within a light crossing time, which incidentally suggests efficient photon generation will already have taken place during the shock crossing phase (neglected here, but explored in Sec.~\ref{sec:shock}). The spectrum is characterized by a Rayleigh-Jeans slope at the lowest energies, a transition to
$I_{\nu} \propto dN/d\ln\nu \propto \nu^0$ at intermediate $x\approx x_{\rm min}$, and the Wien-like peak near $x \sim \theta$.

After the first dynamical time, significant photon generation subsides, and subsequent evolution of the spectral peak is mainly governed by adiabatic cooling. Around the same time, Comptonization exits the saturated regime,
however the spectrum roughly maintains the shape established in the previous stages, showing a very subtle softening in the intermediate range $x_{\rm min} < x < \theta$ that corresponds to optically this free-free emission without efficient Comptonization. The photons can diffuse out of the ejecta after $t \sim 1000$~s, after which the intrinsic radiation field is rapidly depleted.

At times when most radiation is still trapped within the ejecta, the escaping (observable) spectrum (Figure~\ref{fig:spec:esc}) is formed within $\Delta R$ of the ejecta surface, defined as the shell within which the diffusion time due to scattering is shorter than the dynamical time (Eq.~\eqref{eq:DeltaR}). Accounting for reabsorption, the {\it effective} optical depth of this shell becomes $\tau_{\rm eff}(\Delta R) = \Delta R/l_{\star}$, where $l_{\star} = [\alphaff (\alphaff + \alphaT)]^{-1/2}$ is the effective mean free path (e.g., \citealt{rybicki_lightman_79}). If $\alphaT\gg \alphaff$, then one finds using Eq.~(\ref{eq:DeltaR}) that $\tau_{\rm eff}(\Delta R) > 1$ is equivalent to the condition $\tauff c/\vej > 1$ (where $\tauff$ is defined for the entire ejecta rather than $\Delta R$). The latter condition is in turn equivalent to $t_{\rm abs, ff} > \tdyn$ (region below the green line in Figure~\ref{fig:theta}, right panel) and defines a photon energy range where the spectral slope roughly obeys
\begin{align}
  L_{\nu} \propto \epsff l_{\star} R^2 \approx \frac{\epsff R^2}{\sqrt{\alphaT\alphaff}}
  \propto n^{1/2} \theta^{1/4} t^2 \nu
  \propto t^{1/4} \nu, 
  \label{eq:Lnu}
\end{align}
or $\nu L_{\nu} \propto t^{1/4}\nu^2$ (dashed line in Figure~\ref{fig:spec:esc}). Here we have used $\epsff \propto n^2 \theta^{-1/2}$, $\alphaff \propto n^2 \theta^{-3/2} \nu^{-2}$, $\alphaT \propto n$, $\theta \propto t^{-1}$, and $R\propto t$. The theoretical slope agrees rather well with the numerical result before $\tdiff$, despite the fact that the above argument assumes the temperature to be (spatially) constant within the diffusion layer, a rather crude approximation.

At higher energies, photons from the entire diffusion layer $\Delta R$ contribute to the escaping emission and the spectrum roughly reflects the intrinsic spectrum within the ejecta at the time, with a peak evolving as $\nu_{\rm pk} \propto t^{-1}$ after the first dynamical time, and likewise for the luminosity (as long as $\tdiff > \tdyn$). The spectrum below the peak
is roughly described by $\nu L_{\nu} \propto \nu^2$ imprinted by moderately saturated Comptonization at earlier stages, transitioning into a softer slope $\propto \nu$ before connecting to the low-energy part of the spectrum discussed above.

Figure \ref{fig:LC:range} shows the light curves in different energy bands. The X-ray emission peaks within the first few dynamical times as rapid photon production and adiabatic cooling bring the spectral peak down from higher energies,
thereafter essentially following the bolometric light curve as the spectral peak remains within the X-ray band. At lower energies, the light curves are relatively flat as long as $\tdiff > \tdyn$, with a subsequent turnover, implying a rising $t L_{\rm esc}$ and indicating that most of the energy in these bands is released near $t \approx\tdiff \approx \tdyn$. At the low-energy end where the argument preceding Eq.~(\ref{eq:Lnu}) applies, the light curve at $t \ll \tdiff$ is roughly consistent with $\sim t^{1/4}$ as suggested by Eq.~(\ref{eq:Lnu}) (dotted line on the left panel of Figure~\ref{fig:LC:range}).

\section{Photon Production and Spectrum Formation in the Radiation-Mediated Shock}
\label{sec:shock}

\begin{figure*}[h]
\begin{center}
\includegraphics[width=0.99\textwidth]{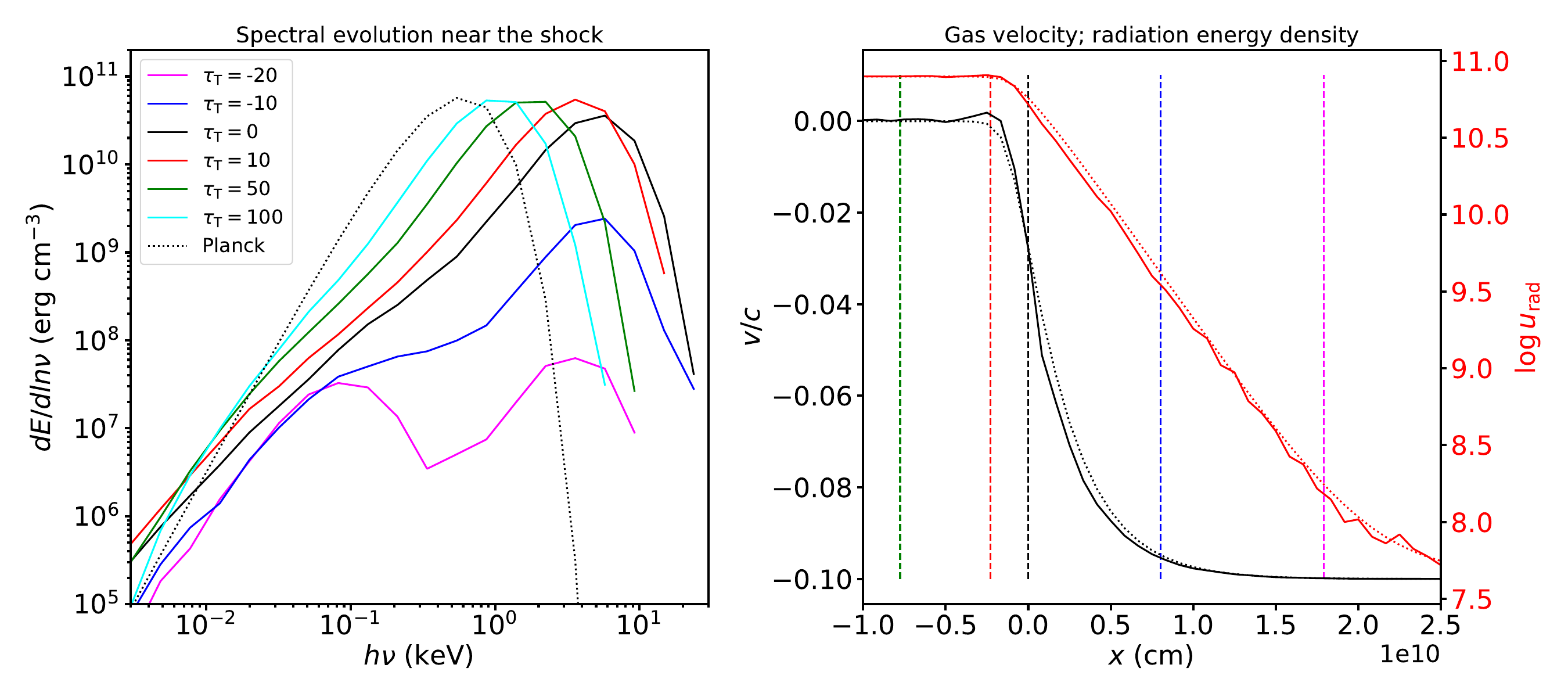}
\end{center}
\vspace{-5mm}
\caption{Spectra at different locations relative to a plane-parallel radiation-mediated shock in steady state (left panel), and the velocity and radiation energy density structure of the shock (right panel). The colors on the left and vertical dashed lines on the right panel correspond to different optical depths relative to the shock ($\tauT>0$ corresponds to the downstream), measured from the location of maximal $|\nabla{v}|$. For reference, the blackbody spectrum corresponding to the downstream energy density is shown by the dotted line (left panel).
The dotted lines on the right panel show the analytical solutions given by Eqs.~(\ref{eq:uanalyt}) and (\ref{eq:vanalyt}). Parameters: upstream density $\rho = 2.5\times 10^{-9}$~g~cm$^{-3}$, relative upstream/downstream velocity $|v_{\rm u} - v_{\rm d}| = 0.1c$, upstream temperature $kT = 25$~eV.}
\label{fig:sh:spec}
\end{figure*}

\begin{figure*}[h]
\begin{center}
\includegraphics[width=0.99\textwidth]{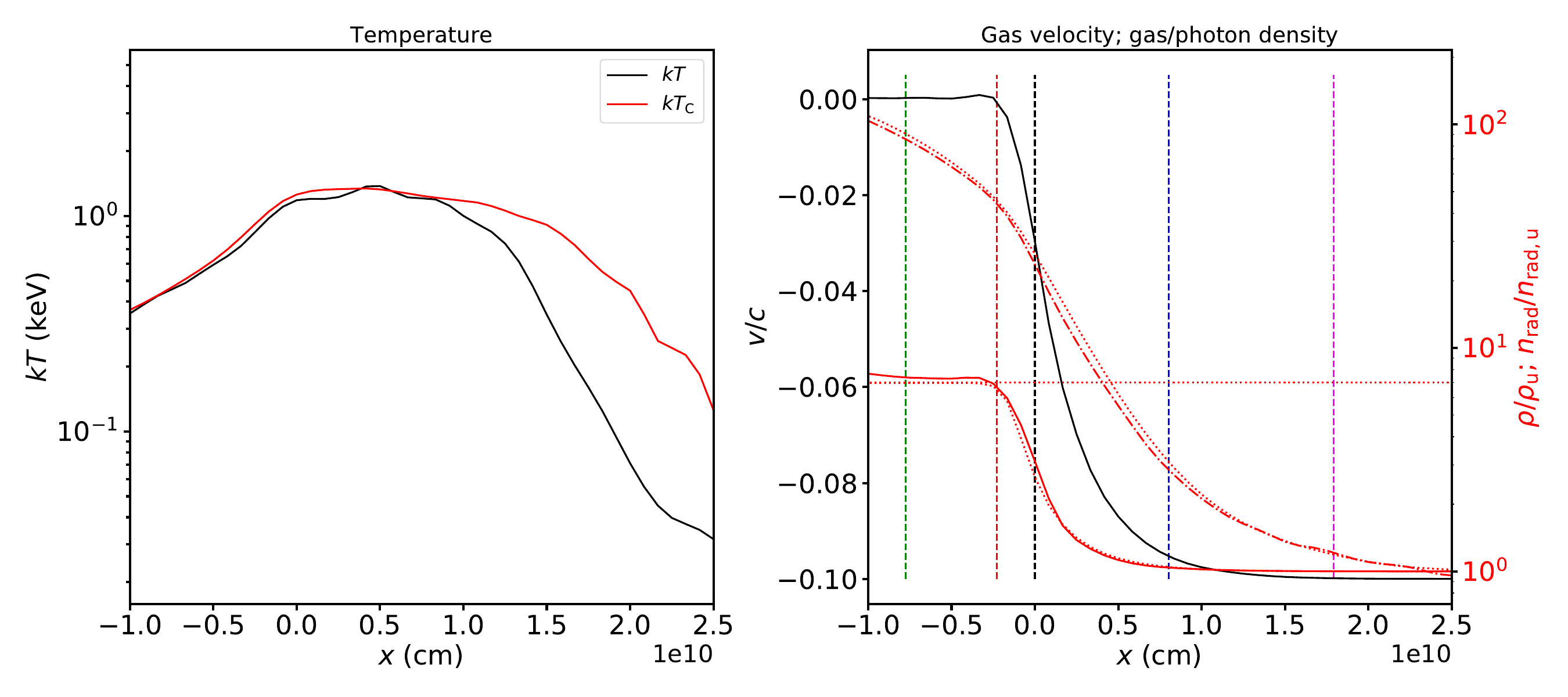}
\end{center}
\vspace{-5mm}
\caption{The matter and Compton temperatures in the shock vicinity (left panel), and matter and photon number densities relative to the upstream values (right panel, red solid and dash-dotted lines, respectively), corresponding to the simulation in Figure \ref{fig:sh:spec}. The red dotted lines on the right panel correspond to semi-analytic solutions for the shock structure (see Appendices \ref{sec:app:sh} and \ref{sec:app:pprod}).}
\label{fig:sh:temp}
\end{figure*}

\begin{figure*}[h]
\begin{center}
\includegraphics[width=0.99\textwidth]{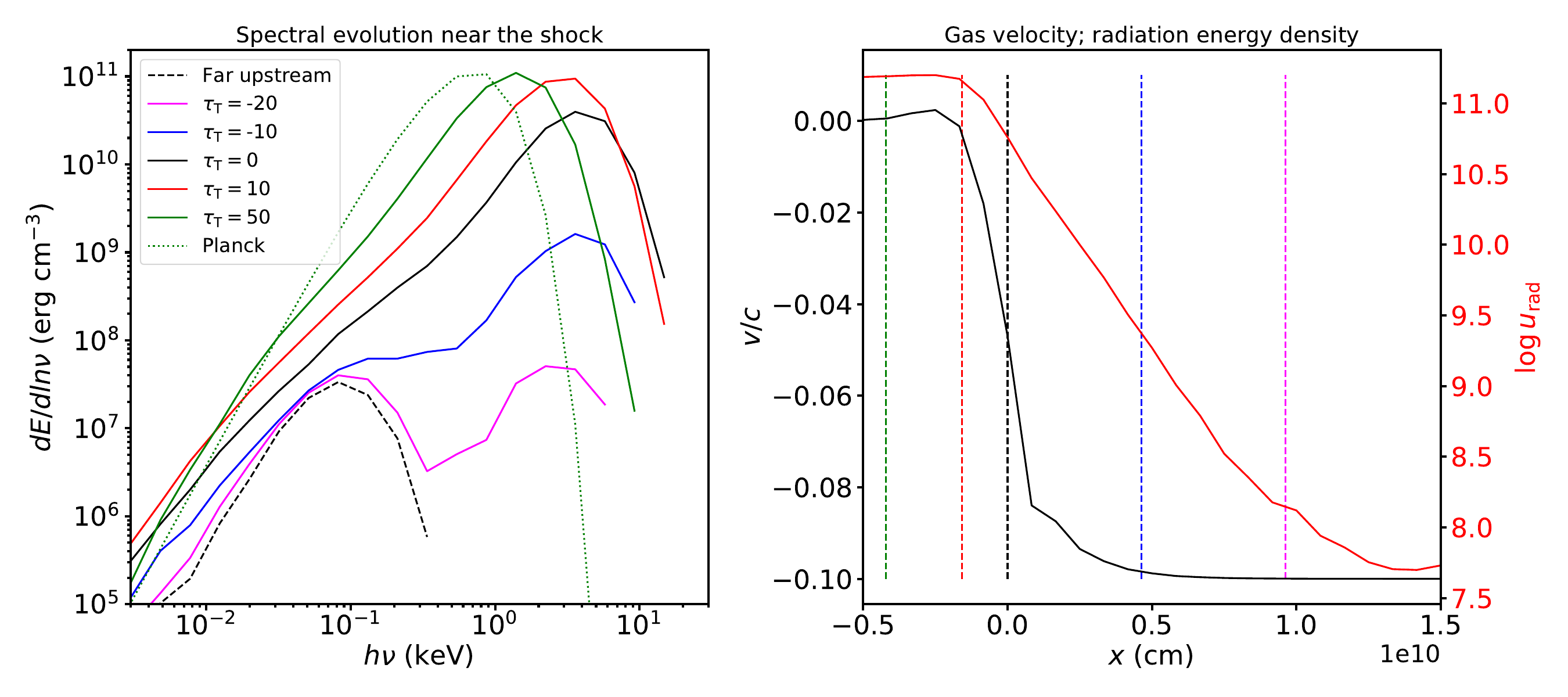}
\end{center}
\vspace{-5mm}
\caption{Same as Figure~\ref{fig:sh:spec}, but for an upstream density twice as large ($\rho = 5\times 10^{-9}$~g~cm$^{-3}$).}
\label{fig:sh:spec2}
\end{figure*}

\begin{figure*}[h]
\begin{center}
\includegraphics[width=0.99\textwidth]{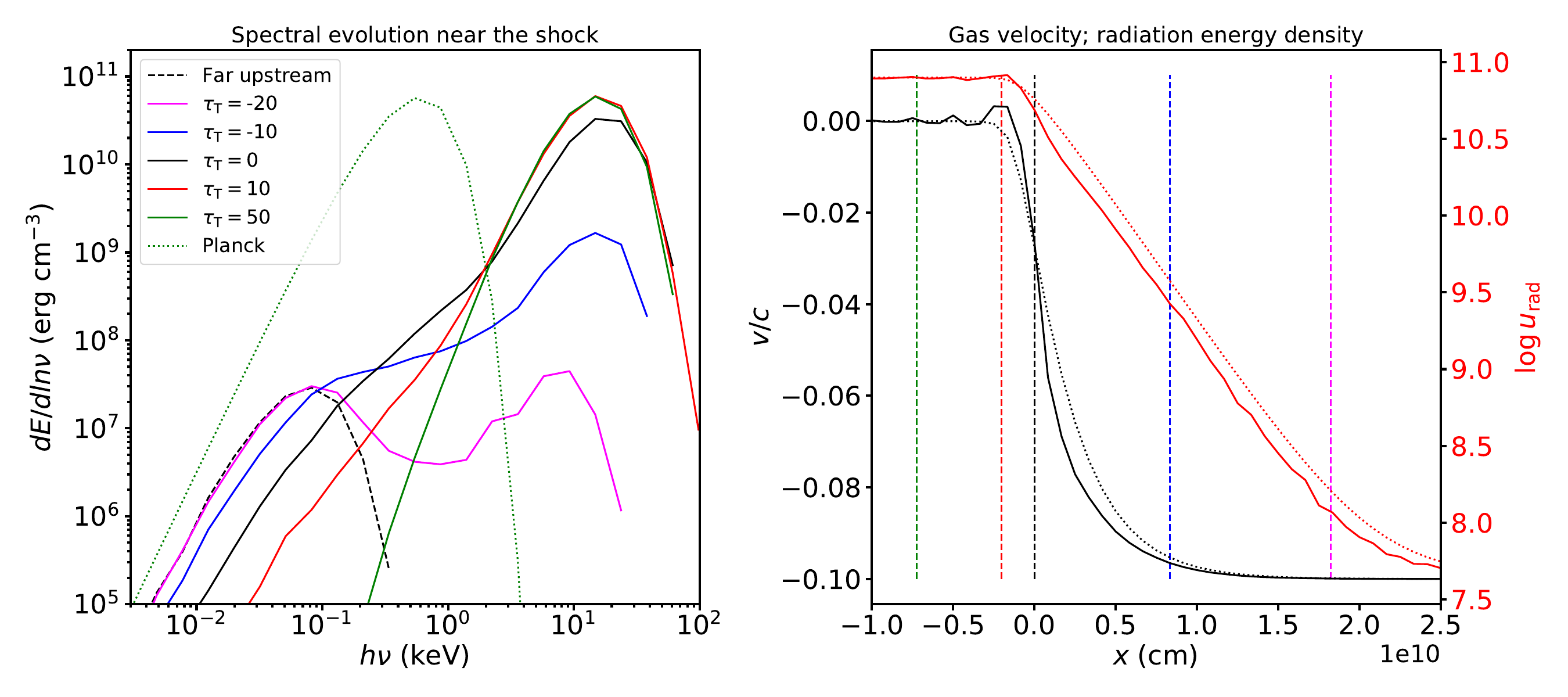}
\end{center}
\vspace{-5mm}
\caption{Same as Figure~\ref{fig:sh:spec}, but with photon production artificially switched off.}
\label{fig:sh:spec0}
\end{figure*}

The passive expansion stage is preceded by an energy deposition phase that 
is responsible for heating up the material and overpressurizing it sufficiently to launch an outflow (phase 1 in Fig.~\ref{fig:schematic}). In this section, we model this phase as a 1-dimensional shock traversing a plane-parallel slab of matter,
mimicking the interaction between a star and the accretion disk material directly ahead of it. While this is at best a crude approximation to the true curved geometry of the shock (which is addressed more accurately by the spherical geometry of the full model presented in Sec.~\ref{sec:model}), it will reveal the key processes at work and is readily amenable to analytic calculations which provide physical insight and offer a check on our numerical results (Appendix \ref{sec:app:sh}, \ref{sec:app:pprod}).

As long as the optical depth of the material being shocked satisfies $\tau_{\rm T} > c/v_{\rm c}$ (see Eq.~(\ref{eq:tauc})) and the photon to baryon ratio $n_{\rm ph,0}/n_{\rm 0}\gg 1$ (Eq.~(\ref{eq:photonbaryonratio0})), a radiation-mediated shock (RMS) is established ahead of the star, in which the ram pressure of the incoming material (in the frame of the star) is balanced by radiation pressure in the shocked downstream region sandwiched between the star and the shock front (see \citealt{Levinson&Nakar20} for a review of RMS). We model this interaction using MCRT simulations, which explicitly follow the energy and momentum exchange between radiation field and matter, while simultaneously tracking photon-production (and absorption) as well as spectral formation within and around the shock. 

Figures~\ref{fig:sh:spec} and \ref{fig:sh:temp} show the steady-state structure of an established high Mach-number RMS, viewed in the downstream (star) frame. The upstream material has a velocity $-0.1c$ (i.e. propagates towards lower $x$) and carries a blackbody radiation field with $kT = 25$~eV.
In terms of the flow velocity (Fig.~\ref{fig:sh:spec}, right panel) the extent of the shock transition region approximately conforms to the nominal value $\Delta \tau \approx c/v_{\rm u} \approx 10$, where $v_{\rm u}$ is the upstream velocity. In contrast, the radiation energy density starts rising above the far upstream value further ahead the shock front, determined by the ability of a small fraction (by number) of energetic photons from the immediate shock downstream to diffuse far ahead of the shock (corresponding to the higher-energy component of the magenta line, left panel of Fig.~\ref{fig:sh:spec}).

This is also reflected in the temperature structure ahead of the shock (Figure~\ref{fig:sh:temp}, left panel). The Compton temperature is biased towards higher photon energies for nonthermal distributions; hence, it starts increasing above the far upstream equilibrium temperature as soon as the energy carried by the high-energy photons diffusing towards the upstream becomes comparable to the far upstream energy density. 

The increase of matter temperature lags the Compton temperature as one approaches the shock, which reflects the finite time it takes to heat the upstream electrons. The temperature equilibrium is re-established close to the shock and maintained in the downstream, owing to the strong radiation field.

The radiation energy density and pressure in the downstream remain approximately constant, at the values determined by the Rankine-Hugoniot shock jump conditions. The evolution of the downstream temperature thus directly reflects ongoing (net) photon production, which eventually ceases once full thermalization is achieved. The Thomson optical depth behind the shock where this occurs depends strongly on the shock velocity, but comparatively more weakly on density (see Appendix~\ref{sec:app:pprod} for more details). However, insofar that the column ahead of a typical shocked fluid element before it exits the disk and begins to decompress increases with the disk surface density $\Sigma$, larger values of $\Sigma$ result in greater thermalization. 

For comparison, Figure~\ref{fig:sh:spec0} shows the structure of a shock otherwise identical to that presented in Figure~\ref{fig:sh:spec}, but with photon production artificially switched off. While the velocity and radiation energy/pressure structure remain almost identical in both cases, the downstream radiation field equilibrates at a significantly higher temperature, reflecting the low initial photon to baryon ratio of the upstream.  

At moderate optical depths ahead of the shock (a few~$\times \, c/v_{\rm u}$) the local spectrum is a superposition of two components: the low-energy thermal radiation advected from the upstream and a high-energy component diffusing towards the upstream from the shock transition region and reflecting the radiation temperature therein. Closer to the shock, the radiation energy density within the high-frequency component becomes sufficiently high to heat up the electrons (via direct Compton scattering) and lock their temperature to the Compton temperature. This initiates thermal Comptonization of lower energy photons, both thermal photons advected from the upstream as well as those generated by the heated electrons via free-free emission. Within the shock transition region the photons are further energized by bulk Compton scattering by repeated shock crossings. In the immediate downstream, the local spectrum is qualitatively similar to the Comptonized spectrum of a low-energy photon source encountered in Section~\ref{sec:expansion:spectra}. However in the present case, thermalization further downstream eventually establishes a blackbody spectrum, provided that the accumulated column before e.g. shock breakout is sufficiently high.

\begin{figure*}[h]
\begin{center}
\includegraphics[width=0.99\textwidth]
{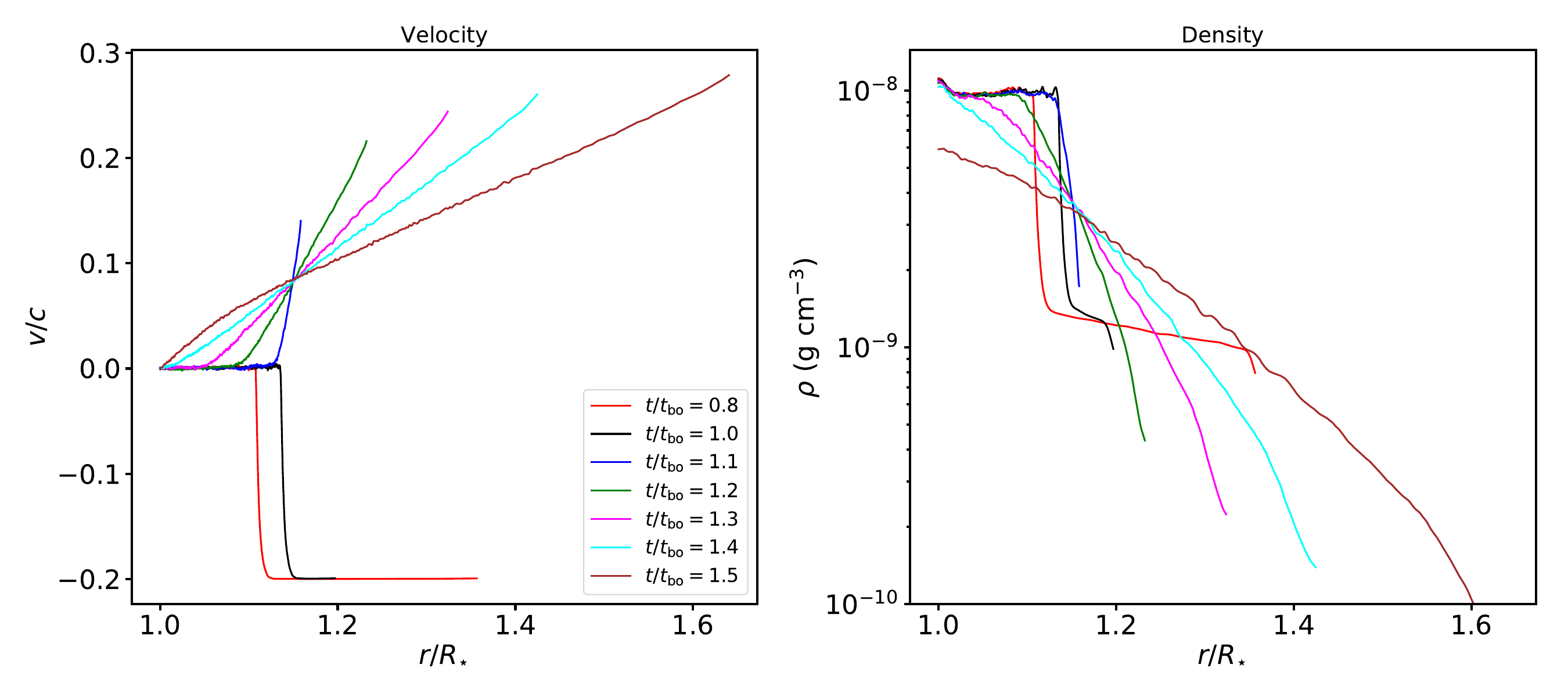}
\end{center}
\vspace{-5mm}
\caption{Velocity (measured in the star frame, left panel) and density (right panel) structure of a 1D radiation-mediated shock ahead of the star during its transition through the disk, around the time the star breaks out from the disk surface. Parameters: ``stellar'' radius $R_{\star} = 10^{12}$~cm, disk thickness $h = 10^{12}$~cm, upstream (disk) density $\rho_0 = 1.5\times 10^{-9}$~g~cm$^{-3}$, impact velocity $v_{\rm K} \approx v_{\star} = 0.2c$. The breakout time is $t_{\rm bo} \approx 6h/7v_{\rm K} \approx 140$~s.
}
\label{fig:bo:vel_dens}
\end{figure*}

\begin{figure*}[h]
\begin{center}
\includegraphics[width=0.99\textwidth]
{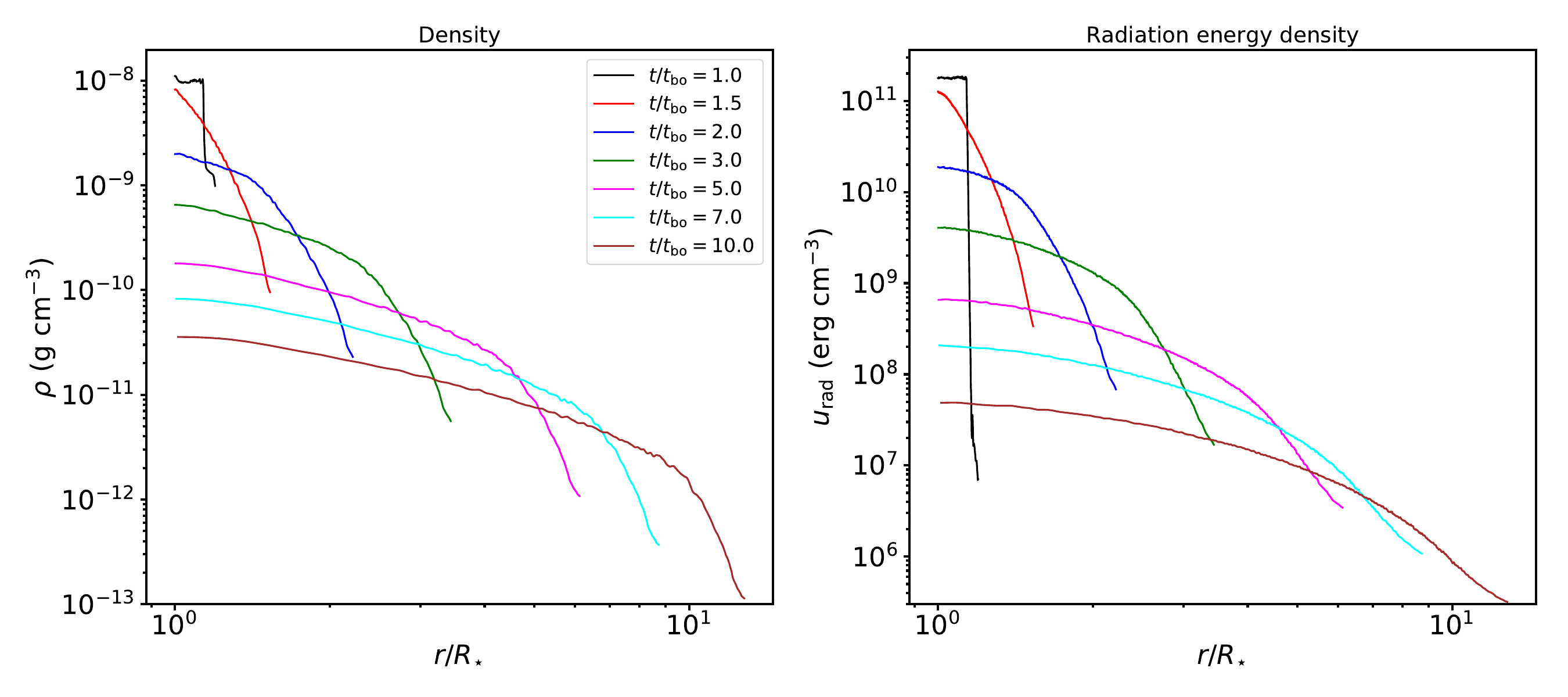}
\end{center}
\vspace{-5mm}
\caption{Matter (left panel) and radiation energy density (right panel) during and after transition into the homologous expansion phase. The parameters are the same as in Figure~\ref{fig:bo:vel_dens}.
}
\label{fig:bo:densrad}
\end{figure*}

\begin{figure*}[h]
\begin{center}
\includegraphics[width=0.49\textwidth]{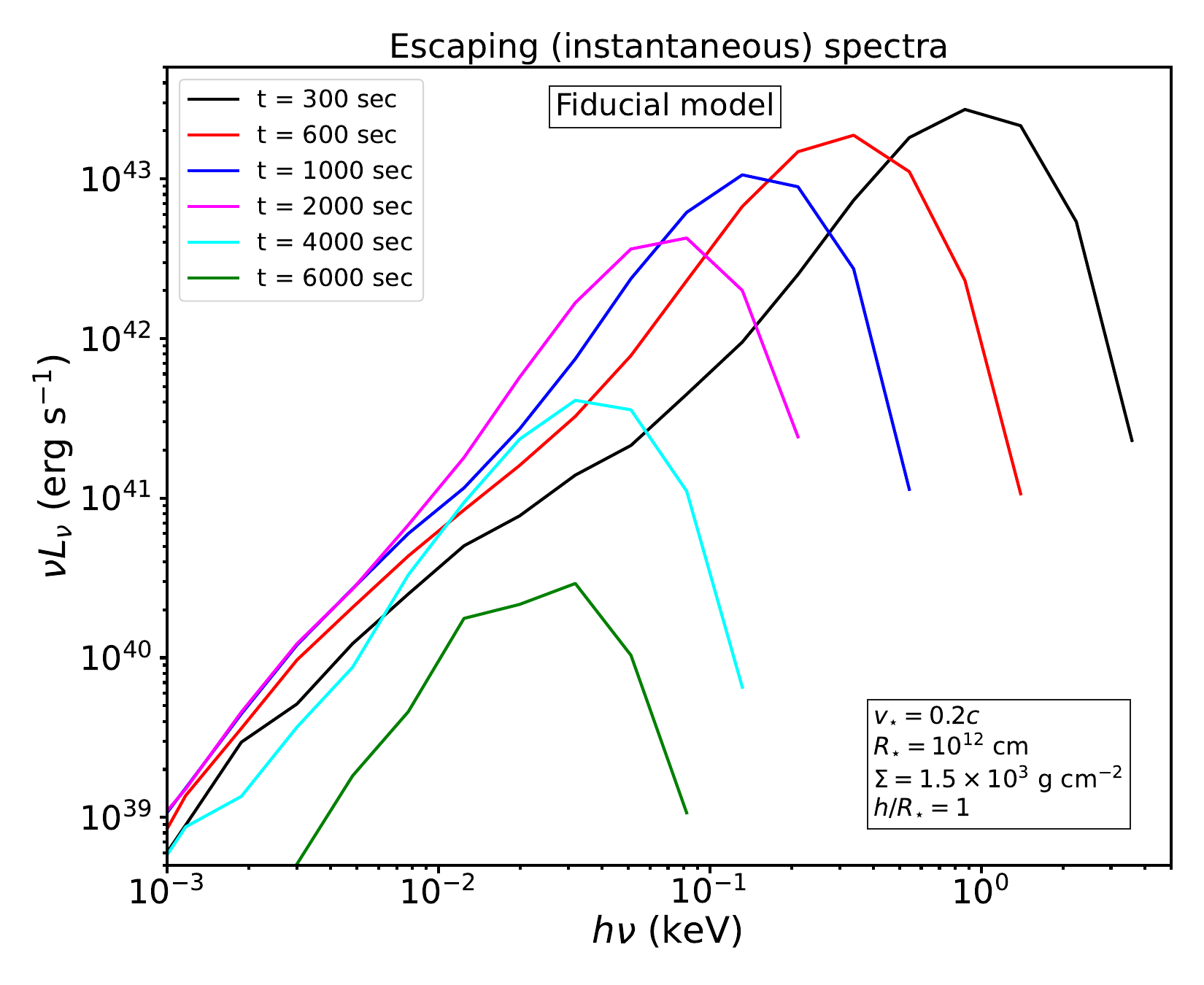}
\includegraphics[width=0.49\textwidth]{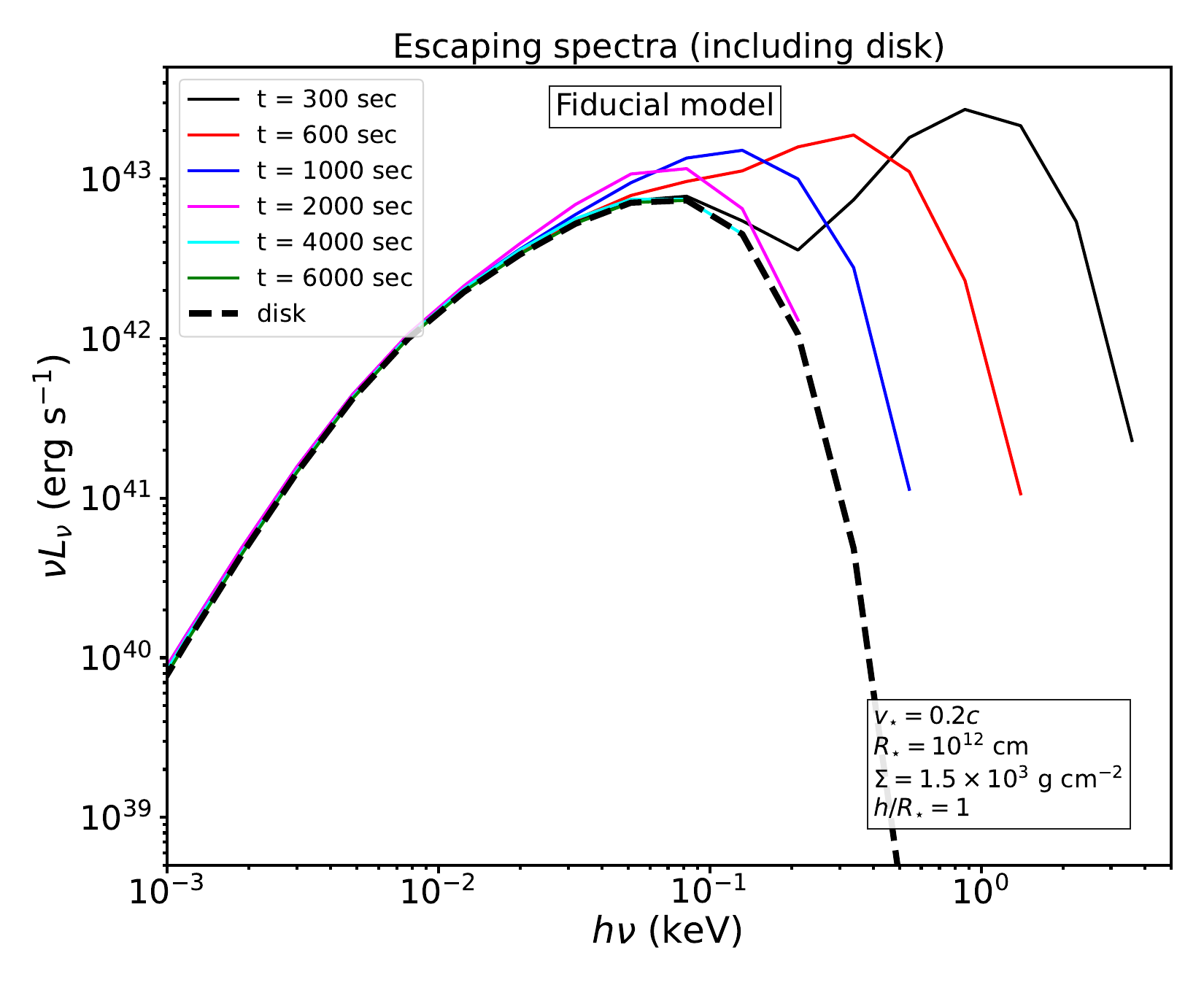}
\includegraphics[width=0.49\textwidth]{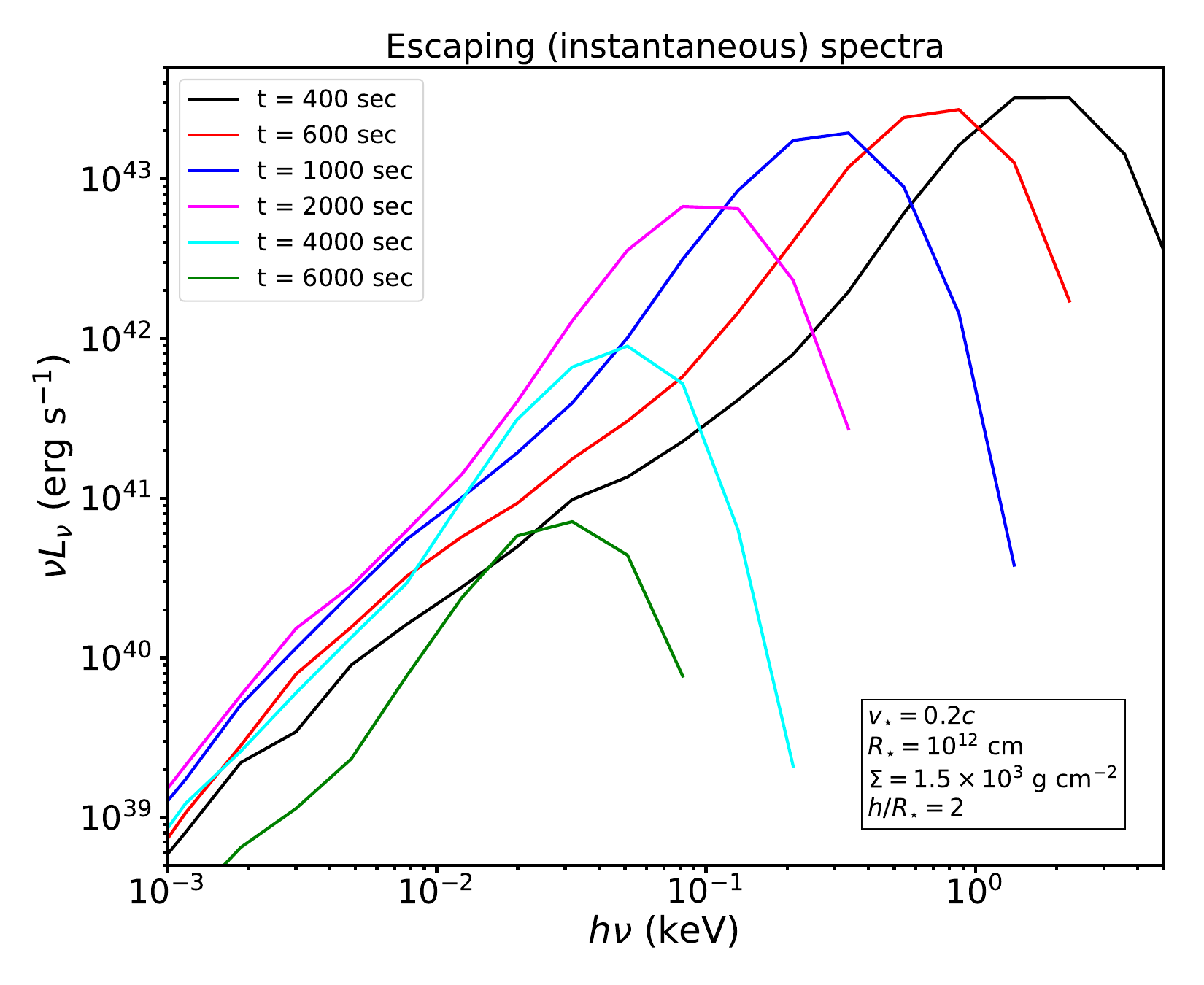}
\includegraphics[width=0.49\textwidth]{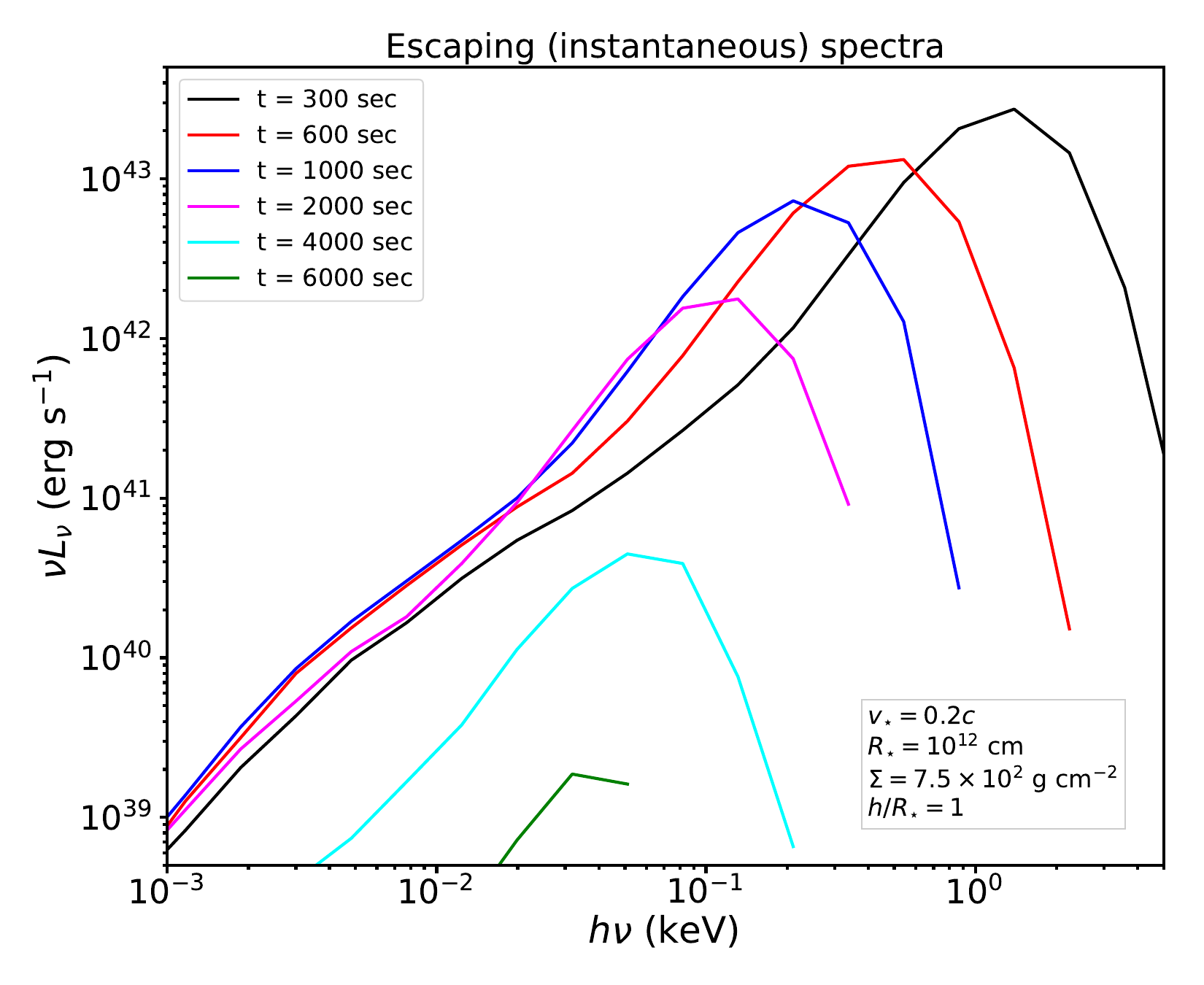}
\includegraphics[width=0.49\textwidth]{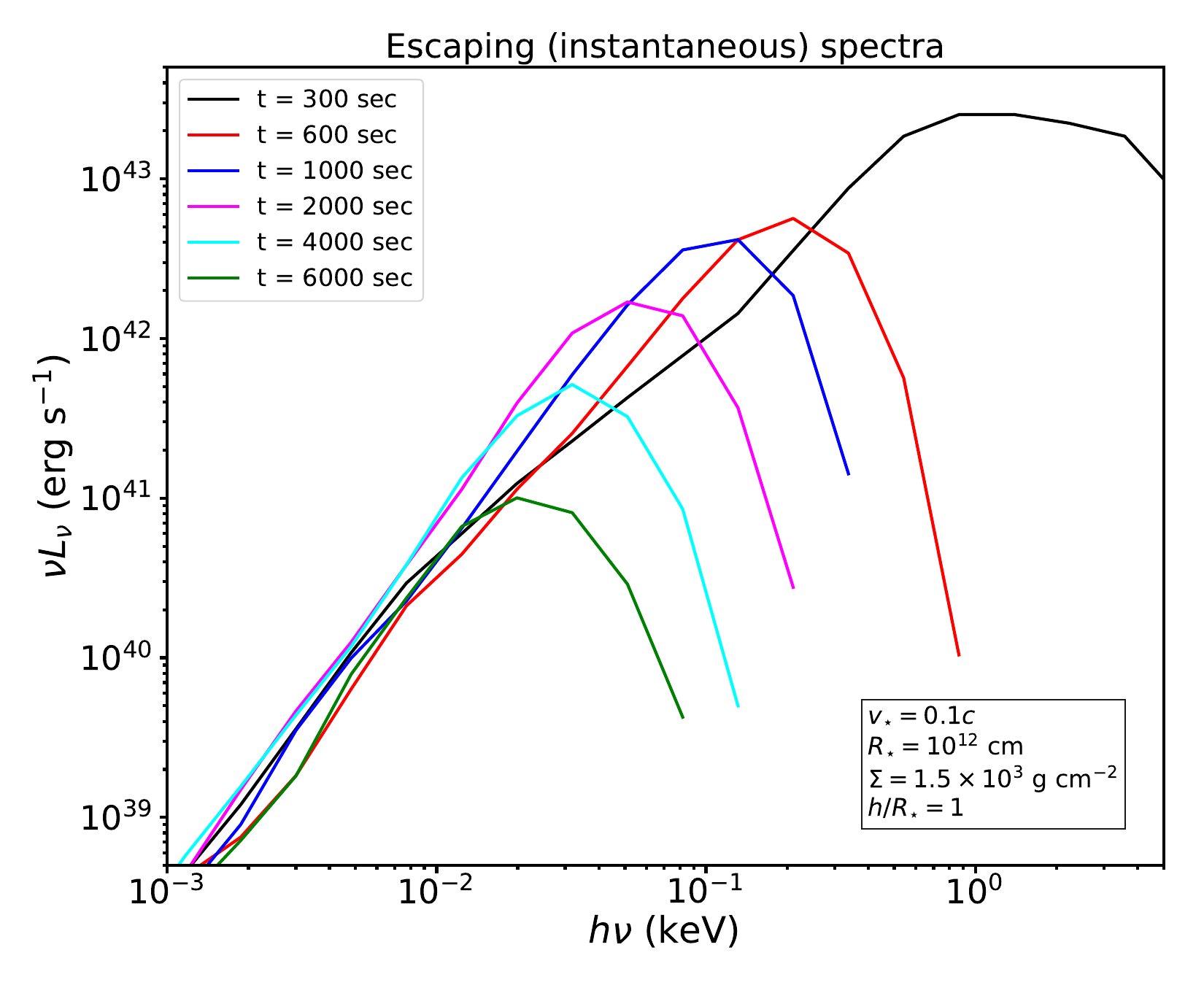}
\includegraphics[width=0.49\textwidth]{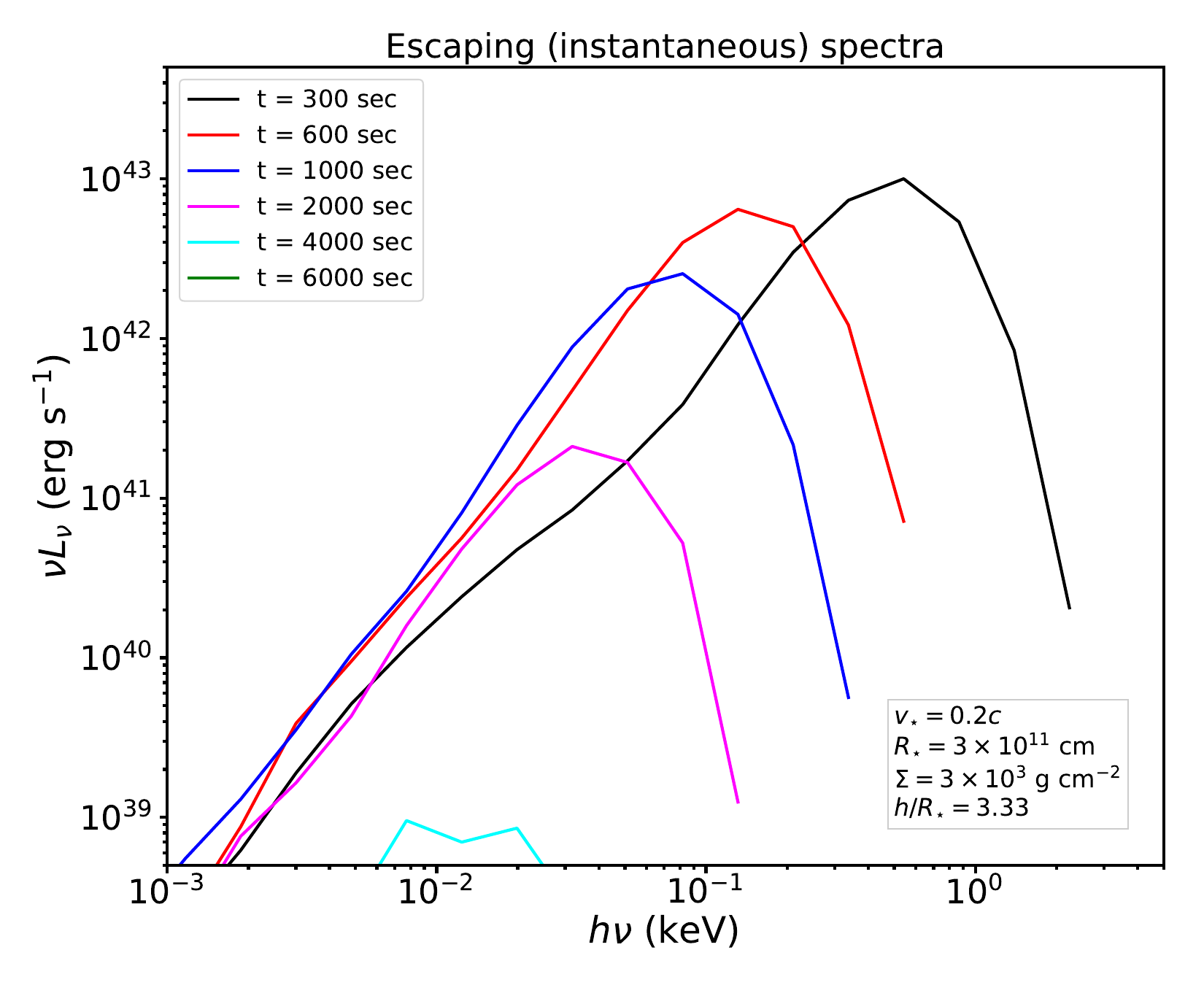}
\end{center}
\vspace{-5mm}
\caption{Escaping spectra at different times following the disk-star collision, shock heating and subsequent expansion of the intercepted material. The top panels show the fiducial model, including the underlying quiescent disk emission (right panel) and without it (left panel). The middle and bottom panels show the results for different parameter sets (excluding the disk).}
\label{fig:spec:esc2}
\end{figure*}

\begin{figure*}[h]
\begin{center}
\includegraphics[width=0.49\textwidth]
{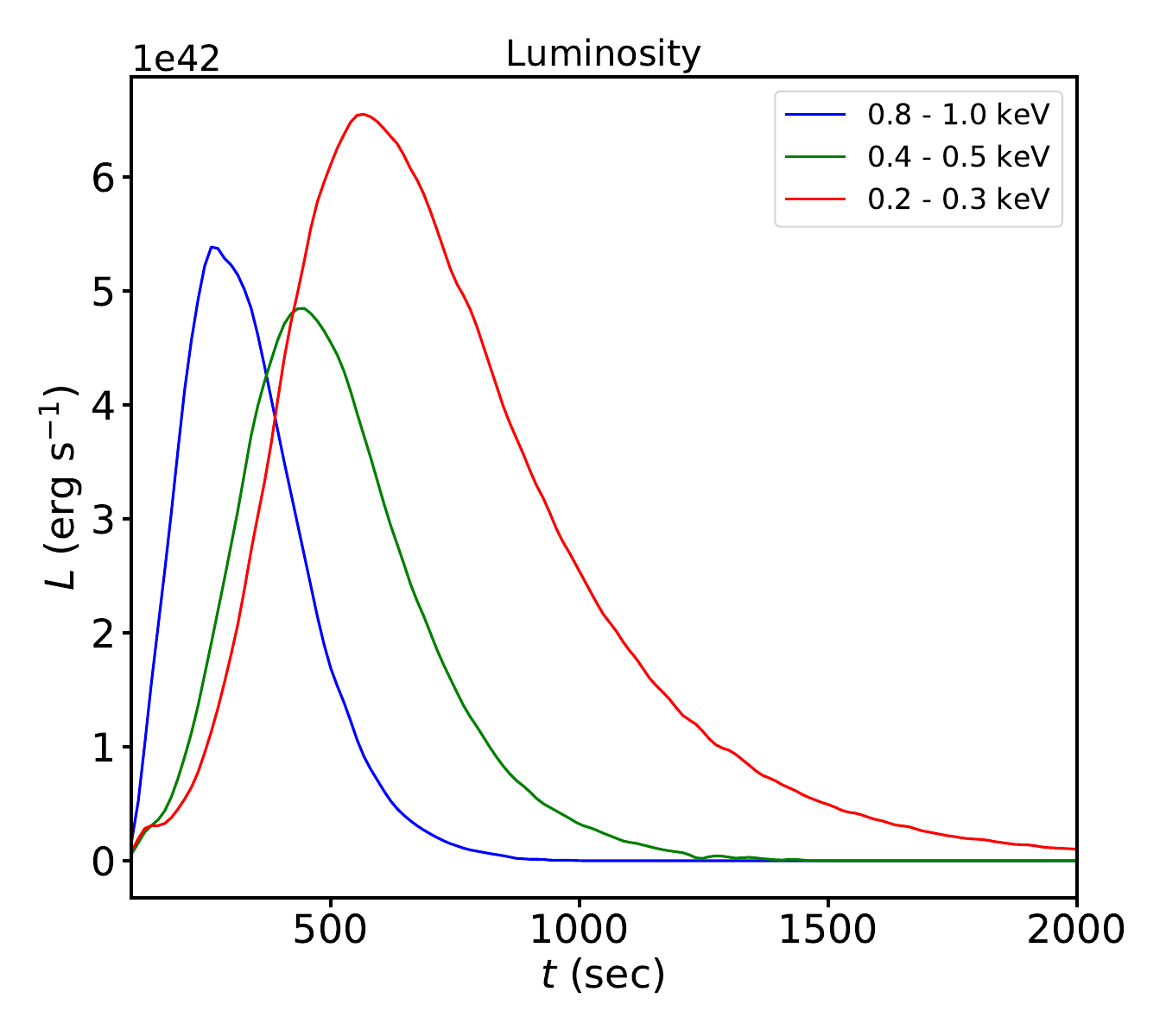}
\includegraphics[width=0.49\textwidth]
{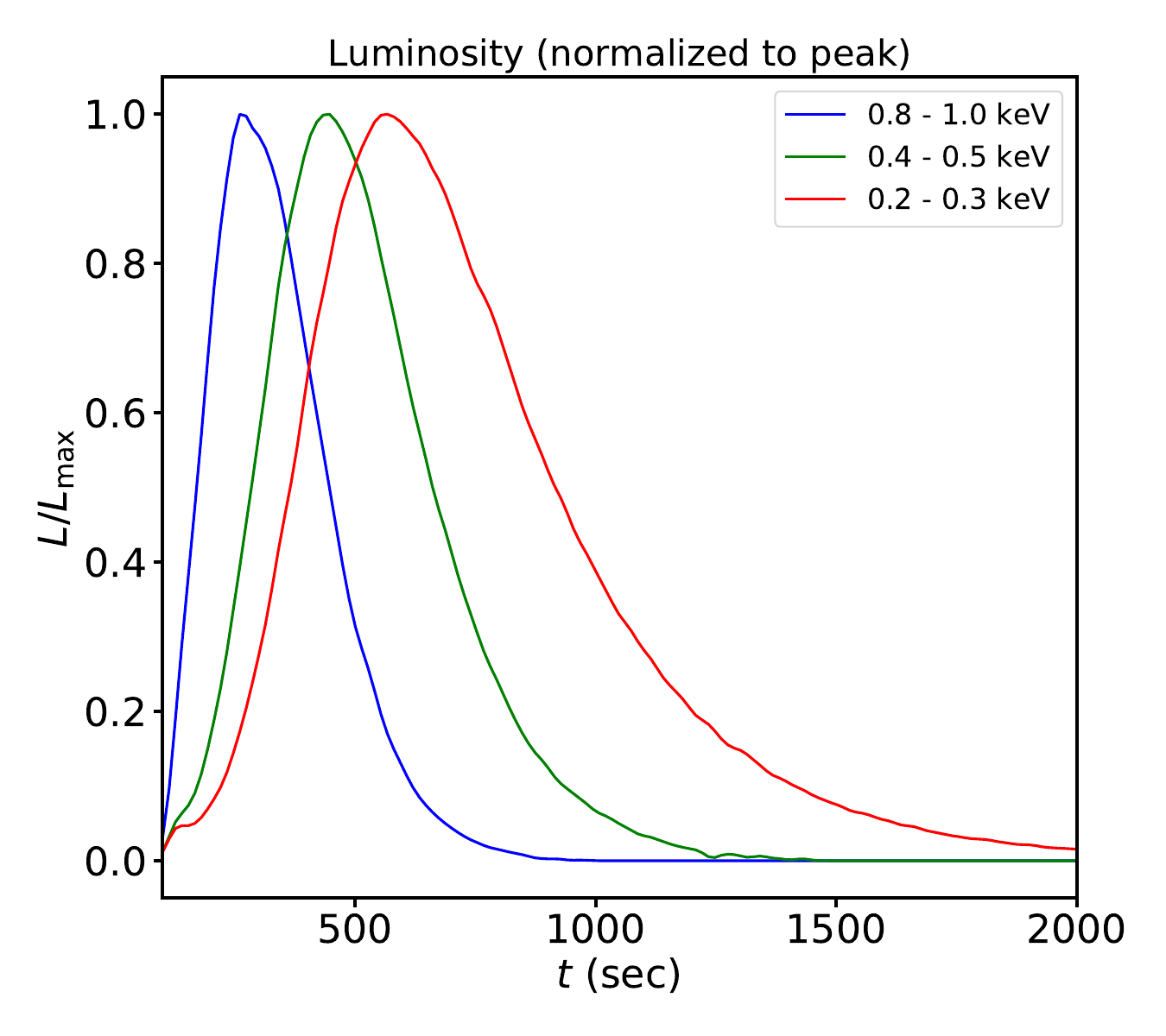}
\end{center}
\vspace{-5mm}
\caption{Unnormalized (left panel) and normalized (right panel) light curves in different bands, corresponding to the escaping spectra shown in the upper left panel of Figure~\ref{fig:spec:esc2}.}
\label{fig:lc:bands}
\end{figure*}

\begin{figure*}[h]
\begin{center}
\includegraphics[width=0.7\textwidth]
{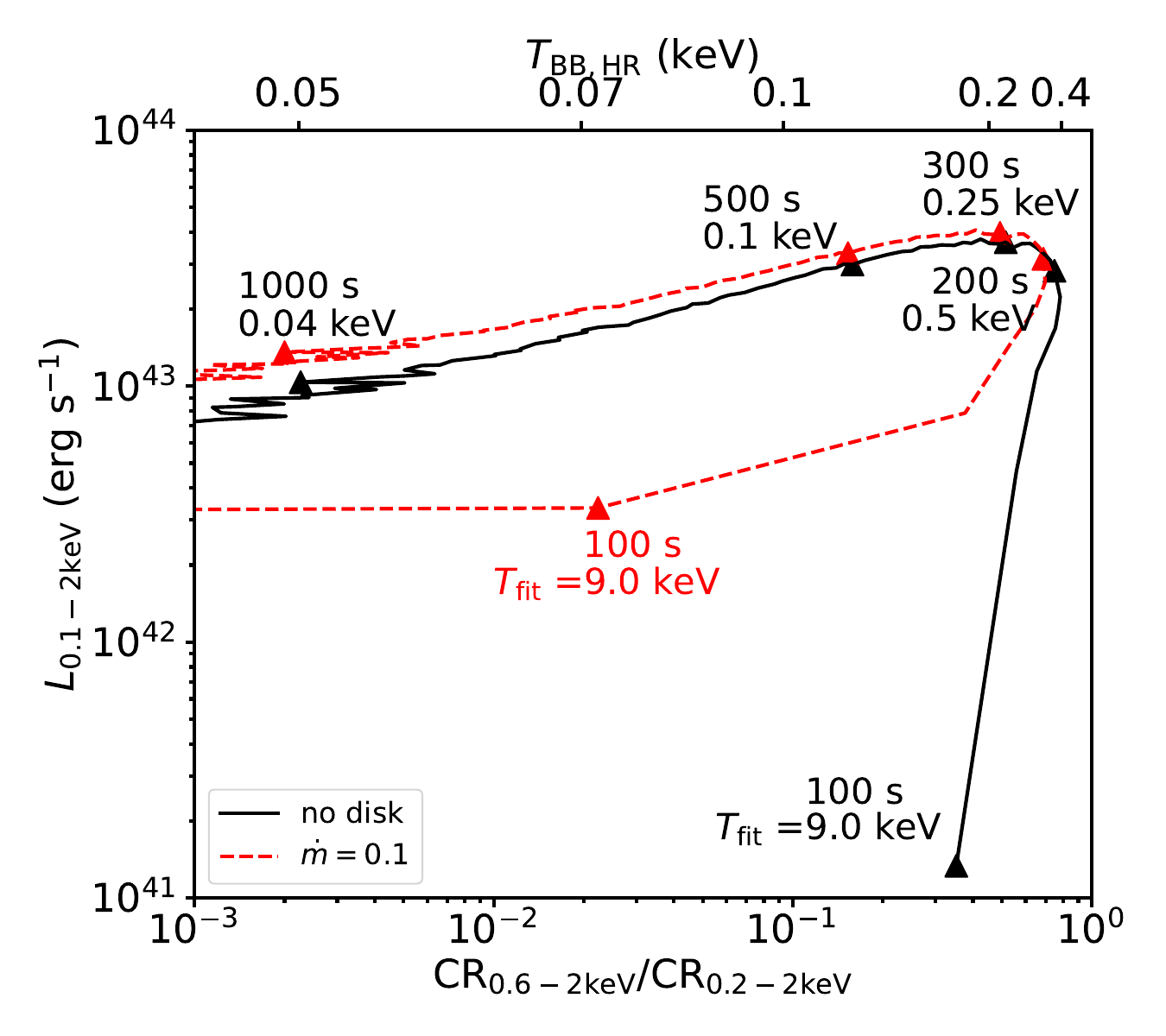}
\end{center}
\vspace{-5mm}
\caption{Hardness ratio between two X-ray bands vs. 0.1-2 keV luminosity for the fiducial model,
both including (red dashed line) and excluding (black solid line) the quiescent disk emission.
We model the latter as a blackbody of luminosity $L_{\rm Q}$ (Eq.~\eqref{eq:LQ}) and temperature $T_{\rm Q}$ (Eq.~\eqref{eq:TQ}),
for an assumed value of $\dot{m} = 0.1$ as marked. Along the top horizontal axis we show the blackbody temperature one would derive from the hardness ratio; importantly, we note that the true spectrum, comprised of both blackbody quiescent disk emission and non-blackbody eruption emission, is clearly {\it not} a single-temperature blackbody. Select times in the eruption are marked, along with the best-fit blackbody temperature to the spectrum at that epoch. Notably, the hardness peaks slightly before the luminosity.}
\label{fig:loops}
\end{figure*}

\section{Full Star-Disk Interaction Model}
\label{sec:model}

Previous sections have outlined the basic physical processes involved in spectral formation behind the radiation-mediated shock driven by the star into the disk (Sec.~\ref{sec:shock}) and the radiative processes that occur during the subsequent re-expansion of this shocked disk material in the idealized set-up of spherical homogeneous ejecta cloud (Sec.~\ref{sec:expansion}).  We now combine these phases into a more complete, albeit still approximate, description of the full
star-shock disk interaction.

The highly supersonic impact of the star with the accretion disk compresses the intercepted material into a narrow shel l, or ``cap'', of density $\rho \approx 7\Sigma/h$ ahead of one side of the star (left panel of Fig.~\ref{fig:schematic}), where the factor of 7 follows from the jump conditions on the disk density for a strong $\gamma = 4/3$ shock.
The build-up of this cap along with the radiative processes within can be roughly approximated within a 1D radial geometry as long as the time it takes for the shocked gas to flow sideways around the edges of the star exceeds the time elapsed from the initial impact.  While an accurate determination would require a fully 3D hydrodynamical simulation, the effective sideways velocity of the shocked disk material around the star, $v_{\perp}$, is likely to fall between the minimum post-shock velocity $\approx \vstar/7$ corresponding to a head-on collision and the post-shock sound speed $c_{\rm s} \approx \vstar$ (if the sideways expansion is better characterized as a rarefaction wave propagating into the lower-pressure surroundings).

Depending on whether the characteristic timescale for this sideways motion $t_{\perp} \approx \Rstar/v_{\perp} = \eta \Rstar/v_{\star}$ is longer or shorter than the passage time $h/\vstar$ of the star through the disk (where $1 \lesssim \eta \lesssim 7$), one can envisage two scenarios. If $h \gg \eta\Rstar$, then a quasi-steady state is established in the shocked layer ahead of the star, where the inflow of material through the shock is approximately balanced by gas escaping sideways around the star. The gas mass in the shocked layer in front of the star at any given time is approximately $\sim \pi\Rstar^2 \vstar\rho t_{\perp} \approx 7\pi\Rstar^3 \rho < \pi\Rstar^2 \Sigma$, i.e. less than the total intercepted mass (our estimate for $M_{\rm ej}$ in Eq.~\eqref{eq:Mej}).  However, this limit may not be applicable to QPE sources insofar that we generally expect $h \lesssim R_{\star}$ at the shock collision radius (Eq.~\eqref{eq:hoverr}), especially if the disk is thicker than naive estimates based on the $\alpha-$viscosity model (see discussion following Eq.~\eqref{eq:tauc}).

Conversely, if $h \ll \eta\Rstar$, then the cap is still being built up when the star exits the accretion diskand most of the intercepted material at that time resides within the compressed layer ahead of the star. As the shock subsides, a rarefaction wave develops at the head of the shocked material and accelerates it to a characteristic velocity $c_{\rm s} \approx \vstar$, i.e. the outflow transitions into an approximately ballistic and homologous quasi-spherical expansion. This is the case that we model in this section, insofar that the geometry of the cap is approximately amendable to 1D spherical symmetry.  We note that a similar evolution should apply even in the other limit ($h \gg \eta\Rstar$) for that material still in front of the star at the exit time, but it is less clear what happens to the material that has escaped sideways.\footnote{If the sideways expanding matter entrains more disk material before exiting, its final velocity will be substantially $\ll \vstar$, making it less likely to contribute to the observed X-ray light curve because of the sensitive dependence of the emission temperature on shock speed.}

We model the star-disk interaction and the subsequent matter ejection in the star frame, whereby the disk material approaches the star at a velocity $\vstar$, undergoes a shock and collects as a heated layer on top of (part of) the stellar surface. Our calculations are performed in spherical geometry centered on the star, which we acknowledge differs from the actual geometry of a spherical star interacting with an approximately plane-parallel disk. We justify this simplification by (1) focusing on a scenario in which the condition $h \ll \eta\Rstar$ holds, whereby the shocked layer is thin and the postshock flow has insufficient time to move appreciably in the sideways direction while the star is passing through the disk, and (2) setting the boundary conditions of the upstream matter which arrives at the (spherical) shock front to follow that set by the vertical structure of the disk encountered by the star.

We follow the evolution through three main phases: (1) accumulation of RMS-heated material ahead of the star while the latter is passing through the disk, (2) breakout of the shock from the disk and subsequent rapid expansion of the post-shock layer, converting most of its internal energy back to bulk motion, and (3) the passive (ballistic) quasi-spherical expansion phase whereby the ejecta dilutes to the point when radiation can escape. In all phases we explicitly follow radiative transfer, photon production and Comptonization as well as the radiation-dominated fluid dynamics, obtaining the observed time-dependent spectra by collecting the photons that escape from the ejecta surface.

\subsection{Fiducial Model}

As a fiducial model we consider the following parameters:
``stellar'' radius $R_{\star} = 10^{12}$~cm, disk thickness $h = 10^{12}$~cm and column density $\Sigma = 1.5\times 10^{3}$~g~cm$^{-2}$, impact velocity $v_{\rm K} \approx v_{\star} = 0.2c$.  

The structure and evolution of the shocked layer accumulating ahead of the star while the latter is still traversing through the disk is well described within the plane-parallel framework discussed in Sections~\ref{sec:shock}, \ref{sec:app:sh} and \ref{sec:app:pprod}, provided that the shocked layer is relatively thin ($h \ll \eta R_{\star}$, see discussion above). While the scenario and simulations described in this section invoke spherical (rather than plane-parallel) geometry, the results of the aforementioned sections largely carry over to the present case prior to breakout and hence will not be repeated here.

Once the star-driven shock reaches to within $\tau \sim c/v_{\rm K}$ from the disk surface, radiation from the immediate downstream starts leaking out into the upstream direction, marking the start of the shock breakout. This is immediately followed by the drop of bulk momentum flux from the upstream, allowing the pressurized shell to start expanding radially, ``detaching'' itself from the star. The shock then subsides, and a rarefaction wave develops that propagates into the shell of earlier swept-up disk material with the local sound speed, driven by the pressure of radiation that is still mostly trapped within the highly opaque shell. 

The velocity and density evolution of the quasi-spherical ``cap'' atop the stellar surface near breakout time is shown in Figure~\ref{fig:bo:vel_dens}. One can see that after the breakout, the outer layers of the shocked shell are rapidly accelerated to velocities comparable to and even exceeding $v_{\rm K}$. The rarefaction wave traverses the shell in roughly $t\sim t_{\rm bo}/3$, after which a quasi-linear velocity structure is established. In general, the high opacity and radiation-dominated pressure means that the bulk of the shell behaves like a $\gamma = 4/3$ fluid throughout its evolution.

Figure~\ref{fig:bo:densrad} shows radial profiles of the matter and radiation energy density structure of the same shell over a longer time span after breakout. The bulk internal (radiation) energy is converted back into kinetic on a timescale comparable to that required for the reverse process prior to breakout, i.e. $\sim t_{\rm bo}$. By this time ($t \sim 2 t_{\rm bo}$), the shell is expanding almost ballistically and has established a roughly power-law density structure throughout most of its extent; for our assumption of a vertically-constant disk density profile, we obtain approximately $\rho \propto v^{-1} \propto r^{-1}$ in the homologous stage. 

The top panels of Figure~\ref{fig:spec:esc2} show the observable spectra at different times in the fiducial model, including or excluding the contribution from the accretion disk (right and left panels, respectively). After shock breakout, the intrinsic emission (i.e. excluding the disk) follows a general hard to soft trend in accordance with the general physical picture presented in preceding sections, resulting from a combination of (1) deeper layers of the ejecta having had longer time to photon-produce, both in the pre-breakout stage when the density is at its maximum (i.e. 'deeper' layers were shocked earlier) as well as in the expansion phase (i.e. radiation from deeper layers is released later), and (2) higher impact of adiabatic cooling for later-released radiation, partially offset by the radially declining (energy) density structure established in the acceleration stage after the breakout.

The spectra maintain a quasi-thermal shape near the peak due to efficient Comptonization;
below the peak, a partially Comptonized free-free emission component may appear as long as full thermalization has not been achieved, transitioning into a self-absorbed slope at even lower energies (see Eq.~\eqref{eq:Lnu} in Section~\ref{sec:expansion:spectra} and the surrounding discussion). In the present context, however, the latter components are swamped by the accretion disk emission (see Fig.~\ref{fig:spec:esc2}, top right panel) and hence challenging to detect.

\section{Discussion}
\label{sec:discussion}

\subsection{Eruption Properties Across Parameter Space}

\begin{figure*}[h]
\begin{center}
\includegraphics[width=0.49\textwidth]
{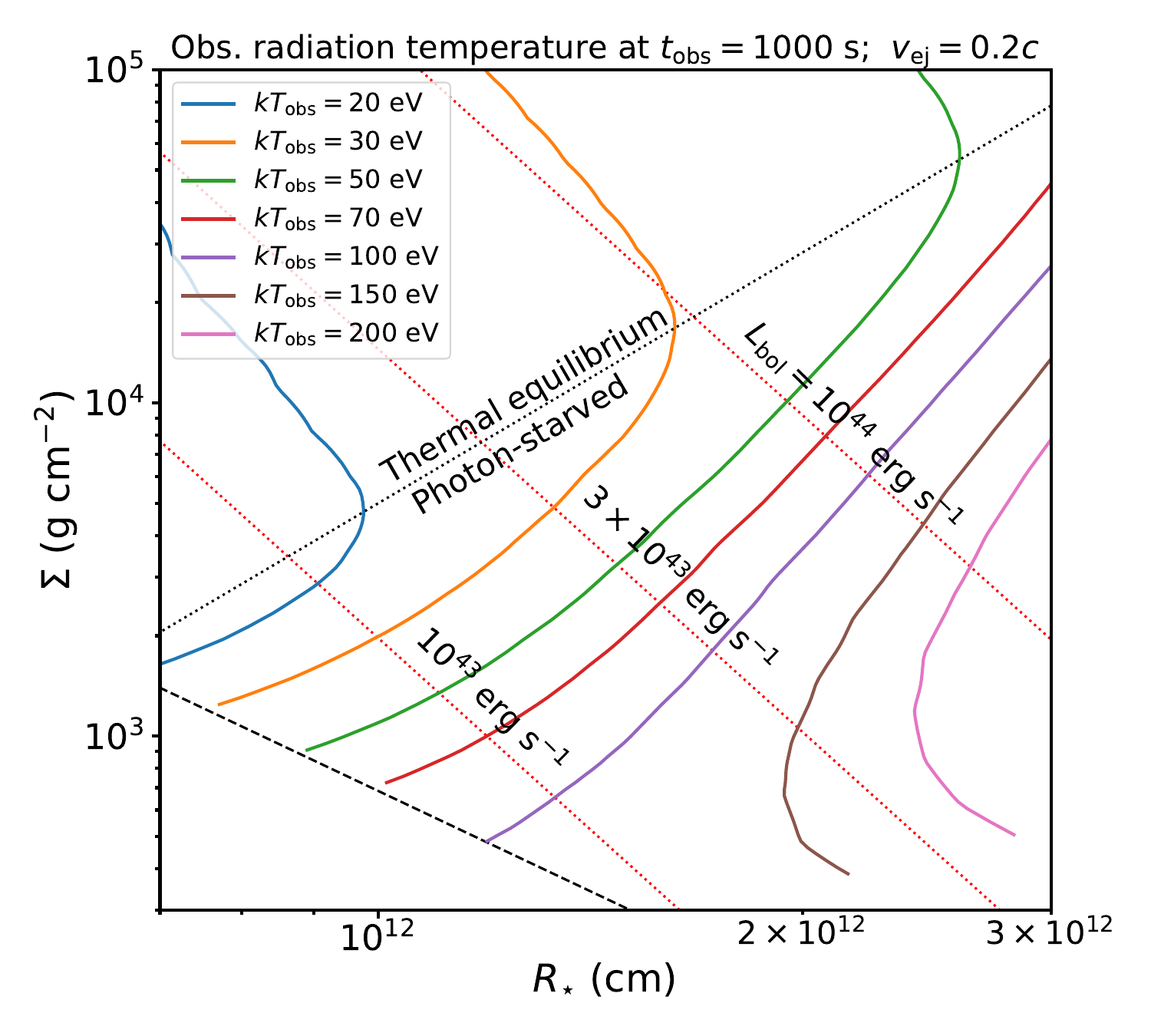}
\includegraphics[width=0.49\textwidth]
{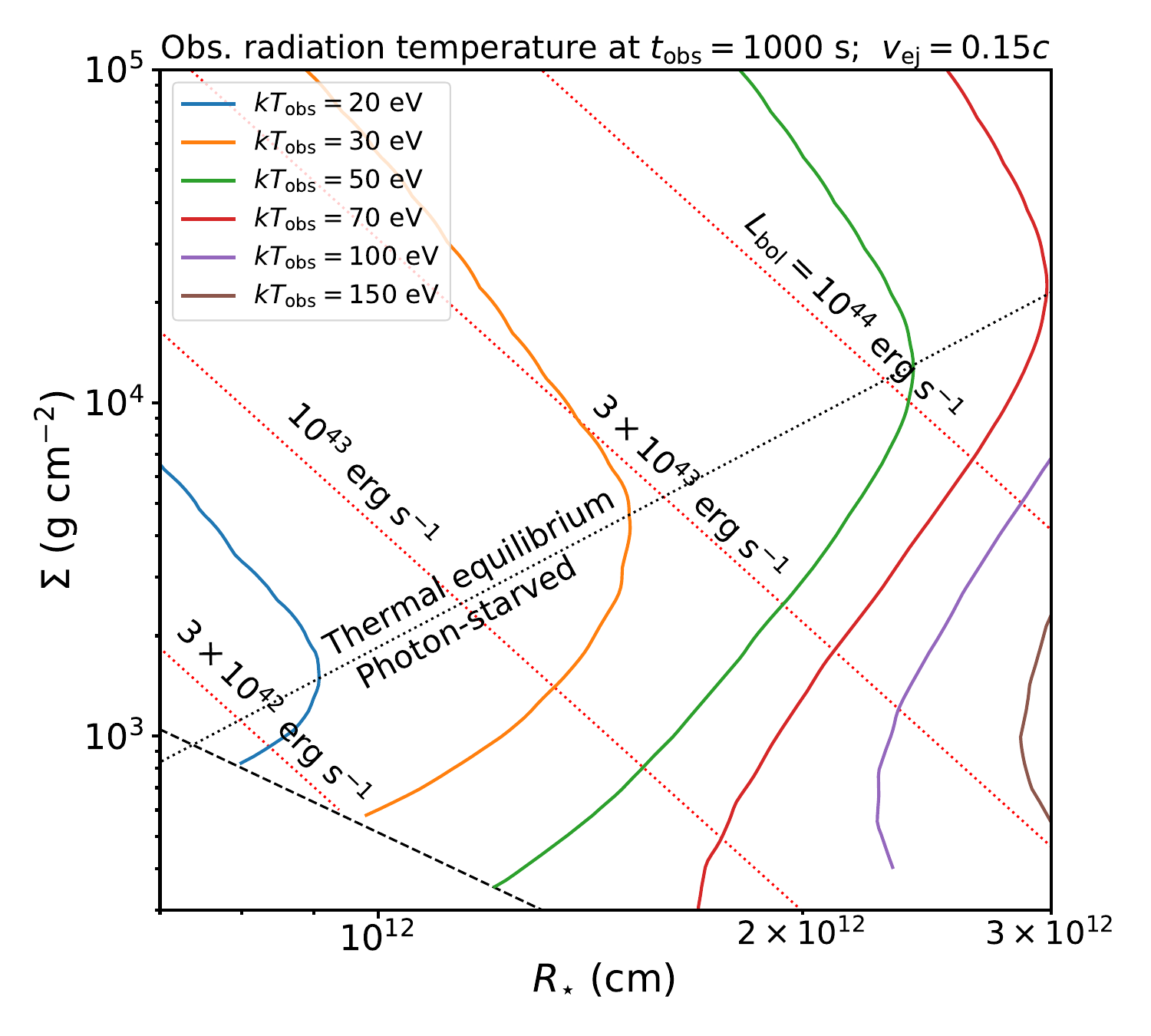}
\includegraphics[width=0.49\textwidth]
{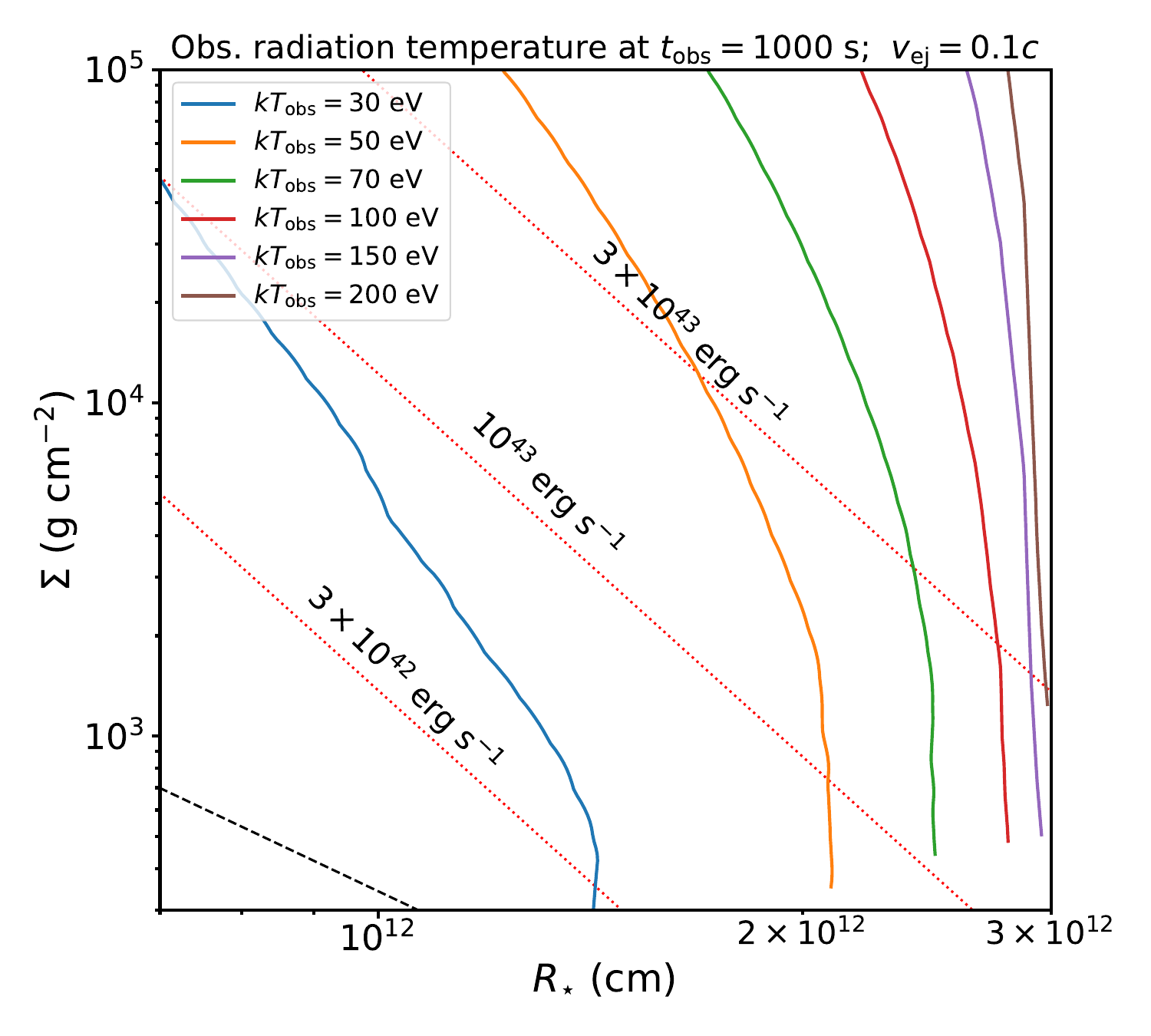}
\includegraphics[width=0.49\textwidth]
{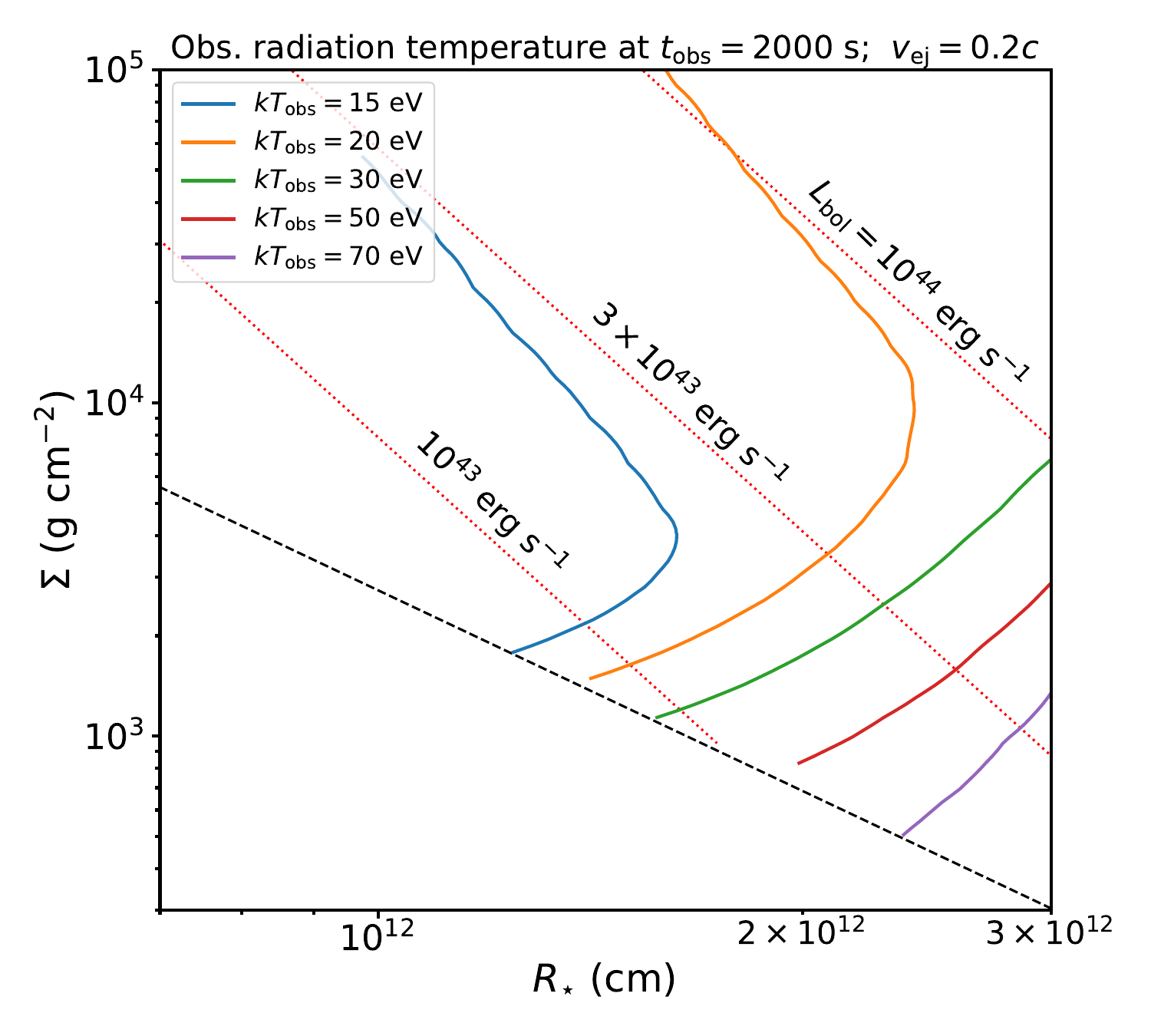}
\end{center}
\vspace{-5mm}
\caption{Isocontours of observed radiation temperature $kT_{\rm obs}$ at a fixed times $t_{\rm obs} = 1000$~s and $2000$~s after the disk-star collision (colored lines), on a plane of
accretion disk column density $\Sigma$ and stellar radius $R_{\star}$.  We assume a disk of thickness $h = R_{\star}.$ The top left, top right, and bottom left panels correspond to $t_{\rm obs} = 1000$~s and collision/ejection speeds $v_{\rm ej} = 0.2c$, $0.15c$, and $0.1c$, respectively. The bottom right panel corresponds to $t_{\rm obs} = 2000$~s and $v_{\rm ej} = 0.2c$.
The dotted line delineates regions of successful thermalization (upper area) and the photon-starved region (lower area). The region below the dashed line corresponds to $t_{\rm obs} > \tdiff$ and is excluded. Diagonal dotted lines show contours of the eruption's bolometric luminosity
at $t_{\rm obs}$ (Eq.~\eqref{eq:Lbol2}, with $\alpha_{\rm t} = 1.1$), adjusted by $f_{\Omega} = 0.5$ corresponding to half-spherical ejecta.
}
\label{fig:kT_iso}
\end{figure*}

Full MCRT simulations of the type presented in Sec.~\ref{sec:model} are costly and hence are not an efficient way to cover the wide parameter space of disk and star properties. As such, it may be helpful to develop approximate semi-analytic prescriptions which connect photon production during the shock-crossing phase (Sec.~\ref{sec:shock}; Appendix \ref{sec:app:sh}, \ref{sec:app:pprod}) with the subsequent spectral and light curve evolution during the free expansion phase (Sec.~\ref{sec:expansion}).

Focusing again on the limit $h \ll \eta R_{\star}$ amenable to an approximate 1D treatment and the simplest case of vertically constant disk density profile, the thin shell of shocked disk material ahead of the star just prior to its breakout from the disk surface is characterized by a steady-state RMS solution (Sec.~\ref{sec:shock}, Appendix~\ref{sec:app:sh}). At this instant, the optical depth through the downstream gas behind the shock is directly proportional to the time each fluid element has spent in the post-shock region, which is relevant to photons carried by each layer and whether the radiation field has had time to thermalize locally. The downstream temperature structure and thermalization optical depth in a steady-state RMS are discussed in Appendix~\ref{sec:app:pprod}.

After shock breakout and subsequent transition to a quasi-spherical expansion phase, one can thus approximately map the postshock optical depth before breakout onto a mass (or radial) coordinate within the homologous outflow. After this transition, the evolution of the radiation field and escaping luminosity are described by the framework of Section~\ref{sec:expansion}, but now with the initial radiation field being radius-dependent as opposed to the case of a homogeneous outflow. The observed radiation temperature at a given instant then reflects the entire history of the fluid element from which photons are just becoming able to diffuse out, and depends on photon production in both the shock and expansion phases.

Using estimates based the above described scenario, Figure~\ref{fig:kT_iso} shows the observed radiation temperature at a fixed time ($t = 1000$~s or $2000$ s, as marked) as a function of the disk surface density $\Sigma$ and stellar radius $R_{\star}$, for $h/R_{\star} = 1 $ and for different star-disk collision speeds. The straight segments of the isotemperature lines at high $\Sigma$ correspond to the high-density regime where thermalization has completed by the time of observation, while below the turnoff point the observed temperature is above the corresponding blackbody temperature (photon-starved regime). A comparison between the first three panels of Figure~\ref{fig:kT_iso} for which $v_{\rm ej} = 0.1 c, 0.15 c, 0.2 c$, respectively, reveals the sensitive dependence of the observed temperature on the star-disk velocity. The reason for this is twofold, relating both to higher characteristic post-shock temperatures for higher shock velocities, resulting in less efficient photon production, and the shorter residence time of matter in the postshock region as well as subsequent expansion stages, leaving less time for photon creation.

\subsection{Analytic Estimates}
\label{sec:comparison}

\begin{figure*}[h]
\begin{center}
\includegraphics[width=0.49\textwidth]{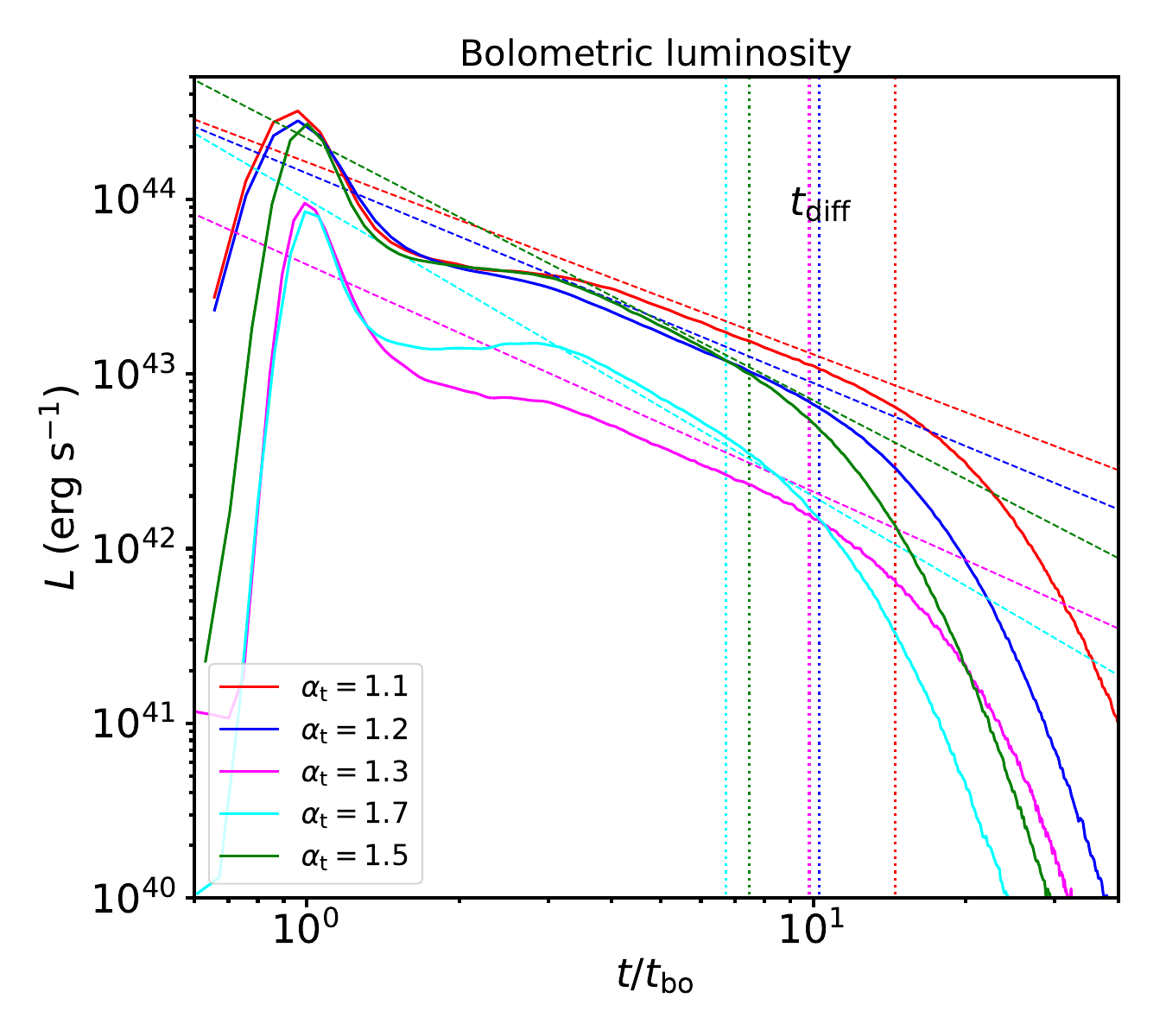}
\includegraphics[width=0.49\textwidth]{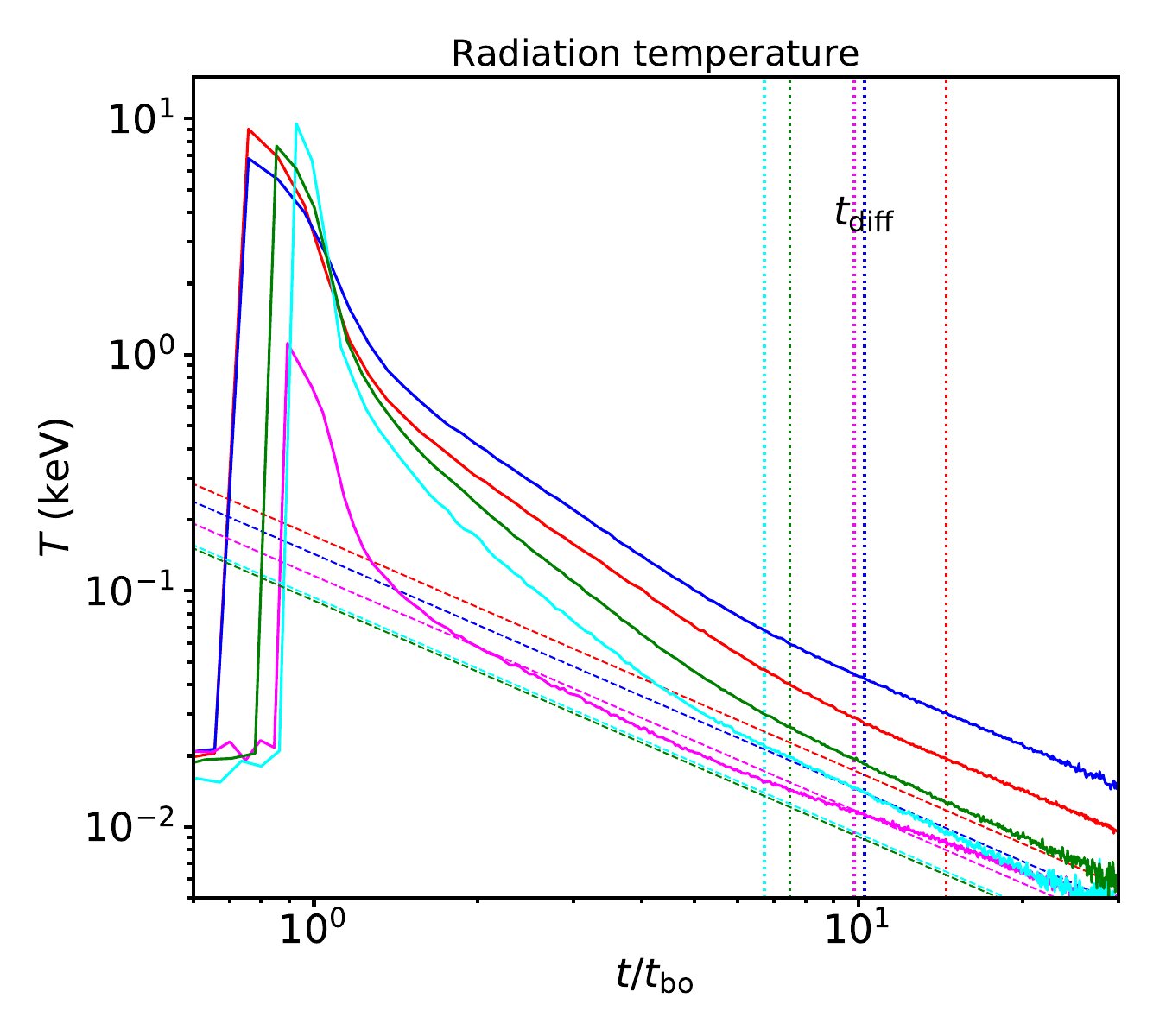}
\end{center}
\vspace{-5mm}
\caption{ Bolometric light curves (left) and escaping radiation temperatures (right) for the simulations shown in Fig.~\ref{fig:spec:esc2}. Red, blue and magenta curves correspond to top left,
middle right and bottom left panels, respectively (all these models assume $h/R_{\star} = 1$,
but exhibit different $\Sigma$ and $v_{\star}$). Green and cyan curves correspond to the middle left
and bottom right panels in Fig.~\ref{fig:spec:esc2}, respectively. The dashed lines on the left panel show the analytic model given by Eq.~\eqref{eq:Lbol2}, for the assumed powerlaw indices $\alpha_{\rm t}$ as labeled. Likewise, the dashed lines on the right panel correspond to
Eq.~\eqref{eq:TempPKBB}, with $\alpha_{\rm BB} = 1$.}
\label{fig:LC:norm}
\end{figure*}

Here we present analytic estimates of the flare luminosity and temperature evolution, building on the estimates provided in \citetalias{Linial&Metzger23} and in Sec.~\ref{sec:expansion}, but now allowing for more accurate calibration of the light curve shape and normalization based on our numerical models.

\subsubsection{Luminosity}

As described in Sec.~\ref{sec:expansion}, the escaping luminosity is set by the thermal energy which remains in the shocked ejecta after expanding adiabatically from the midplane to the diffusion surface (where $\tau \sim c/v_{\rm ej}$ for a layer of depth $\tau$). The evolution of the bolometric luminosity can be derived following a similar argument used for deriving Eq.~\eqref{eq:Lbol}. However, in the more realistic scenario considered here neither $\rho$ or $u_{\rm rad}$ are spatially constant, nor can their radial profiles even be fit to a single power-law in the regions carrying the bulk of the mass and energy. These complications preclude obtaining simple closed analytic expressions from the above procedure. In this section we therefore adopt a more phenomenological approach, {\it assuming} a physically motivated general form of $L_{\rm esc}$ similar to Eq.~\eqref{eq:Lbol} and calibrating it on numerical results. Let us write
\begin{align}
 L_{\rm esc} &\approx 
 L_{\rm diff} \left(\frac{t}{t_{\rm diff}} \right)^{-\alpha_{\rm t}}
= \zeta \frac{M_{\rm ej}v_{\rm ej}^{2}}{2} \frac{t_0}{t_{\rm diff}^2} \left(\frac{t}{t_{\rm diff}} \right)^{-\alpha_{\rm t}} \nonumber \\
 &= \frac{2\pi c \, \zeta R_{\star} v_{\rm ej}^2}{\kappa} \left(\frac{t}{t_{\rm diff}} \right)^{-\alpha_{\rm t}} \nonumber \\
 &= 5 \times 10^{42} \, {\rm erg} \; \zeta \, R_{\star, 12} \, v^2_{-1} \left(\frac{t}{t_{\rm diff}} \right)^{-\alpha_{\rm t}}, 
 \label{eq:Lbol2}
\end{align}
where 
\be
t_0 = \frac{R_{\star}}{v_{\rm ej}}
\approx 3.3\times 10^{2}\,{\rm s}\,R_{\star,12}\,v_{-1}^{-1},
\ee
is the initial expansion timescale and
\begin{eqnarray}
    t_{\rm diff} &=& \left(\frac{\kappa M_{\rm ej}}{4\pi c v_{\rm ej}}\right)^{1/2} = 2.4\times 10^{3}\,{\rm s}\,M_{\rm ej,-5}^{1/2}v_{-1}^{-1/2} \nonumber \\
    &\approx& 1.5\times 10^3 \, {\rm s} \; R_{\star,12}\dot{m}^{-1/2} \alpha_{-1}^{-1/2} M_{\bullet,6}^{-2/3} P_{\rm QPE,4}^{2/3} \,,
    \label{eq:tdiff2}
\end{eqnarray}
is the diffusion time (Eq.~\eqref{eq:tdiff}). In the second line of each expression, we have substituted in physical quantities following the estimates from Sec.~\ref{sec:LM23}. 

The factor $\zeta$ in Eq.~\eqref{eq:Lbol2} denotes the fraction of the total dissipated energy that remains in radiation after the initial approximately planar expansion of the shocked shell from thickness $h/7$ to $R_{\star}$ after shock breakout,
i.e. $\zeta = E_{\rm rad,0}/(M_{\rm ej}v_{\rm ej}^{2}/2) \approx
(h/7R_{\star})^{1/3}$. We note that for $\alpha_{\rm t} = 1$ and $R_{\star} = R_0$, Eq.~\eqref{eq:Lbol2} reduces to Eq.~\eqref{eq:Lbol} (with $\xi_{\rm diff} = 1$).

Fig.~\ref{fig:LC:norm} (left panel) shows the bolometric light curves for the simulations shown in Fig~\ref{fig:sh:spec2}, with time normalized to $t_{\rm bo} \approx 6h/7v_{\rm ej}$.
The light curves are qualitatively similar in all cases, exhibiting an initial breakout flash, followed by a brief plateau and then a power-law decay up to $t \approx t_{\rm diff}$. The latter are reasonably well approximated by Eq.~\eqref{eq:Lbol2} (shown by dashed lines in Fig.~\ref{fig:LC:norm}), for empirically determined slopes ranging from $\alpha_{\rm t} \approx 1.1$ to $1.7$ and the normalization deviating from the numerical result by $\lesssim 30\%$. The total radiated energy scales roughly as $\tdiff L_{\rm diff} \propto \zeta\,\Mej^{1/2}\vej^{3/2} R_{\star} \propto \zeta\,\Sigma^{1/2}\vej^{3/2} R_{\star}^2$.

To first approximation, the energy emitted by the breakout flash per unit surface area is comparable to the energy in radiation from the immediate downstream with density $u_{\rm rad, d} = 2\rho_{\rm u}v_{\rm u}^2/(\gamma^2 - 1)$ that can penetrate into the shock transition layer $\Delta R_{\rm sh} \approx D/v$ and is released (locally) over time $\Delta R_{\rm sh}/v$ when this layer breaks through the surface.  Thus one can approximately write
\begin{align}
  \Delta E_{\rm bo} &\approx 4\pi f_{\Omega} R^2 \, \Delta R_{\rm sh} u_{\rm rad, d} = \frac{8\pi c f_{\Omega} \, R_{\star}^2  \, v_{\rm u}}{3(\gamma^2 - 1)\, \kappa} \nonumber \\
  &\approx 2.8\times 10^{45} \, {\rm erg} \, R_{\star, 12}^2 \, v_{-1} \, f_{\Omega},
  \label{eq:Ebo}
\end{align}
where $\gamma = 4/3$ is the adiabatic index and $1/4 \lesssim f_{\Omega} \lesssim 1/2$ is the fraction of the total solid angle subtended by the emitting surface\footnote{Due to the imposed spherical symmetry, our numerical results correspond to $f_{\Omega} = 1$ by construction; the reported isotropic luminosities should thus be corrected appropriately if the ejecta subtends only a portion of the total solid angle, e.g. $f_{\Omega} \approx 0.5$ in cases when the star-disk interaction produces a quasispherical plume of ejecta confined to one side of the disk.}.

Equation~\eqref{eq:Ebo} neglects contributions from radiation that starts diffusing out of the shocked layer immediately after breakout on a timescale that is still short compared with the dynamical time (i.e., while $\Delta R_{\rm sh}/v < t - t_{\rm bo} < R_{\star}/v$). On the other hand, the rarefaction wave propagating into the shocked layer overtakes the diffusion surface on a timescale $\sim \Delta R_{\rm sh}/v$, resulting in significant adiabatic degradation of the downstream photons before they have had a chance to escape. Overall, Eq.~\eqref{eq:Ebo} underestimates energy emitted near breakout by a factor of $\sim 2$ compared with the numerical result. Our numerical results near breakout should be taken with caution due to the imposed spherical symmetry and lack of a realistic disk surface structure. Nevertheless, the preceding arguments still shed light on the dependencies on different parameters (or lack thereof, e.g. on $\rho$) and aid in interpreting Fig~\ref{fig:LC:norm}.

While in the idealized set-up of our simulation the breakout luminosity is approximately $\Delta L_{\rm bo} \sim \Delta E_{\rm bo}/(R_{\star}/c)$, in reality the breakout signal is likely to be smeared out over the longer timescale $R_{\star}/v$ the star takes to fully emerge from the disk surface; in the latter case, the breakout luminosity and the luminosity at the diffusion time, $L_{\rm diff}$, scale similarly as $\propto R_{\star} v^2$ (where $v \approx v_{\rm ej}$).

Following the breakout, the diffusion surface from which photons escape initially samples the outermost ejecta where the matter and radiation density gradients are steep, thus creating the brief plateau phase in the light curve that precedes the steeper powerlaw decay. In models with fixed $h \simeq R_{\star}$ and hence a thin shocked shell, the decay powerlaw indices are all similar ($\alpha_{\rm t} \approx 1.1 - 1.3$, shown by red, blue and magenta lines in Fig.~\ref{fig:LC:norm}), whereas models with larger $h/R_{\star}$ and hence thicker shells exhibit steeper slopes.

\subsubsection{Emission Temperature}

Continuing the above phenomenological approach, the blackbody temperature evolution can be written as
(Eq.~\eqref{eq:thetaBB})
\be
kT_{\rm BB}(t) = kT_{\rm BB,diff}\left(\frac{t}{t_{\rm diff}}\right)^{-\alpha_{\rm BB}},
\ee
where
$kT_{\rm BB,diff} \equiv kT_{\rm BB}(t_{\rm diff})$ is found by equating the radiation energy carried by the ejecta at $t_{\rm diff}$, $E_{\rm diff} \approx \zeta M_{\rm ej}v_{\rm ej}^{2} t_0/2t_{\rm diff}$, with $4\pi R_{\rm diff}^3 aT_{\rm BB, diff}^4/3$, which yields
\begin{eqnarray}
kT_{\rm BB,diff} &\approx&
k\left[ \frac{6\pi\zeta c^2 R_{\star}}{a\kappa^2 M_{\rm ej}} \right]^{1/4} \approx
15.2\,{\rm eV}\,\frac{\zeta^{1/4} R_{\star,12}^{1/4}}{M_{{\rm ej}, -5}^{1/4}} \nonumber \\
&\approx&
16.3\,{\rm eV}
\frac{\alpha_{-1}^{1/4}\dot{m}^{1/3}M_{\bullet,6}^{1/3}}{R_{\star,12}^{1/3}\mathcal{P}_{\rm QPE,4}^{1/4}}, 
\label{eq:TempPKBB}
\end{eqnarray}
where again physical quantities are normalized to values motivated in Sec.~\ref{sec:LM23}.

The right panel of Fig.~\ref{fig:LC:norm} shows the observed temperature evolution corresponding to luminosities on the left panel. The initial rapid rise corresponds to shock breakout, with the peak temperature reflecting the efficiency of photon production within the immediate downstream, i.e. within $\tau \sim c/v_{\rm d} \approx 7c/v_{\rm u}$ from the shock front from which photons are able to diffuse into the shock transition layer. The subsequent rapid decline in temperature arises due to (1) adiabatic losses due to the rapid expansion of the overpressured shell after breakout, especially of the outer layers which are rapidly accelerated to $v \gtrsim v_{\rm coll}$ at the expense of radiation (see Figs.~\ref{fig:bo:vel_dens}  and \ref{fig:bo:densrad}), (2) sampling of the temperature structure in the immediate downstream established by photon production before breakout (Fig.~\ref{fig:1Dphprod}), and (3) moderate residual photon production during the decompression/acceleration phase.

This early stage is followed by a slower approximately powerlaw decay phase with indices spanning the range $\alpha \approx 1.1 - 1.9$ across different models. When thermalization is efficient (e.g., the $v = 0.1c$ model shown by magenta lines in Fig.~\ref{fig:LC:norm}), one has $\alpha = \alpha_{\rm BB} \approx (\alpha_{\rm t} + 3)/4 \approx 1$. By contrast, in the photon-starved cases the temperature starts out higher, but the decline is steeper and reflects the diffusion surface sampling successively deeper layers of material with different photon to baryon density ratios. In the idealized case of a uniform ejecta cloud (Sec.~\ref{sec:expansion}) negligible photon production occurs in the later stages. However, in the more physical stratified case considered here, the photon to baryon ratio is determined predominantly prior to breakout when the matter is densest; the observed radiation temperature of the flare at any time is determined by this ratio mapped from the postshock (mass) coordinate to the instantaneous location of the diffusion surface, accounting for adiabatic losses due to expansion in the interim. This phase ends near $\tdiff$ when photons start diffusing through the entire ejecta, and the decay index changes to $\alpha \approx 1$, reflecting the adiabatic losses of the photons still remaining within the cloud.

 When photon production is not rapid enough to achieve thermal equilibrium in any ejecta layer, the observed temperature remains above $T_{\rm BB}$ at all stages of evolution (e.g., \citealt{Weaver76,Katz+10,Nakar&Sari10}), i.e. $\Upsilon \equiv \theta_{\star}/\thetaBB \gg 1$ (Eq.~\eqref{eq:thetaratio}).
 In this regime, it is worth comparing the results of the full model (Fig.~\ref{fig:LC:norm}) with our earlier estimate for a uniformly expanding ejecta (Eq.~\eqref{eq:thetaratio}),
 which will highlight the necessity for understanding and properly modeling the early RMS and post breakout phases of the event.

Taking $v=0.2c$ and $\Mej \approx 2.5 \times 10^{-6} M_{\odot}$ appropriate for the model shown by red lines in Fig.~\ref{fig:LC:norm}, one obtains $\Upsilon \approx 11$ (Eq.~\eqref{eq:thetaratio}) and $kT_{\rm esc}(t_{\rm diff}) \approx 200$~eV (Eqs.~\eqref{eq:thetaBB}, \eqref{eq:thetaratio}),
predicting a strongly photon-starved regime. However, the above estimates are several times above the value $\Upsilon \sim 2$ obtained within the full model (e.g. by comparing the solid and dashed lines at $\tdiff$).
This result is corroborated by the analytical results in Fig~\ref{fig:1Dphprod}, which for the above model (with upstream density $n_{\rm b, u} = \Sigma/\mpr h \approx 10^{15}$~cm and $v_{\rm u} = 7v/6 \approx 0.23 \, c$) would predict complete thermalization within $\tau \lesssim 900$ downstream of the shock, whereas in the numerical model the shocked shell has $\tau \approx 500$ and thermalization is thus only marginally not achieved.

The main reason for this discrepancy lies in the fact that
in the full model, photon production within a given fluid element begins as soon as it crosses the shock and proceeds while the shell is still being collected (i.e., before breakout). This phase is absent in the simple spherical expansion model and in the case $h/R_{\star} \sim 1$ contributes an extra dynamical time $\sim R_{\star}/v$ for accumulating photons at constant (matter) density (in contrast, in the spherical cloud scenario, density halves already at $t/t_0 \approx 2^{1/3}$, i.e. after $\approx 0.26$ dynamical times).
Furthermore, immediately after breakout, the deeper layers of the geometrically thin shocked shell that become transparent near $\tdiff$ initially expand somewhat slower than $\rho \propto t^{-3}$ (appropriate for the spherical cloud). Finally, the photon production near the RMS is not strictly local within a co-moving fluid element, being influenced by radiation diffusing from further downstream due to the photon density gradient, which has the effect of lowering the local temperature (and hence speeding up photon production, see Eq.~\eqref{eq:nff}).

Motivated by our finding that most photon production occurs on a timescale $t_{\rm cross} \sim h/v_{\star}$ of the midplane crossing, a more accurate--if still imprecise--estimate for the emission temperature ratio $\Upsilon \equiv \theta_{\star}/\thetaBB$ (Eq.~\eqref{eq:thetaratio}) can be obtained by computing the number of accumulated photons for a fluid element in the shock downstream before breakout. Neglecting diffusion for this simple estimate,
 Eq.~\eqref{eq:pprod} can be cast in the form
\begin{align}
\nph^{-1/2} \frac{d\nph}{dt} = \frac{\dot{n}_{\rm ff}}{\nph^{1/2}},
\end{align}
where right hand side is independent of $\nph$ (since for free-free emission $\dot{n}_{\rm ff} \propto \theta^{-1/2} \propto (\uph/3\nph)^{-1/2}$ and $\uph$ is known from the shock jump conditions). Integrating over $t_{\rm cross}$ and neglecting the number of photons advected from the upsteam, one obtains
\begin{align}
    \frac{\nph}{n_{\rm BB}} = \frac{1}{4} \left[\frac{\dot{n}_{\rm ff}(\TBB) \, t_{\rm cross}}{n_{\rm BB}}\right]^2 \equiv \frac{1}{4\eta_{\star}^2},
\end{align}
or $\Upsilon = 4 \eta_{\star}^2$, where $\eta_{\star}$ is analogous to parameter $\eta$ in LM23 (their Eq.~20). Normalizing to the parameters of our fiducial simulation, one obtains
\begin{align}
    \Upsilon &= \frac{\theta_{\star}}{\thetaBB} = 2.8 \, h_{12}^{1/4} \left( \frac{\Sigma}{1.5\times 10^3 \, \mbox{g}\, \mbox{cm}^{-2}} \right)^{-9/4} \left( \frac{v_{\rm coll}}{0.2c} \right)^{11/2} \nonumber \\
    &\times \left(\frac{\gff}{2}\right)^{-2}\left(\frac{\overline{E}_1}{3}\right)^{-2}
    \approx 0.9 \,  \alpha_{-1}^{9/4} \dot{m}^{5/2} M_{\bullet,6}^{13/3} P_{\rm QPE,4}^{-49/12},
    \label{eq:UpsSh}
\end{align}
where we have used the Rankine-Hugoniot shock jump conditions \eqref{eq:RHu} and \eqref{eq:RHrho} for downstream fluid quantities, and have chosen $\overline{E}_1(x_{\rm min}/\theta) \approx 3$ appropriate for the postshock conditions. This estimate agrees reasonably well with the full model (where $\Upsilon \approx 2$); a somewhat higher value is to be expected as Eq.~\eqref{eq:UpsSh} neglects the residual photon production in the expanding phase.  

In the limit $\theta_{\star} \gg \thetaBB$ ($\Upsilon \gg 1)$, the emission temperature at the diffusion timescale $t = \tdiff$ can thus be estimated as:
\begin{align}
kT_{\rm esc}(t_{\rm diff})  &\approx 2.1\,{\rm keV} \, R_{\star,12}^5 M_{\rm ej,-6}^{-5/2}\left(\frac{h}{R_{\star}} \right)^{1/3}
\nonumber \\
&\times \left( \frac{v_{\rm coll}}{0.2c} \right)^{11/2} 
\left(\frac{\gff}{2}\right)^{-2}\left(\frac{\overline{E}_1}{3}\right)^{-2}
\label{eq:TempPK}
\end{align}
The parameter dependencies of Eq.~\eqref{eq:TempPK} agree with a similar estimate performed in \citetalias{Linial&Metzger23} (their Eq.~21), but our normalization disagrees significantly with theirs. This is in part because \citetalias{Linial&Metzger23} underestimate the shock-crossing time, taking $t_{\rm cross} = (h/7)/v_{\star}$ instead of $h/v_{\star}$ and make different assumptions for $\gff$ and $\overline{E}_1$.

\subsection{Application to Observed QPE Flares}
\label{sec:QPE}

\begin{figure}
\begin{center}
\includegraphics[width=0.48\textwidth]{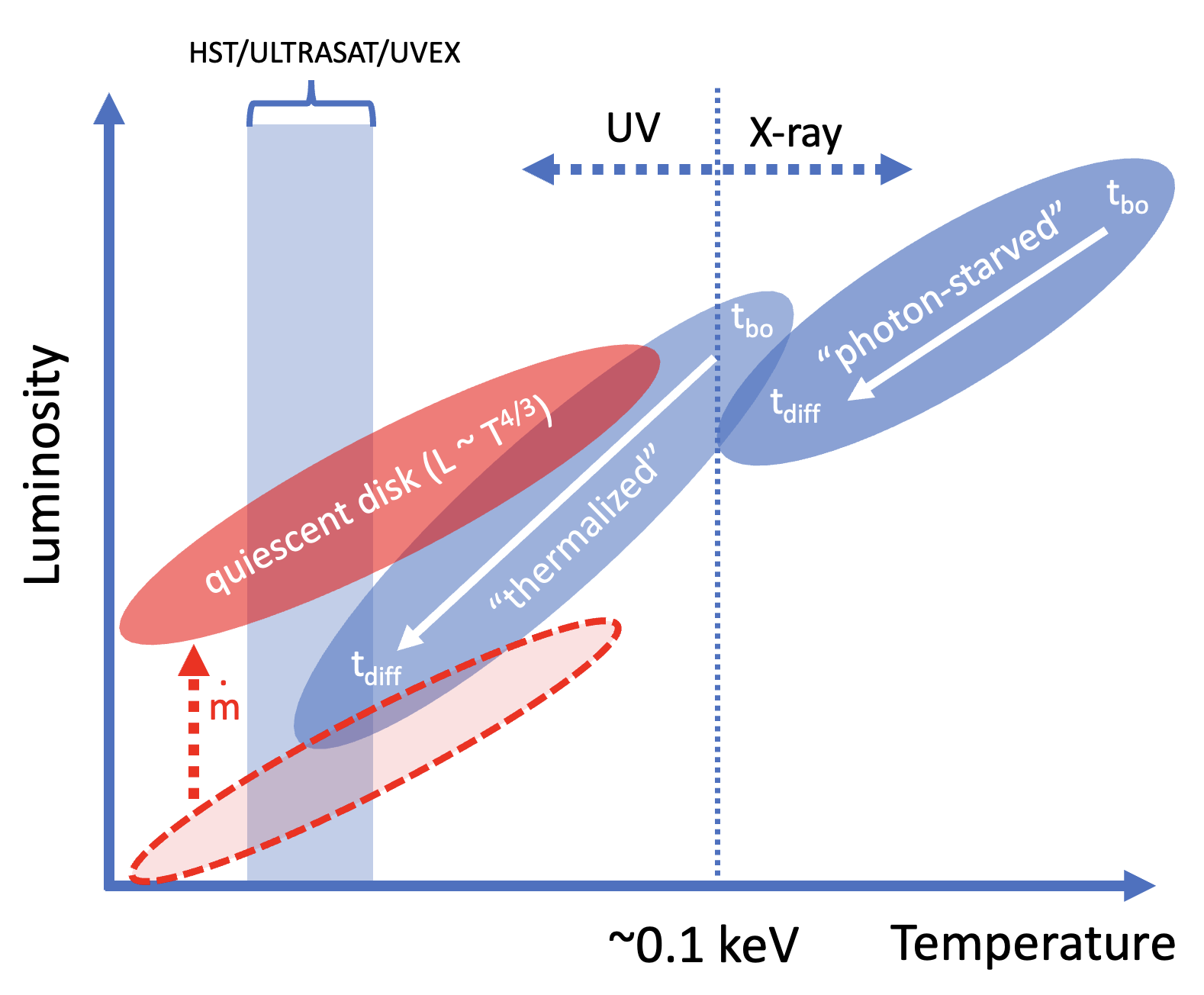}
\end{center}
\caption{Schematic illustration of two scenarios for X-ray QPE emission in the space of eruption luminosity vs. temperature (Sec.~\ref{sec:QPE}). A given eruption evolves in time from the high luminosity and high temperature (upper right) at shock break-out ($t \approx t_{\rm bo}$) to lower luminosity and temperature (lower left) on the bulk diffusion time ($t \approx t_{\rm diff}$). In the ``photon-starved'' scenario (Sec.~\ref{sec:photon_starved}) achieved for low-density disks and/or high collision velocities, breakout occurs in the hard X-ray band but most of the total radiated energy takes place in the soft X-ray band, powering QPE emission. By contrast, in the ``thermalized'' scenario (Sec.~\ref{sec:thermal_equilibrium}) achieved for high density disks and/or low collision velocities, the break-out phase takes place in the soft X-ray band and generates the observed eruption, but the bulk of the emission occurs in the UV band and goes unobserved. Shown for comparison is the temperature/luminosity relation for the quiescent accretion disk emission. In some cases the eruptions may outshine the quiescent disk emission in the UV bands accessible to space telescopes such as HST/ULTRASAT/UVEX (e.g., \citealt{Linial&Metzger24a}).}
\label{fig:schematic_emission}
\end{figure}

Observed QPE flares are characterized by peak luminosities $L_{\rm pk} \approx 10^{42}$ erg s$^{-1}$, peak temperatures $k T_{\rm obs} \approx 100-200$ eV, and durations $t_{\rm pk} \approx 0.3-3$ hr (e.g., \citealt{Miniutti+19,Arcodia+21,Arcodia+22}). As illustrated by two branches in Figure~\ref{fig:kT_iso}, there are two possible regimes of star-disk interactions potentially capable of explaining these properties (see Fig.~\ref{fig:schematic_emission} for a schematic illustration).

\subsubsection{Photon Starved Scenario}
\label{sec:photon_starved}

\begin{figure*}[h]
\begin{center}
\includegraphics[width=0.49\textwidth]{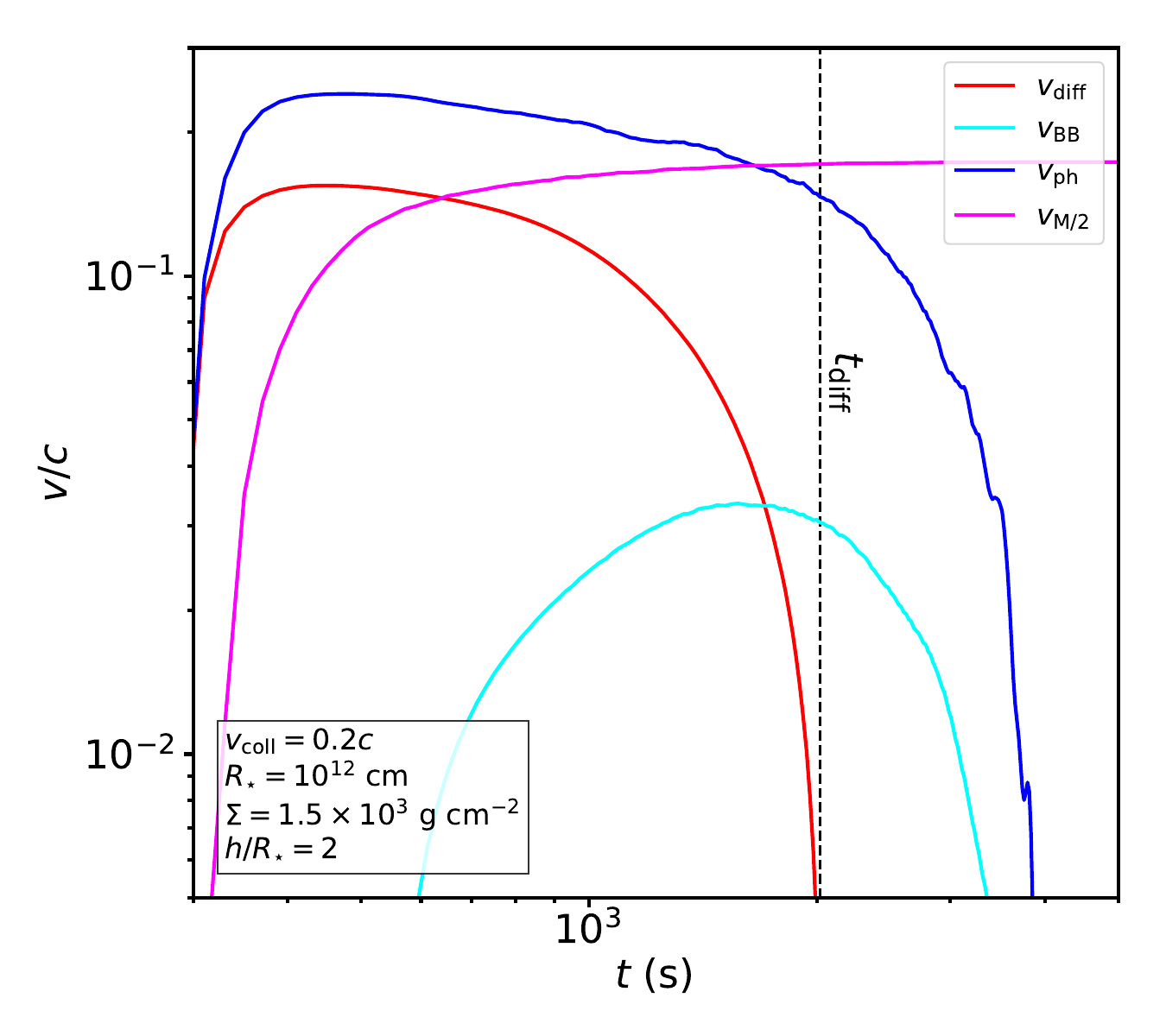}
\includegraphics[width=0.49\textwidth]{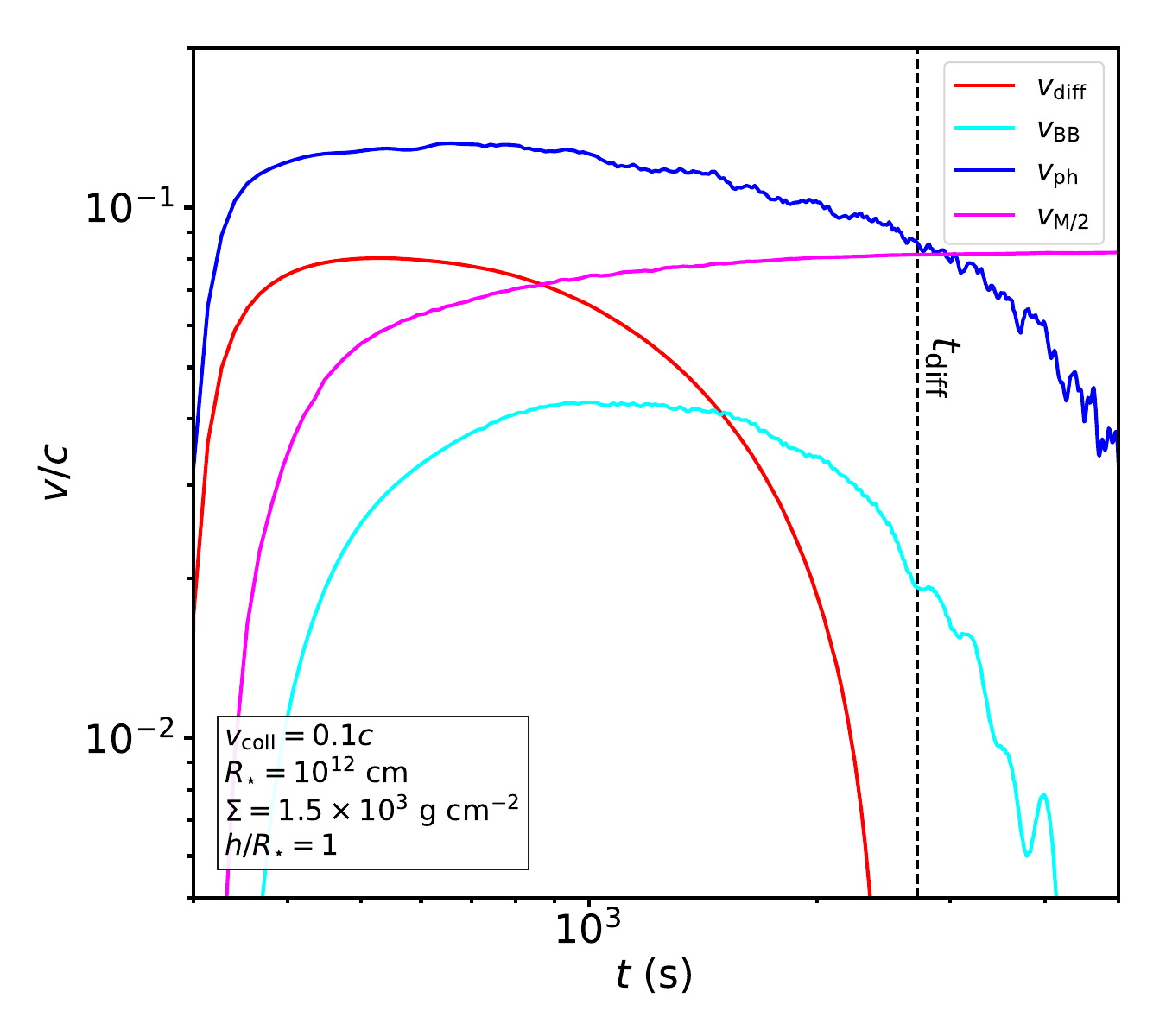}
\end{center}
\vspace{-5mm}
\caption{Radial velocities of different characteristic surfaces within the ejecta in the photon-starved and (almost) completely thermalized regimes (left and right panels, respectively, corresponding to the middle and bottom left panels of Fig.~\ref{fig:spec:esc2}). The red, blue, magenta and cyan lines correspond to the diffusion surface, Thomson photosphere, half-mass radius, and the inferred blackbody radius, respectively.  The inferred expansion speed of the blackbody radius can underestimate the true ejecta speed by almost an order of magnitude.
}
\label{fig:charR}
\end{figure*}

For low disk densities and/or high shock velocities, the ejecta is photon-starved (the lower region of the $\Sigma-R_{\star}$ parameter space; top panels and the bottom right panel of Fig.~\ref{fig:kT_iso}).
The observed soft X-ray band emission in this case can occur close to the diffusion timescale $t_{\rm diff}$, over which most of the total fluence is emitted, similar to what sets the duration of supernova light curves.  However, because of photon starvation the emission temperature is substantially higher than if radiation and matter were in equilibrium at the same blackbody temperature.  This is the scenario explored by our MCRT simulations in Sec.~\ref{sec:model} and shown to produce flare properties which are quantitatively consistent with observations (see also \citetalias{Linial&Metzger23}).  

Fig.~\ref{fig:loops} shows that the predicted emission during an eruption undergoes counterclockwise loops in hardness-luminosity space, qualitatively similar to observed QPE flares. In the quiescent state prior to an eruption, the emission is dominated by the accretion disk, which is soft with a low luminosity in the 0.1-2 keV band (lower left hand region of Fig.~\ref{fig:loops}).  As the eruption starts, the transient suddenly brightens and hardens. 
The peak temperature is achieved somewhat ahead of the peak luminosity, also consistent with observed QPE eruptions (e.g., \citealt{Arcodia+22}).  As shown in Fig.~\ref{fig:spec:esc2}, the spectral energy distribution exhibits an overall thermal-like shape, but is somewhat flatter than a single-temperature blackbody spectrum below the peak.  

One potential challenge to the photon-starved scenario for QPEs is the strong sensitivity of the emission temperature to the parameters of the system, particularly the density of the disk and the velocity of the star (Fig.~\ref{fig:kT_iso}), compared to the relatively narrow range of inferred eruption temperatures $kT_{\rm obs} \approx 100-200$ eV.  However, this apparent ``fine tuning'' is alleviated by the fact that the transient flare is predicted to sweep down across a wide range of emission temperatures following the shock break-out (see Figs.~\ref{fig:sh:temp} and \ref{fig:LC:norm}). 

At select times during the eruption, we mark in Fig.~\ref{fig:loops} the best-fit blackbody temperature one would attribute to the theoretically predicted spectra, which importantly does not generally match the blackbody temperature one would infer from the hardness ratio alone (shown along the top horizontal axis of Fig.~\ref{fig:loops}). Although the earliest phases of the eruption immediately after shock breakout are very hard (e.g. $kT_{\rm fit} = 9$ keV at $t = 100$ s), the hardness-inferred temperature remains much softer, $kT_{\rm Q} \approx 200$ eV.  Including also emission from the quiescent disk (shown as a red curve in Fig.~\ref{fig:loops} for a SMBH accretion rate $\dot{m} = 0.1$) further decreases the inferred temperature. By the peak of the eruption at $t = 300$ s (as defined by the $0.1-2$ keV luminosity) the eruption emission is cooler and the best-fit spectral temperature is $kT_{\rm fit} \approx 250$ eV, similar to the inferred temperatures of QPE flares. 


We speculate that the reason why most QPE eruptions are inferred to possess temperatures just a factor of a few times hotter than that of the quiescent disk emission may at least in part be an artefact fitting the eruption to a single-temperature blackbody. In reality, the Comptonized Wien spectrum of the flare can deviate substantially from the blackbody shape, particularly below the peak.  Furthermore, any contribution to the flux from the quiescent disk emission (if not perfectly subtracted off) would also reduce the inferred eruption temperature, particularly during the early rise of the flare.\footnote{The quiescent disk emission can in principle vary on timescales as short as the light-crossing time near the ISCO radius of the SMBH (typically seconds), and hence may not be perfectly constant over the eruption duration.} 
 Also keep in mind that our 1D model, which by construction predicts a single break-out time across the surface of the star and assumes a disk with an infinitely sharp edge, is an idealization to the true break-out geometry. In reality, breakout will take place at different times across the disk surface and into the more vertically-extended atmosphere of the disk.\footnote{Similar to the way that shock break-out through the turbulent atmosphere of a red supergiant differs substantially from a spherically-symmetric star with a single well-defined radius (\citealt{Goldberg+22}).} This more realistic scenario will ``smear out'' the break-out signal temporally compared to our predictions. In particular, soft emission from regions of the disk surface where break-out occurs first may precede the hardest emission generated by the bulk of the break-out, thus also softening the early rise of the eruption relative to our predictions.

In principle the spectral evolution of the eruption can be used to diagnose the kinematics of the ejecta cloud (e.g., \citealt{Franchini+23,Chakraborty+24}), for example through the evolution of the inferred blackbody radius $R_{\rm BB} \equiv (L_{\rm BB}/4\pi T_{\rm BB}^{4}\sigma)^{1/4}$ based on the best-fit blackbody luminosity $L_{\rm BB}$ and temperature $T_{\rm BB}$.  Figure~\ref{fig:charR} shows the time evolution of various velocity scales within the ejecta (Sec.~\ref{sec:expansion}). In both the photon-starved and thermal equilibrium regimes, the velocity one would infer from the derivative of the blackbody radius $v_{\rm BB} = dR_{\rm BB}/dt$, substantially underestimates the true velocity of either the ejecta or the photosphere (defined as the location where $\tau = 1)$.  As discussed in Appendix~\ref{sec:app:Rbb}, $R_{\rm BB}(t)$ underestimates the true ejecta radius because (1) photon starvation results in a higher temperature at a given luminosity than predicted for a Planckian spectrum; (2) even when gas and radiation are in equilibrium, an electron scattering atmosphere deviates from a pure blackbody (a similar physical effect necessitates the ``hardening factor'' applied to the interpretation of accretion disk spectra; e.g., \citealt{Shimura&Takahara95}). This strongly cautions against interpreting the low $v_{\rm BB}$ values measured from some QPE eruptions to deduce low star-disk collision speeds \citep{Franchini+23,Miniutti+23b,Chakraborty+24}.

\subsubsection{Thermalized Scenario}
\label{sec:thermal_equilibrium}

If the gas and radiation are in equilibrium, the temperature with which the bulk of the shocked disk material radiates (on the timescale $t_{\rm diff}$) is too low to explain QPE emission (e.g., \citetalias{Linial&Metzger23}).  However, the outer ejecta layers which emit first after shock break-out are more compact and experience fewer adiabatic losses before radiating, thus producing substantially harder emission even if those layers enter thermal equilibrium.  In such a scenario, the bulk of the total radiated energy still occurs in the (unobserved) far-UV band, with X-ray instruments only detecting the earliest hottest phase of the eruption.  In exact opposite to the photon starved scenario described above, this ``thermal shock break-out'' scenario for QPE emission instead favors a {\it high} surface density and a {\it low} collision speed.  

Quantitatively, in the case of thermal equilibrium, the observed temperature is the blackbody spectrum, but as observed at times $t_0 \ll t_{\rm diff}$.  In this scenario, the luminosity, temperature, and flare duration obeys: 
\be
L_{\rm QPE} \approx L_{\rm esc}(t_{0}) \approx 1\times 10^{43}\,{\rm erg\,s^{-1}}M_{\rm ej,-6}^{1/2}v_{-1}^{5/2},
\ee
\be
kT_{\rm QPE} \approx kT_{\rm esc}(t_0) \approx 64\,{\rm eV}\,M_{\rm ej,-6}^{1/4}v_{-1}^{1/2}R_{0,12}^{-3/4}
\ee
\be
t_{\rm QPE} \approx t_0 \approx 3.3\times 10^{2}\,{\rm s}\,R_{0,12}\,v_{-1}^{-1},
\ee

These relationships make clear that to simultaneously produce QPE luminosities $L_{\rm QPE} \sim 10^{42}$ erg/s, temperatures $kT_{\rm QPE} \approx 100-200$ eV while also explaining their long timescales $t_0 \gtrsim 10^{3}$ s, one requires small collision speeds $v_{\rm coll} < 0.03c$, very large ejecta masses $M_{\rm ej} \gtrsim 10^{-5}M_{\odot}$, and very large stellar radii $R_{\star} \gtrsim 50R_{\odot}$. This regime corresponds to the right side of the lower-left hand panel in Fig.~\ref{fig:kT_iso}.  Although such large radii are in tension with the expected size of main-sequence stars, or potentially even their Hills radii (Eq.~\eqref{eq:rH}), they may be consistent with the cross section of the colliding body being dominated by a comet-like stream of debris stripped from the (much smaller) star in previous disk encounters, instead of the stellar surface itself.  In such a scenario, the QPE duration can also be set by the finite spread in the stream arrival time instead of the break-out time (see \citealt{Yao+24} for discussion).

\subsection{Other Predictions}

Across much of the parameter space of star-disk collisions, particularly for small stellar radii and large disk surface densities, the predicted emission peaks at temperatures below the X-ray band, in the ultraviolet $kT_{\rm obs} \lesssim 30$ eV (Fig.~\ref{fig:kT_iso}).  Provided that the eruptions are also sufficiently luminous to outshine the quiescent disk emission at these lower frequencies (see Fig.~\ref{fig:schematic_emission}), this opens the possibility of QPE-like periodic flaring phenomena but manifesting in the UV band (\citealt{Linial&Metzger24a}).  Such ``UV QPEs'' present promising targets for wide-field impending satellite missions such as ULTRASAT \citep{Sagiv+14} or UVEX \citep{Kulkarni+21}.  

If the properties of either the star's orbit or the accretion disk evolve with time, this can imprint secular timescale changes in the eruption properties.  For example, growing evidence hints at a connection between QPEs and TDEs (e.g., \citealt{Miniutti+19,Chakraborty+21,Quintin+23,Nicholl+24}), in which case the surface density of the gaseous disk fed by the disrupted star is generally expected to drop in time as the disk accretes and its outer edge viscously spreads outwards in radius (e.g., \citealt{Mummery+24}).  Interestingly, the non-monotonic evolution of $kT_{\rm obs}$ with decreasing $\Sigma$ (at fixed $R_{\star}$) suggests that X-ray QPEs generated when $\Sigma$ is maximal (via the shock break-out scenario; Sec.~\ref{sec:thermal_equilibrium}) could in principle disappear as $\Sigma$ drops, only to reappear again once the collisions enter the photon-starved regime at low-$\Sigma$ (Sec.~\ref{sec:photon_starved}).

\section{Conclusions}
\label{sec:conclusions}

We have presented one-dimensional Monte Carlo radiation hydrodynamic simulations of the shock interaction between a star passing supersonically through the midplane of a SMBH accretion disk, which self-consistently follows the formation of a radiation-mediated shock, photon production via free-free emission, Comptonization and  the eventual escape of radiation.  Our main conclusions are summarized as follows: 

\begin{itemize}
    \item If the disk is thin relative to the star, $h \lesssim few \times R_{\star}$, the shocked disk material does not have time to flow around the star before the star emerges from the midplane.  In this limit, on which our modeling focuses, the star-disk interaction can be approximately described in one spherical dimension and occurs in three phases (Figs.~\ref{fig:schematic}, \ref{fig:bo:vel_dens}, \ref{fig:bo:densrad}): (1) passage of the star through the midplane, during which the disk material passes through a radiation-mediated shock, collecting in a thin ``cap'' ahead of the star of roughly constant density; (2) emergence of the star from the disk, and resulting shock acceleration down the disk's vertical density gradient and associated shock break-out from the photosphere; (3) expansion of the shocked material into a quasi-spherical ejecta cloud, which subsequently decompresses and enables a fraction of the shock-deposited energy to escape as radiation, generating a transient UV/X-ray flare (a single QPE eruption).  

    A similar sequence of events will take place if the disk is thicker $h \gtrsim few \times R_{\star}$.  However, in this case a greater fraction of the shocked disk material will have time to flow around the star, allowing this high pressure gas to mix/entrain more disk material and to escape also through the ingress side of the disk (similar to the ejecta geometry from black hole-disk collisions; e.g., \citealt{Ivanov+98}).  Such a configuration of two diametrically opposite plumes of hot gas, enables emission which may be more clearly visible from both sides of the optically-thick disk and hence is capable of generating the twice-per-orbit eruptions for typical viewing angles.  
    
    \item 
    
In general gas and radiation in the shocked ejecta will not be in equilibrium, as there may not be sufficient time to generate the photons required for a blackbody spectrum. At the typically high shock velocities and temperatures of interest, free-free processes dominate the photon production. Although the shocked material spends most of its optically-thick evolution expanding to large radii, the total photon yield is dominated by the comparatively short time the shocked matter spends at the highest densities when the star is still passing through the disk midplane. Photon production behind the effectively one-dimensional radiation-mediated shock can be largely understood analytically (Appendix \ref{sec:app:sh}).

    \item Photons which first escape at shock break-out are typically the hardest and most luminous, but as time goes on deeper and deeper layers contribute to the escaping emission, which softens in time (Figs.~\ref{fig:spec:esc}, \ref{fig:spec:esc2}). The bulk of the total radiated energy comes from the greatest depths which expand at roughly the original star/shock speed and emit over the bulk diffusion timescale $t_{\rm diff}$ (Eq.~\eqref{eq:tdiff}).  
    Similar to how supernovae peak faster in bluer bands than in redder bands, the light curves at higher X-ray energies are narrower and peak earlier than at lower X-ray energies (Fig.~\ref{fig:lc:bands}).  When full thermalization is not established, the escaping radiation exhibits a partially Comptonized spectrum, characterized by a power-law shape $L_{\nu} \propto \nu^{\beta}$ with $\beta \approx 1$ below a Wien-like exponential cutoff above the peak frequency (Fig.~\ref{fig:spec:esc2}). 
    
    Simple semi-analytic descriptions for the luminosity and temperature evolution are not trivial to construct, because different ejecta layers experience not only different initial thermal content and different adiabatic losses but also different photon production yields, depending on their thermodynamic expansion histories (Figs.~\ref{fig:sh:temp}, \ref{fig:LC:norm}). The outer ejecta layers are typically more photon-starved, causing the escaping temperature to drop faster than the naive $T_{\rm obs} \propto t^{-1}$ prediction for a fully thermalized ejecta cloud. The dynamics of the shock break-out process is also sensitive to the disk's vertical density profile, which we approximate for simplicity as being constant with a sharp cut-off above the assumed scale-height.  
    
    Also as a result of both photon starvation and the scattering-dominated opacities, the blackbody radius one derives from the luminosity and best-fit spectral temperature can under-estimate the true photosphere radius by almost an order of magnitude (Fig.~\ref{fig:charR}, Appendix \ref{sec:app:Rbb}). This cautions against using expansion rates inferred from time-resolved blackbody spectral fits to infer ejecta speeds and hence the dynamics of the star-disk collision.

    \item The predicted eruptions execute counter-clockwise loops in hardness-luminosity space (Fig.~\ref{fig:loops}), with the hardness peaking somewhat ahead of the total count rate/luminosity, similar to observed QPE sources. 
    Although our idealized set-up (1D spherically-symmetric geometry, sharp disk edge) predicts a very rapid rise to the eruption, several physical effects not captured by our model, particularly non-simultaneous break-out across the disk surface, will likely act soften the rise phase relative to our predictions. We also caution that a single-temperature blackbody is not a good fit to the true eruption spectra in the photon-starved case, particularly below the peak. Even though the peak of the eruption spectrum sweeps down across a wide range of photon energies with time (Fig.~\ref{fig:spec:esc2}), we speculate that this may bias observed QPE temperatures to be clustered around values $kT_{\rm obs} \approx 100-200$ eV, just a factor of a few above the quiescent disk temperature. 

    \item The dual requirements to produce eruption temperatures $kT_{\rm obs} \gtrsim 100$ eV and durations $t_{\rm obs} \gtrsim 1000$ s compatible with QPE observations limits the allowed parameter space of disk-star collision speed, disk surface density, and the stellar radius (Fig.~\ref{fig:kT_iso}).  In photon-starved scenarios for QPE emission high collision velocities and low surface densities are favored (Sec.~\ref{sec:photon_starved}).  Alternatively, if QPE eruptions arise from thermalized radiation but represent just the early shock break-out phase, then low collision velocities and high surface densities are instead required (Sec.~\ref{sec:thermal_equilibrium}). In this latter case in particular, the bulk of the emitted fluence would occur in the UV instead of the X-ray band.   
    
    In both photon-starved and thermal break-out scenarios, large stellar radii $R_{\star} \gtrsim 10-50R_{\odot}$ are required to match QPE observations. This naively appears to disfavor low-mass main sequence stars as the colliding bodies, despite their likelihood to be the most abundant in galactic nuclei from the initial mass function. However, large stellar radii could be expected if the stars' atmospheres are inflated by shock heating from repeated midplane passages, or if the colliding body is not a star at all, but rather a ballistic stream of dense matter stripped from the star over previous orbits (e.g., \citealt{Yao+24}). On the other hand, if the colliding body were a black hole instead of a star, the high required cross section (in this case, set by the Bondi radius, $R_{\rm B} \simeq G m_{\bullet}/v_{\rm coll}^{2} \simeq 2R_{\odot}(m_{\bullet}/10^{4}M_{\odot})(v_{\rm coll}/0.1c)^{-2}$) would necessitate an intermediate-mass black hole $m_{\bullet} \gtrsim 10^{4}M_{\odot}$; however, such massive EMRIs are disfavored as contributing the majority of QPE sources by rate arguments (see \citetalias{Linial&Metzger23}).

    \item Future work should target two- or three-dimensional radiation hydrodynamic simulations, which allow for a range of disk-star collision angles and explore more realistic disk density profiles (which may be modified due to heating from the repeated star collisions; e.g., \citealt{Linial&Metzger24a}).  Despite these important details, we expect that many of the relevant hydrodynamic and radiative processes are captured at least semi-quantitatively by the 1D spherical model presented here. Likewise, our model should be broadly applicable also to collisions with the disk of a body without a solid surface, such as a black hole; however, the details of the shock compression during the midplane crossing phase, and hence of the photon production, as well as the dynamics of the re-expansion, are likely to differ than the stellar collision case. Yet another challenge is to understand the sometimes complex long-term evolution of the QPE emission and timing properties, which may arise due to secular evolution of the disk surface density or its inclination with respect to the stellar orbit; the latter can arise from general relativistic precession of the stellar orbit and/or precession or warping of the disk.
    
\end{itemize}

\begin{acknowledgments}
We thank Riccardo Arcodia and Joheen Chakraborty for helpful comments on an early version of the manuscript. IV acknowledges support by the ETAg grants PRG2159 and PRG1006. This work was partially supported by the ETAg CoE grant ``Foundations of the Universe" TK202 and the European Union's Horizon Europe research and innovation programme (EXCOSM, grant No. 101159513).
    IL acknowledges support from a Rothschild Fellowship and The Gruber Foundation. BDM was supported in part by the National Science Foundation (grant No. AST-2009255). The Flatiron Institute is supported by the Simons Foundation.
\end{acknowledgments}

\appendix

\section{Structure of the radiation-mediated shock}

\label{sec:app:sh}

The conservation laws for fluid quantities in a steady-state radiation-dominated flow are
\begin{align}
\rho v &= \rho_{\rm u} v_{\rm u}  
\label{eq:cont} \\
\rho v^2 + P &= \rho_{\rm u} v_{\rm u}^2 + P_{\rm u}  
\label{eq:mom}\\
\left( \frac{\rho v^2}{2} + \frac{\gamma}{\gamma-1} P\right) v - D\frac{du}{dx} &= \left( \frac{\rho_{\rm u} v_{\rm u}^2}{2} + \frac{\gamma}{\gamma-1} P_{\rm u}\right) v_{\rm u},
\label{eq:en}
\end{align}
where $\gamma = 4/3$ is the adiabatic index and subscript u refers to quantities in the far upstream\footnote{We note that the above equations neglect radiative viscosity. Comparison of the analytical and numerical solutions suggests that the error made by this is relatively minor.}. The radiative diffusion coefficient is defined as (assuming Compton-dominated opacity)
\begin{align}
D = \frac{c\lambda}{3} = \frac{c}{3\sigmaT n_{\rm e}},
\end{align}
where $\lambda$ is the photon mean free path and $n_{\rm e}$ is the electron density.

It is convenient to scale the quantities in Eqs.~(\ref{eq:cont})-(\ref{eq:en}) to their values in the far upstream by defining dimensionless mass density, velocity and energy density as  $\tilde{\rho} = \rho/\rho_{\rm u}$, $\tilde{v} = v/v_{\rm u}$ and $\tilde{u} = u/u_{\rm u}$, respectively, where $u = P/(\gamma - 1)$ is the (radiation) energy density. The conservation laws can now be written as
\begin{align}
\tilde{\rho}\tilde{v} &= 1 
\label{eq:cont2} \\
\tilde{v} + \frac{\tilde{u}}{\gamma \Msq} &= 1 + \frac{1}{\gamma \Msq}
\label{eq:mom2} \\
\frac{\tilde{v}^2}{2} + \frac{\tilde{v}\tilde{u}}{(\gamma - 1) \Msq} - \frac{\tilde{v} \, d\tilde{u}/d\tau}{3\gamma(\gamma-1) \Msq \beta_{\rm u}} &= \frac{1}{2} + \frac{1}{(\gamma - 1) \Msq},
\label{eq:en2}
\end{align}
where 
\begin{align}
{\cal M}_{\rm u} = \sqrt{\frac{\rho_{\rm u} v_{\rm u}^2}{\gamma(\gamma-1) u_{\rm u}}}
\end{align}
is the upstream Mach number, and we have defined $d\tau = \sigmaT n_{\rm e, u} dx$ (in terms of upstream rather than local electron density).

Substituting $\tilde{v}$ from Eq.~(\ref{eq:mom2}) into Eq.~(\ref{eq:en2}) and rearranging terms, one obtains
\begin{align}
\frac{1}{\beta_{\rm u}}\frac{d\tilde{u}}{d\tau} = -\frac{3 (\gamma + 1) (\tilde{u} - \tilde{u}_{\rm d})(\tilde{u} - 1)}{2(\gamma\Msq + 1 - \tilde{u})},
\label{eq:dudt}
\end{align}
where
\begin{align}
\tilde{u}_{\rm d} = \frac{2\gamma\Msq + 1 - \gamma}{\gamma + 1}
\label{eq:RHu}
\end{align}
is the (relative) energy density in the far downstream, consistent with Rankine-Hugoniot shock jump conditions.

Equation (\ref{eq:dudt}) can be integrated straightforwardly, yielding an implicit algebraic equation for $\tilde{u}$:
\begin{align}
\frac{\tilde{u} - 1}{(\tilde{u}_{\rm d} - \tilde{u})^{1/\tilde{\rho}_{\rm d}}} = C_u\exp\left[ \frac{3(\Msq - 1)}{\Msq} \, \beta_{\rm u} \tau \right],
\label{eq:uanalyt}
\end{align}
where
\begin{align}
\tilde{\rho}_{\rm d} = \frac{\Msq(\gamma + 1)}{\Msq(\gamma-1) + 2}
\label{eq:RHrho}
\end{align}
is the relative density far downstream. The integration constant $C_u$ depends on our choice of origin for the independent variable $\tau$. Setting $\tau = 0$ where $d^2 \tilde{v}/d\tau^2 = 0$ and $d\tilde{v}/d\tau \ne 0$ (i.e. at $\max|\nabla{v}|$), one obtains
\begin{align}
C_u = \left[\frac{\gamma\Msq(\sqrt{\tilde{\rho}_{\rm d}} - 1)}{\tilde{\rho}_{\rm d}}\right]^{1-1/\tilde{\rho}_{\rm d}} \sqrt{\tilde{\rho}_{\rm d}}.
\label{eq:Cu}
\end{align}
The above analytical solution is shown by the red dotted line in Figure~\ref{fig:sh:spec} (right panel).

The corresponding solution for $\tilde{\rho}$ is
\begin{align}
\frac{\tilde{\rho} - 1}{\tilde{\rho}}\left(\frac{\tilde{\rho}}{\tilde{\rho}_{\rm d} - \tilde{\rho}}\right)^{1/\tilde{\rho}_{\rm d}} = C_{\rho} \exp\left[ \frac{3(\Msq - 1)}{\Msq} \, \beta_{\rm u} \tau \right],
\label{eq:vanalyt}
\end{align}
where
\begin{align}
C_{\rho} = \frac{\left(\sqrt{\tilde{\rho}_{\rm d}} - 1\right)^{1-1/\tilde{\rho}_{\rm d}}}{\sqrt{\tilde{\rho}_{\rm d}}}.
\label{eq:Crho}
\end{align}
The corresponding solution for velocity can obtained from $\tilde{\rho}\tilde{v} = 1$.

\section{Free-free photon production in a radiation-mediated shock}
\label{sec:app:pprod}

The advection-diffusion equation for the evolution of the photon number density in a steady-state plane-parallel flow can be written as
\begin{align}
\frac{d}{dx} \left(\nph v - \frac{c}{3\sigmaT n_{\rm e}} \frac{d\nph}{dx} \right) = \dot{n}_{\rm ff},
\label{eq:pprod}
\end{align}
where $\dot{n}_{\rm ff}$ is given by Eq.~(\ref{eq:nff}).
Dividing by the constant $\nb v$, where $\nb = \rho/\mpr$ is the baryon density, and defining a new independent variable as
\begin{align}
dq = \frac{3v}{c\lambda} dx = \frac{3\sigmaT \nel v}{c} dx,
\end{align}
one can cast Eq.~(\ref{eq:pprod}) in the form
\begin{align}
-\frac{d}{dq}\left[ \frac{e^q}{\nb} \, \frac{d}{dq} \left( \nph e^{-q} \right) \right] = 
\frac{c \, \dot{n}_{\rm ff}}{3\sigmaT \nel \nb v^2}.
\label{eq:pprod2}
\end{align}
To the lowest order, the only quantity in Eq.~(\ref{eq:pprod2}) that depends on the number of photons (besides $\nph$ itself) is temperature (within $\dot{n}_{\rm ff}$). Hence we can regard
$\nb = \rho/\mpr$ and $v$ as known (via the solutions in Section~\ref{sec:app:sh}) when integrating the above equation. Furthermore, one can approximately relate temperature to the photon number and the energy densities (the latter is again given in Section~\ref{sec:app:sh}), hence closing Eq.~(\ref{eq:pprod2}).

The formal solution of Eq.~(\ref{eq:pprod2}) in a semi-infinite medium is
\begin{align}
\nph(q) = \frac{n_{\rm ph, u}}{n_{\rm b, u}} \int_q^{\infty} \nb \, e^{-(q^{\prime} - q)} \, dq^{\prime}
+ \frac{1}{n_{\rm b, u} v_{\rm u}} \int_q^{\infty} S(q^{\prime}) \, \nb \, e^{-(q^{\prime} - q)} \, dq^{\prime},
\label{eq:pprod3}
\end{align}
where
\begin{align}
S(q) = \frac{c}{3\sigmaT n_{\rm e, u} v_{\rm u}} \int_{0}^{q} \dot{n}_{\rm ff} \, dq^{\prime} = \int_{0}^{x} \dot{n}_{\rm ff} \, dx^{\prime}
\end{align}
characterizes the cumulative photon production between the upstream boundary and $q$ (or $x$).
Since $\dot{n}_{\rm ff}$ depends on temperature and hence the solution of Eq.~(\ref{eq:pprod3}), one can solve the latter by iterating till convergence.

As it stands, the solution given by (\ref{eq:pprod3}) is valid for any $\nb = \nb(q)$ and relies only on the steady-state condition $\nb v = n_{\rm b, u} v_{\rm u} = \mbox{constant}$.
Coupled with the solution for the radiation-mediated shock structure given by Eqs.~(\ref{eq:uanalyt}) and (\ref{eq:vanalyt}), Eq.~(\ref{eq:pprod3}) yields the radiation density and temperature structure of such shocks.

\begin{figure*}[h]
\begin{center}
\includegraphics[width=0.49\textwidth]{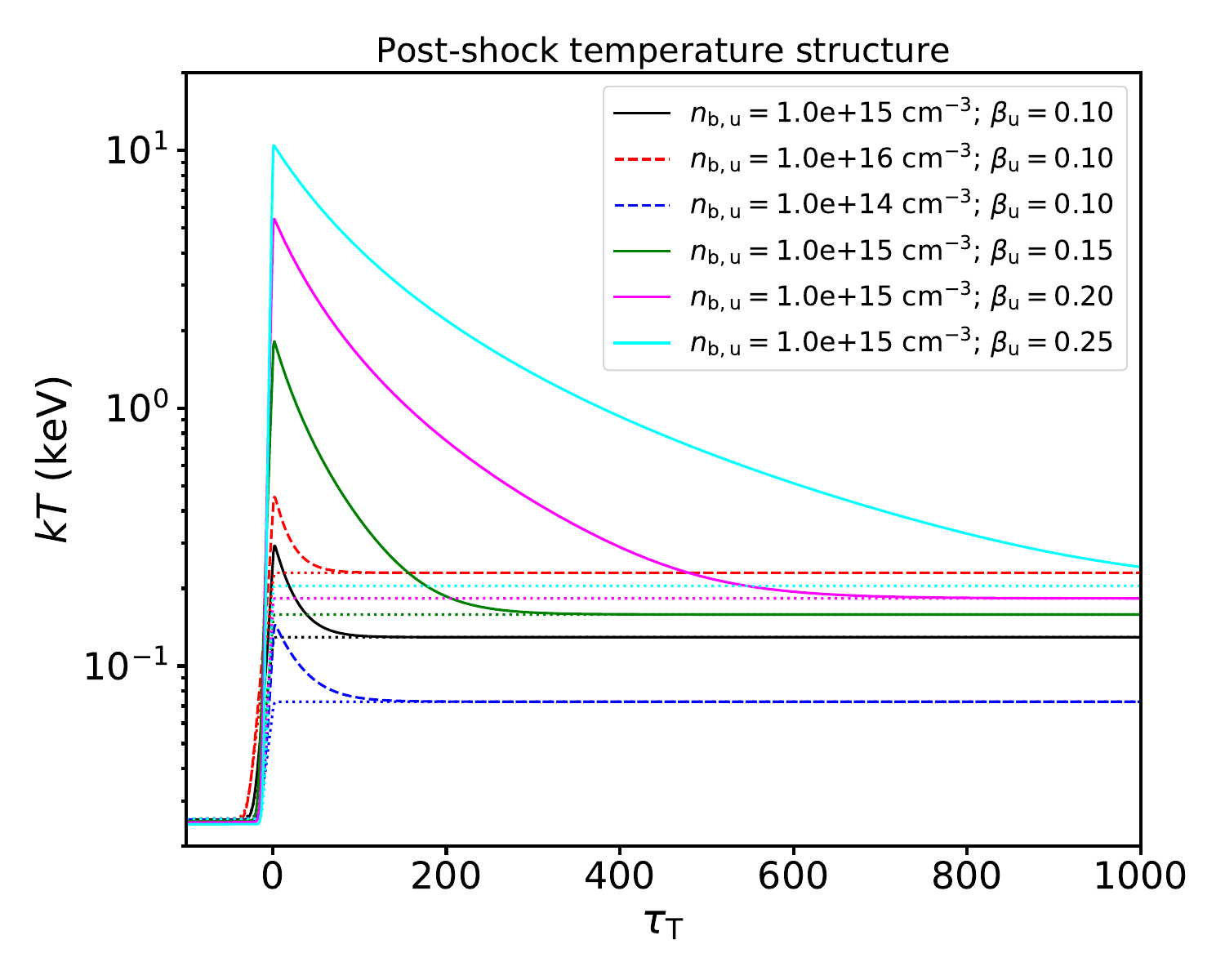}
\includegraphics[width=0.49\textwidth]{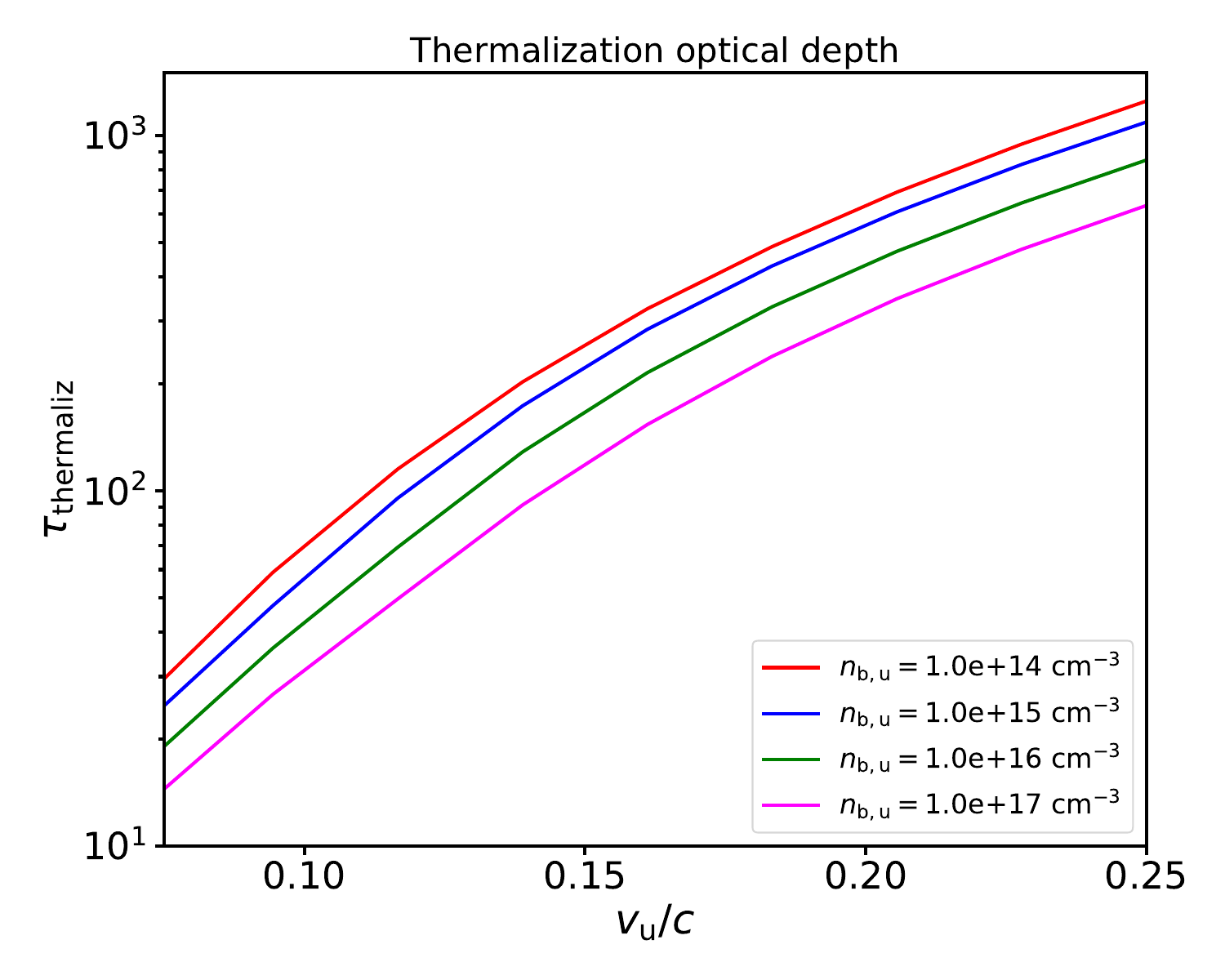}
\end{center}
\vspace{-5mm}
\caption{Left panel: temperature structure downstream of the 1D radiation-mediated shock (left), for different upstream densities and flow velocities (in the shock frame). The dotted lines show the blackbody temperature corresponding to the local radiation energy density. Right panel: Thomson opacity downstream from the shock front where thermalization is achieved. The upstream temperature is fixed at $25$~eV in all cases.}
\label{fig:1Dphprod}
\end{figure*}

Figure~\ref{fig:1Dphprod} illustrates the dependence of the downstream photon production and thermalization efficiency on the upstream density and velocity. The pre-shock flow velocity has a strong impact on thermalization, with an approximate (empirical) scaling $\tau_{\rm thermaliz} \propto \beta_{\rm u}^3$. Somewhat counterintuitively, the upstream density only has a minor effect on the post-shock thermalization, provided one measures the latter as a function of $\tau_T$. This is partly because in Eq.~(\ref{eq:pprod2}) the explicit density-dependence on the RHS cancels out (since $\dot{n}_{\rm ff}\propto \nb^2$). However, note that in the specific application of star-disk collisions, the degree of thermalization experienced by a typical shocked fluid element before it begins to expand out of the disk midplane still depends sensitively on the disk column density $\Sigma \propto \tau_T$.

\section{Inferred blackbody radius in scattering-dominated, photon-starved ejecta}
\label{sec:app:Rbb}

The effective blackbody radius is defined via the relation
\begin{align}
    L_{\rm BB} = 4\pi R_{\rm BB}^2 \sigma T^4 = \pi R_{\rm BB}^2 c u_{\rm BB}(T),
\end{align}
where $T$ is the observed radiation temperature.
On the other hand, the escaping bolometric luminosity from a uniform, spherically expanding and scattering-dominated cloud is given by Eq.~(\ref{eq:Lbol}),
\begin{align}
    L_{\rm esc} = \frac{ 4\pi R^3 \uph \, \xi_{\rm diff}^2}{3\tdiff} = 
    \frac{4\pi R^2}{3} \,  \vej \uph \, \xi_{\rm diff}^2\, \frac{t}{\tdiff},
    \label{eq:app:Lesc}
\end{align}
where we have used $R = \vej t$. Setting $L_{\rm BB} = L_{\rm esc}$ and assuming complete thermalization, $\uph = u_{\rm BB}(T)$, one obtains
\begin{align}
\frac{R_{\rm BB}}{\xi_{\rm diff} R} = \frac{R_{\rm BB}}{R_{\rm diff}} = \frac{2}{\sqrt{3}}  \left(\frac{\vej}{c}\right)^{1/2} \left(\frac{t}{\tdiff}\right)^{1/2}.
\end{align}

In the photon-starved regime one can write
\begin{align}
\uph \approx 2.7kT \, \nph = \frac{\nph}{n_{\rm BB}(T)} \, 2.7kT \, n_{\rm BB}(T) =  \frac{\nph}{n_{\rm BB}(T)} \, u_{\rm BB}(T).
\end{align}
Inserting this into Eq.~(\ref{eq:app:Lesc}) and again setting $L_{\rm BB} = L_{\rm esc}$ yields
\begin{align}
\frac{R_{\rm BB}}{R_{\rm diff}} = \frac{2}{\sqrt{3}} \left(\frac{\vej}{c}\right)^{1/2} \left(\frac{t}{\tdiff}\right)^{1/2} \left(\frac{\nph}{n_{\rm BB}(T)}\right)^{1/2}.
\label{eq:app:RBBs}
\end{align}

It is worth stressing that $n_{\rm BB}(T)$ is the blackbody radiation density that corresponds to the {\it observed} temperature, which is different from the value $n_{\rm BB}(\TBB)$ one would obtain at a given energy density $\uph$ if thermalization was allowed to complete. The two are related as $n_{\rm BB}(\TBB)/n_{\rm BB}(T) = (\TBB/T)^3$, whereby the last factor in Eq.~(\ref{eq:app:RBBs}) becomes
\begin{align}
\left(\frac{\nph}{n_{\rm BB}(T)}\right)^{1/2} = \left(\frac{\TBB}{T}\right)^2 = \Upsilon^{-2},
\end{align}
where we have used $\nph/n_{\rm BB}(\TBB) = \TBB/T$.

\newpage

\bibliography{ms}
\bibliographystyle{aasjournal}

\end{document}